\documentclass{jfm}

\usepackage{graphicx}
\usepackage{natbib}
\usepackage{epstopdf}
\usepackage{amsmath}
\usepackage{xfrac}
\usepackage{multirow}
\usepackage{colortbl}

\usepackage{ amssymb, amsfonts}

\usepackage{xcolor}

\usepackage{subfig}

\ifCUPmtlplainloaded \else
  \checkfont{eurm10}
  \iffontfound
    \IfFileExists{upmath.sty}
      {\typeout{^^JFound AMS Euler Roman fonts on the system,
                   using the 'upmath' package.^^J}%
       \usepackage{upmath}}
      {\typeout{^^JFound AMS Euler Roman fonts on the system, but you
                   dont seem to have the}%
       \typeout{'upmath' package installed. JFM.cls can take advantage
                 of these fonts,^^Jif you use 'upmath' package.^^J}%
      }
  \else
  \fi
\fi

\ifCUPmtlplainloaded \else
  \checkfont{msam10}
  \iffontfound
    \IfFileExists{amssymb.sty}
      {\typeout{^^JFound AMS Symbol fonts on the system, using the
                'amssymb' package.^^J}%
       \usepackage{amssymb}%
       \let\le=\leqslant  
       \let\ge=\geqslant  
      }{}
  \fi
\fi

\ifCUPmtlplainloaded \else
  \IfFileExists{amsbsy.sty}
    {\typeout{^^JFound the 'amsbsy' package on the system, using it.^^J}%
     \usepackage{amsbsy}}
    {\providecommand\boldsymbol[1]{\mbox{\boldmath $##1$}}}
\fi

\newsavebox{\astrutbox}
\sbox{\astrutbox}{\rule[-5pt]{0pt}{20pt}}

\newcommand{\dd}{\ensuremath{\mathrm{d}}}
\newcommand{\id}{\ensuremath{\hphantom{.}\mathrm{d}}}
\newcommand{\f}[2]{\ensuremath{\frac{#1}{#2}}} 
\newcommand{\df}[2]{\ensuremath{\f{\dd #1}{\dd #2}}}

 \mathchardef\mhyphen="2D
\newcommand{\pf}[2]{\ensuremath{\frac{\partial{#1}}{\partial{#2}}}}

\newcommand{\Retwo}{$0.59\pi$}

\definecolor{myred}{RGB}{200, 13, 13}
\definecolor{myblue}{RGB}{0, 0, 180}
\definecolor{mygreen}{RGB}{0, 102, 0}
\definecolor{LGrey}{rgb}{.5,.5,.5}
\definecolor{myblack}{rgb}{.0,.0,.0}

\title[The minimal-span channel for rough-wall turbulent flows]{The minimal-span channel for rough-wall turbulent flows}

\author[M. MacDonald, D. Chung, N. Hutchins, L. Chan, A. Ooi and R. Garc\'{i}a-Mayoral]
{M. MacDonald$^1$\thanks{Email address for correspondence: michael.macdonald@unimelb.edu.au},\ns
D. Chung$^1$,
N. Hutchins$^1$,
L. Chan$^2$,
A. Ooi$^1$
and 
R. Garc\'{i}a-Mayoral$^3$
}

\affiliation{
$^1$
Department of Mechanical Engineering, University of Melbourne, Victoria 3010, Australia \\
$^2$
Department of Mechanical Engineering, Universiti Tenaga Nasional, Kajang 43000, Malaysia\\
$^3$
Department of Engineering, University of Cambridge, Cambridge CB2 1PZ, UK
}

\pubyear{2016}
\volume{999}
\pagerange{998--999}
\date{?; revised ?; accepted ?}
\begin{document}

\maketitle

%
%
\begin{abstract}
Roughness predominantly alters the near-wall region of turbulent flow while the outer layer remains similar with respect to the wall shear stress. 
This makes it a prime candidate for the minimal-span channel, which only captures the near-wall flow by restricting the spanwise channel width to be of the order of a   few hundred viscous units.
Recently, Chung \emph{et al.} (\emph{J.\ Fluid Mech.}, vol. 773, 2015, pp. 418--431) showed that a minimal-span channel can accurately characterise the hydraulic behaviour of roughness.
 Following this, we aim to investigate the fundamental dynamics of the minimal-span channel framework with an eye towards further improving performance.
The streamwise domain length of the channel is investigated with the minimum length found to be three times the spanwise width or 1000 viscous units, whichever is longer.
The outer layer of the minimal channel is inherently unphysical and as such alterations to it can be performed so long as the near-wall flow, which is the same as in a full-span channel, remains unchanged.
Firstly, a half-height (open) channel with slip wall is shown to reproduce the near-wall behaviour seen in a standard channel, but with half the number of grid points.
Next, a forcing model is introduced into the outer layer of a half-height channel. This reduces the high streamwise velocity associated with the minimal channel and allows for a larger computational time step.
Finally, an investigation is conducted to see if varying the roughness Reynolds number with time is a feasible method for obtaining the full hydraulic behaviour of a rough surface. Currently, multiple steady simulations at fixed roughness Reynolds numbers are needed to obtain this behaviour. The results indicate that the non-dimensional pressure gradient parameter must be kept below $0.03$--$0.07$ to ensure that pressure gradient effects do not lead to an inaccurate roughness function.
An empirical costing argument is developed to determine the cost in terms of CPU hours of minimal-span channel simulations \emph{a priori}. This argument involves counting the number of eddy lifespans in the channel, which is then related to the  statistical uncertainty of the streamwise velocity. For a given statistical uncertainty in the roughness function, this can then be used to determine the simulation run time.
Following this, a finite-volume code with a body-fitted grid is used to determine the roughness function for square-based pyramids using the above insights. Comparisons to experimental studies for the same roughness geometry are made and good agreement is observed.

\end{abstract} 


%
%
\section{Introduction}
Conventional Direct Numerical Simulations (DNS) of wall-bounded turbulent flows represent a challenging computational problem, as both small and large scales need to be represented. The former requires a fine grid to resolve the small viscous scales, while the latter requires a large domain to capture the large outer-layer motions. However, pioneering work by \cite{Jimenez91} and \cite{Hamilton95} into minimal-flow units showed that a small computational domain can be used to exclusively capture the turbulent near-wall cycle, independent of the large outer scales. This is achieved by simply restricting the domain of the channel to a small size, where the spanwise and streamwise lengths are prescribed in terms of viscous units. \cite{Jimenez91} showed that turbulence could be maintained in the form of the near-wall cycle of the buffer layer when the spanwise domain width was only on the order of $100 \nu/U_\tau$; here $\nu$ is the kinematic viscosity and $U_\tau = \sqrt{\tau_w/\rho}$ is the friction velocity, defined using the wall shear stress $\tau_w$ and the fluid density $\rho$ . This was further supported by future studies into minimal-flow units \citep{Jimenez99,Flores10,Hwang13,Chung15,Hwang16}. In particular the work of \cite{Flores10} showed that the minimal-span channel can also capture the logarithmic layer of turbulent flows. 

An important aspect of minimal-span channels is that the near-wall flow is accurately captured up 
to
 a critical wall-normal location, $z_c$. Above this point, the streamwise velocity increases compared to a full-span channel. This unphysical increase occurs as the narrow spanwise domain width of the minimal-span channel acts as a  filter which limits the largest spanwise scale of energy-containing eddies. The flow does, however, retain turbulent scales smaller than the spanwise domain width, so it is not laminar \citep{Jimenez99}. 
 The critical value where the minimal-span channel departs from the full-span channel has been shown to scale with the spanwise domain width,  $z_c\approx 0.4 L_y$ \citep{Chung15}, although a constant of 0.3 is suggested in \cite{Flores10} and 0.3--0.4 in \cite{Hwang13}. Following \cite{Flores10}, we will refer to the flow below $z_c$ as `healthy' turbulence, as it is in a full-span channel.

In the context of roughness, the central question is how the geometry of a rough surface is related to its hydraulic behaviour; namely, what value the (Hama) roughness function, $\Delta U^+$, takes \citep{Hama54}. This quantity reflects the flow retardation, or velocity shift, that the roughness imposes on the flow when compared to a smooth wall, for matched friction Reynolds numbers, $Re_\tau = U_\tau h/\nu$ (here $h$ is the half-channel height, boundary layer thickness, or pipe radius). For a given surface, $\Delta U^+$ can be obtained from various semi-empirical models and approximations, or directly from full-scale experiments and numerical simulations. The former are of varying accuracy and depend on the model selected and the rough surface in question, while the latter can be prohibitively expensive. In particular, full-scale numerical simulations suffer from the drawbacks mentioned above, in which both small- and large-scale motions need to be captured. However, roughness is thought to primarily alter only the near-wall flow, in a region called the roughness sublayer which typically extends 3--5$k$ from the wall, where $k$ is some characteristic height of the roughness \citep{Raupach91}. Well outside the roughness sublayer the flow is only changed insofar as it depends on the friction velocity $U_\tau$. This is the basis of Townsend's outer-layer similarity hypothesis \citep{Townsend76} which has received significant attention and has been supported by several rough-wall studies \citep{Flack05,Leonardi10,Chan15} and also numerical simulations with modified boundary conditions \citep{Flores06,Mizuno13,Chung14}. If we assume Townsend's hypothesis holds, then it follows that the roughness only alters the near-wall region of the flow which therefore makes it a prime candidate for use in the minimal-span channel framework.

The equivalent sandgrain roughness, $k_s$, is a single dynamic parameter that is used to describe the hydraulic behaviour of roughness. It is defined so that the roughness function collapses for all data in the fully rough regime, when the friction factor no longer depends on the bulk-velocity Reynolds number (\citealt{Jimenez04}, who called this $k_{s\infty}$). However, each rough surface will have a unique behaviour in the transitionally rough regime.
It is an expensive process to determine the full hydraulic behaviour, as a range of simulations need to be conducted for the different roughness Reynolds numbers, each requiring a different body-fitted grid and its own initialisation period.  It would therefore be desirable to conduct a simulation in which only a single computational grid is used, and the bulk velocity is changed over time to sweep through a range of roughness Reynolds numbers. This is a similar approach to how experimental studies are performed in that a single rough surface with fixed $k/h$ is tested at multiple flow speeds, so that the roughness Reynolds number varies. In order to obtain statistics at a desired friction Reynolds number in a temporally evolving flow, statistics are averaged over a small window in which the instantaneous friction Reynolds number is close to the desired one. Within this window, the flow is assumed to be quasi steady. This would therefore generate a near-continuous curve of $\Delta U^+$ versus $k^+$, rather than the conventional (steady) approach which only generates a few data points.
An important consideration which will be investigated here is whether acceleration effects from the changing mass flux become significant and distort the estimation of $\Delta U^+$, that is, how quickly can the bulk velocity be varied such that the quasi-steady assumption remains approximately valid.

Recently, we have applied the idea of minimal-span channels to the roughness problem in a proof-of-concept study \citep{Chung15}.
This study demonstrated the feasibility of using the minimal-span channel to accurately compute the roughness function, when compared to full-span channels.
However, only the spanwise width was investigated, with the roughness function being the primary quantity of interest. In this paper, we aim to further investigate the fundamental dynamics of the minimal-span channel framework for roughness.  This will include analysing higher order flow statistics to identify and understand the essential features of the minimal-span channel.
Firstly, the streamwise domain length of the minimal channel is investigated in \S\ref{sect:slength} to identify the minimum streamwise length required to maintain healthy turbulence.
Second, given that the outer-layer of minimal channels are inherently unphysical, then alterations to this region should not alter the healthy near-wall flow.
To assess this claim, two alterations consisting of a half-height (open) channel and outer-layer damping are investigated in \S\ref{sect:half} and \S\ref{sect:outerForcing}, respectively.
Finally, an investigation in \S\ref{sect:sweepres} is conducted in which the bulk velocity (and hence roughness Reynolds number) is varied with time, to see if the full hydraulic behaviour of the rough surface can be obtained in one unsteady simulation.
 A costing argument is then developed in \S\ref{sect:cost} using the characteristic time and  length scales of eddies, so that the CPU hours can be estimated \emph{a priori}. The insights gained in this paper are then used in \S\ref{sect:pyramids} to simulate the flow over square-based pyramids using a finite-volume code with a body-fitted grid, which has been experimentally studied in the literature by \cite{Schultz09} and others.

%
%

\section{Numerical Method}
\label{sect:method}
\begin{figure}
\begin{center}
\includegraphics[trim=0 0 0 0,clip = true,scale=0.75]{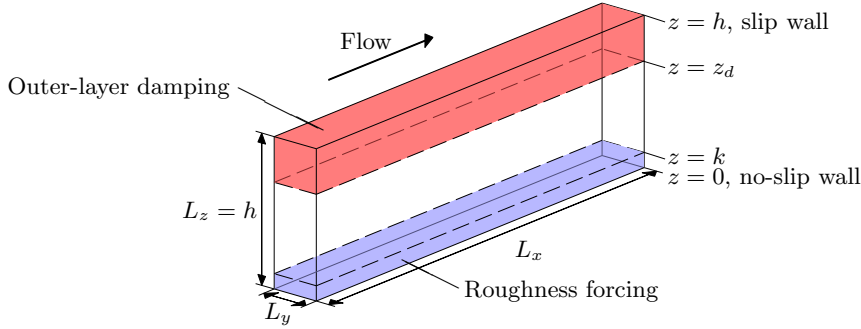}
\put(-2.8,0.9){$L_x$}
\put(-7.25,-0.2){$L_y$}
\put(-8.7,1.6){$L_z = h$}
\put(-0.10,2.2){$z = 0$, no-slip wall}
\put(-0.10,2.55){$z = k$}
\put(-0.10,4.15){$z = z_d$ }
\put(-0.10,4.95){$z = h$, slip wall}
\put(-5.9,4.6){Flow}
\put(-11.8,3.8){Outer-layer damping}
\put(-7.7,3.9){\line(2,-1){1.5}}
\put(-3.7,0.2){Roughness forcing}
\put(-3.8,0.35){\line(-2,1){1.0}}
\end{center}
\vspace{-0.5\baselineskip}
\caption{(Colour online) Half-height minimal-span channel, showing the roughness forcing zone (\ref{eqn:roughF}) and the outer-layer damping zone (\ref{eqn:outerForcing}). No-slip wall at $z=0$ and free-slip wall at $z=h$.
}
\label{fig:forcingZones} 
\end{figure}

The majority of Direct Numerical Simulations in this paper are conducted using the same finite difference code as in \cite{Chung15}, which uses a fully conservative fourth-order staggered-grid scheme. Time integration is performed using the third-order low-storage Runge--Kutta scheme of \cite{Spalart91}, and the fractional-step method of \cite{Kim85} is used.  The Navier--Stokes equations are
\begin{equation}
\label{eqn:nsmmtm}
\nabla\cdot\mathbf{u}=0,  \hspace{1.5cm}
\pf{\mathbf{u}}{t}+\nabla\cdot\left(\mathbf{u}\mathbf{u}\right) = -\f{1}{\rho}\nabla
p+\nabla\cdot\left(\nu\nabla\mathbf{u}\right) + \mathbf{F} + \mathbf{G} + \mathbf{K},
\end{equation}
where $\mathbf{u} = (u,v,w)$ is the velocity in the streamwise ($x$), spanwise ($y$) and wall-normal ($z$) directions, $t$ is time, and $p$ is pressure. $\mathbf{G} = G_x(t)\mathbf{i}$ is the spatially invariant, time-varying streamwise forcing term which drives the flow at constant mass flux. The flow is solved in a reference frame in which the mass flux is zero at all times, although all equations in this paper are given in the stationary-wall reference frame. Solving in a zero-mass-flux reference frame permits a larger computational time step \citep{Lundbladh99}, as well as reducing high-wavenumber convective disturbances produced by finite difference schemes \citep{Bernardini13}.
The roughness model, $\mathbf{F} = -\alpha F(z,k)u|u|\mathbf{i}$, is based on the work of \cite{Busse12}, which applies a forcing in the streamwise direction that opposes the flow. The roughness factor $\alpha = 1/(40k)$ is kept constant for all simulations, with the roughness height $k = h/40$. The function $F(z,k)$ is a simple step function which applies the roughness forcing adjacent to the top and bottom no-slip channel walls at $z = 0$ and $z = 2h$,
\begin{equation}
\label{eqn:roughF}
F(z,k) = \begin{cases}
1, & \text{if $z < k$ or $2h - k < z$}. \\
0, & \text{otherwise}.
\end{cases}
\end{equation}
This roughness model removes any spanwise or streamwise roughness length scales from the problem and allows for simpler computations as the same smooth-wall grid can be used. This is a similar model to \cite{Borrell15thesis}, although there the author used an $\mathbf{F}\propto- u\mathbf{i}$ scaling.
The final term in (\ref{eqn:nsmmtm}), $\mathbf{K}$, is a forcing function which damps the velocity in the outer layer of the minimal channel. 
It has the form
\begin{equation}
\label{eqn:outerForcing}
K_ i = -\gamma \Gamma(z,z_d) \Big(u_i(x,y,z,t) - \langle u_ i \rangle(z_d,t)\Big)\Big|u_i(x,y,z,t) - \langle u_ i \rangle(z_d,t)\Big|,
\end{equation}
(no summation over $i$) where angled brackets denote the spatial average of the instantaneous velocity over a wall-parallel plane. The factor $\gamma$ has units of inverse time and in this study is set according to $\gamma h/U_\tau\approx 1$. The parameter $\Gamma(z,z_d)$ is similar to the function $F(z,k)$ of the roughness forcing model in that it indicates where the damping is applied, namely in the outer layer of the channel,
\begin{equation}
\label{eqn:outerForcingRegion}
\Gamma(z,z_d) = \begin{cases}
1, & \text{if $z > z_d$ and $z < 2h-z_d$}. \\
0, & \text{otherwise}.
\end{cases}
\end{equation}
The value of $z_d$ should be greater than the critical wall-normal location, $z_c$, of the minimal-span channel, so that the forcing does not contaminate the healthy turbulence of the near-wall flow.
A step function for $\Gamma$ is used to minimise the parameter space, as well as to ensure the location of $z_d$ is unambiguous.
 Several different values of $z_d$ are tested and will be presented in the following section. In (\ref{eqn:outerForcing}), the term $\langle u_i\rangle(z_d,t)$ is the wall-parallel spatially averaged velocity at $z_d$. This is present to reduce the magnitude of the streamwise velocity; the minimal channel has a very high streamwise velocity which presents a time step restriction from the $\textit{CFL}$ number. The forcing is in this form so that the velocity is forced to remain at the same level as at $z=z_d$.  This outer-layer damping is similar to the masks employed in \cite{Jimenez99}, which damp fluctuations in the outer-layer region of minimal channels. The current forcing model  simply sets the streamwise velocity to be approximately that at $z=z_d$, however it will be shown to work reasonably well for the purpose of obtaining $\Delta U^+$. Note that the average velocity in the spanwise and wall-normal directions is zero, so that the forcing in these directions can simply be $K_i = -\gamma \Gamma(z,z_d)u_i|u_i|$, for $i = 2,3$. Figure \ref{fig:forcingZones} shows the two forcing regions that are employed in the current study.

\setlength{\unitlength}{0.75cm}

The streamwise and spanwise grid is evenly spaced, while the wall-normal grid is stretched with a cosine mapping. Periodic boundary conditions are applied in the spanwise and streamwise directions. In the case of standard-height channels, no-slip walls are located at $z = 0$ and $z = 2h$. For clarity, the word `full' will be used to refer to the span (full span), while the case with two no-slip walls will be referred to as `standard' height. For half-height (open) channels, a no-slip wall is still positioned at $z=0$, however now the top domain surface is a free-slip wall with $\partial u/\partial z = \partial v/\partial z=0$, positioned at $z=h$. This slip wall maintains the impermeability constraint, $w=0$. The spanwise width, $L_y$, of the minimal-span channel satisfies the guidelines of \cite{Chung15}, namely that $L_y\gtrsim\max(100\nu/U_\tau,k/0.4,\lambda_{r,y})$, where $\lambda_{r,y}$ is the spanwise length scale of roughness elements. Because the homogeneous roughness forcing model has no spanwise length scale then the final constraint can be ignored. Simulations are conducted at a friction Reynolds number of $Re_\tau\approx590$, with a few additional simulations conducted at $Re_\tau\approx 2000$. Relevant simulations are introduced at the start of each section for each of the investigations conducted in this study.

\subsection{Temporal sweep}
\label{sect:sweep}
As discussed in the introduction, instead of conducting multiple steady simulations of different roughness Reynolds numbers in order to determine the equivalent sandgrain roughness, a single unsteady simulation is to be performed in which the bulk velocity is varied. In particular, we will investigate different rates of change of the bulk velocity and the effects that this acceleration has on the flow.
The sweep will start with the highest friction number to be tested, $Re_{\tau,start}$  and will then be decelerated via an adverse pressure gradient. This ensures an adequate grid resolution at the start of simulation. At the final friction Reynolds number, $Re_{\tau,end}<Re_{\tau,start}$, the grid resolution would be finer than necessary, which for the current simulations would have 4--5 times more cells than if a conventional steady simulation were conducted at this final Reynolds number.

\setlength{\unitlength}{1cm}
\begin{figure}
\centering
 \captionsetup[subfigure]{labelformat=empty}
	\subfloat[]{
		\includegraphics[width=0.49\textwidth]{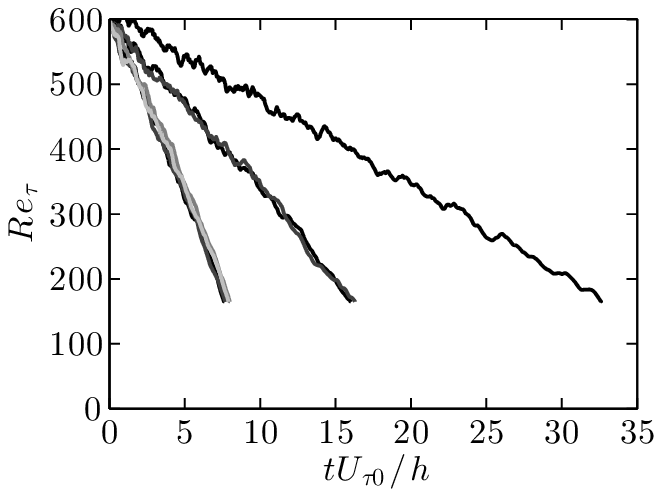}
		\label{fig:sweepRetau}
	}
	\subfloat[]{
		\includegraphics[width=0.49\textwidth]{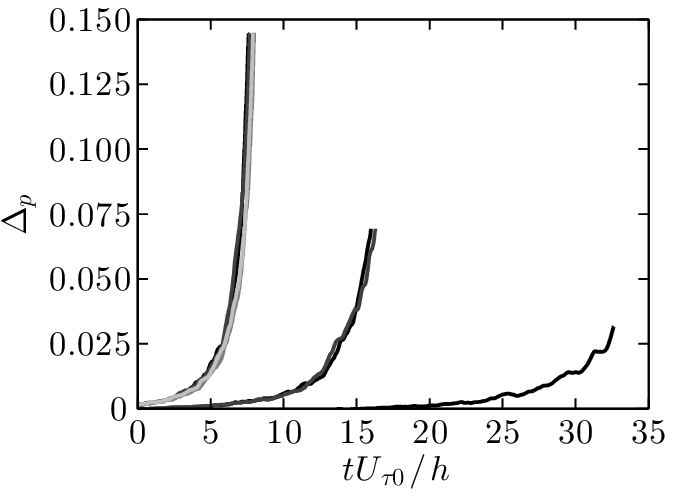}
		\label{fig:sweepDeltap}
	}
	\put(-13.6,4.6){(\emph{a})}
	\put(-7.8,1.6){$A$}
	\put(-10.3,1.6){$2A$}
	\put(-11.6,1.6){$4A$}
	\put(-6.8,4.6){(\emph{b})}
	\put(-2.9,1.75){$\varDelta_{p,Re_\tau=180}$$\approx$$0.03$}
	\put(-2.95,2.5){$\varDelta_{p,Re_\tau=180}$$\approx$$0.07$}
	\put(-4.2,4.2){$\varDelta_{p,Re_\tau=180}$$\approx$$0.15$}
	\vspace{-1.8\baselineskip}
	\caption{(\emph{a}) Instantaneous Reynolds number and (\emph{b}) pressure gradient parameter (\ref{eqn:deltap}) of the smooth-wall temporal sweep, against viscous time.
		The cases with faster sweeps have multiple runs (shown by grey lines) and are ensemble averaged. $A$ is the linear rate of change of $U_b$  (\ref{eqn:SweepA}).
}
	\label{fig:sweept}
\end{figure}

Presently, we vary the bulk velocity $U_b(t)$ with time,
\begin{equation}
	\df{U_b}{t} = A = \f{U_{b,end} - U_{b,start}}{\Delta T} = const.,
\label{eqn:SweepA}
\end{equation}
where $U_{b,end}$ is the desired end point of the simulation (the bulk velocity corresponding to $Re_\tau = 180$), and $\Delta T$ is a time scale which should be sufficiently long enough that acceleration effects are not significant to the flow. The initial bulk velocity, $U_{b,start}$, is set corresponding to a friction Reynolds number of $Re_{\tau,start}=590$ and different values of $A$ are tested. An initial value of $A$ is selected  such that over the course of the sweep $\Delta T U_{\tau 0}/h=30\Rightarrow U_{\tau 0}^2/hA\approx30$, with subsequent runs using $2A$ and $4A$. In these cases of a higher rate of change of $U_b$, multiple sweeps are run from different initial conditions with the results ensemble averaged, to ensure that statistics are obtained over the same amount of simulation time. For example, the case where the gradient is quadrupled, four sweeps are conducted with four different initial conditions. This is visualised in figure \ref{fig:sweept}(\emph{a}), which shows the change in the friction Reynolds number as a function of time. An important consideration with this technique is that the friction velocity and hence the normalised spanwise domain width $L_y^+$ will vary with time. The domain width must remain larger than 100  viscous units at all times to ensure the turbulent flow is sustained, which may necessitate multiple stages in the sweep if $L_y^+$ becomes too small. Each stage in the sweep should be set up such that $L_y^+\gtrsim100$.

A dimensionless measure of the flow acceleration is the pressure gradient parameter,
\begin{equation}
\label{eqn:deltap}
\varDelta_p = \f{\nu}{\rho U_\tau^3}\df{p}{x}.
\end{equation}
Various experimental studies of decelerating boundary layers show that mild decelerations of $\varDelta_p < 0.01$ have small ($<3\%$) errors in calculating $U_\tau$ from methods based on the assumption of zero pressure gradients \citep{Patel65,Jones01}. While we are interested in channel flow as opposed to boundary layers, this value should still be useful in providing an indication as to whether acceleration effects will be significant. 
\cite{Seddighi14}, meanwhile, compared a step acceleration with a slow ramp up in which the bulk velocity was linearly varied in a channel  such that $-\varDelta_p\approx 0.73$ (favourable pressure gradient). Acceleration effects were still seen in this slow ramp up. 
Note that for steady channel flow, the pressure gradient parameter is $\varDelta_{p}=-1/Re_\tau<0$, which for the current simulations at $Re_\tau=590$ takes a value of $-0.0017$.
In the current study, three different sweep rates are studied. As shown in figure \ref{fig:sweept}(\emph{b}), the linear variation in $U_b$ results in the pressure gradient parameter having a maximum value at the end of the simulation, when $Re_\tau\approx180$. 
The different sweeps will therefore be referred to by their maximum pressure gradient parameter, which are $\varDelta_{p,Re_\tau=180} \approx  (0.03,0.07,0.15)$. These values place the sweeps considered in this study into the range between negligible ($\varDelta_p<0.01$) and noticeable ($-\varDelta_p>0.73$) pressure gradient effects, and are substantially larger than the steady value ($-\varDelta_p=0.0017$).

%
%

\section{Results}
\label{sect:res}
%
%
\subsection{Varying streamwise domain length (table \ref{tab:sims1v2})}
\label{sect:slength}
\begin{table}
\centering
\begin{tabular}{ c c  c c c c   c c c  c c c c  c c}
ID	& $Re_\tau$	& $\tfrac{L_x}{h}$	&  $L_x^+$	& $L_y^+$	& $\tfrac{L_z}{h}$	& $N_x$	& $N_y$	& $N_z$	& $\Delta x^+$	& $\Delta y^+$	& $\Delta z_w^+$	& $\Delta z_h^+$	& $\Delta t^+_{S}$	& $\Delta t^+_{R}$\\[0.5em] 
  \multicolumn{15}{c}{Full-span channel}\\[0.5em] 
FS	& 590		& $2\pi$	& 3707	& 1854	&2	& 384		& 384		& 256		& 9.7		& 4.8		& 0.04	& 7.2		& 0.26	& 0.29	\\[0.5em] 
  \multicolumn{15}{c}{Minimal-span channel with varying streamwise length} \\[0.5em] 
MS1	&  590		& $0.1\pi$	& 190		& 118		&2	& 24		& 24		& 256		& 7.7		& 4.9		& 0.04	&7.2		& 0.28	& 0.32	\\
MS2	&  590		& $0.16\pi$	& 300		& 118		&2	& 48		& 24		& 256		& 6.2		& 4.9		& 0.04	& 7.2		& 0.24	& 0.27	\\
MS3	&  590		& $0.25\pi$	& 460		& 118		&2	& 48		& 24		& 256		& 9.7		& 4.9		&0.04		&7.2		& 0.36	& 0.42	\\
MS4	&  590		& $0.5\pi$	& 930		& 118		&2	& 96		& 24		& 256		& 9.7		& 4.9		& 0.04	& 7.2		& 0.36	& 0.42	\\	
MS5	&  590		& $\pi$	& 1850	& 118		&2	& 192		& 24		& 256		& 9.7		& 4.9		& 0.04	&7.2		& 0.36	& 0.41	\\[0.5em] 
  \multicolumn{15}{c}{Minimal-span channel with varying streamwise length (wider span)} \\[0.5em] 
MSW1	&  590		& $0.1\pi$	& 190		& 354		&2	& 24		& 72		& 256		& 7.7		& 4.9		&0.04		&7.2		& 0.37	& 0.45	\\
MSW2	&  590		& $0.16\pi$	& 300		& 354		&2	& 48		& 72		& 256		& 6.2		& 4.9		& 0.04	& 7.2		& 0.31	& 0.40	\\
MSW3&  590		& $0.25\pi$	& 460		& 354		&2	& 48		& 72		& 256		& 9.7		& 4.9		& 0.04	& 7.2		& 0.46	& 0.52	\\
MSW4	&  590		& $0.5\pi$	& 930		& 354		&2	& 96		& 72		& 256		& 9.7		& 4.9		&0.04		&7.2		& 0.43	& 0.46	\\
MSW5	&  590		& $\pi$	& 1850	& 354		&2	& 192		& 72		& 256		& 9.7		& 4.9		&0.04		&7.2		& 0.38	& 0.40	\\
\end{tabular}
\caption{Description of the simulations performed with the finite difference code to investigate streamwise domain length. All simulations done as both smooth wall and modelled rough wall.
Entries:
$Re_\tau$, nominal friction Reynolds numbers;
$L$, domain length;
$N$, number of grid points;
$\Delta$, grid-spacing in  viscous units, subscript $w$ and $h$ refers to the wall-normal grid spacing at the wall and channel centre;
$L_z/h=2$ denotes standard-height (two no-slip walls), 
$\Delta t_S^+$ and $\Delta t_R^+$ is the smooth- and rough-wall time step.
Roughness height $k=h/40$ so $k^+\approx15$ at $Re_\tau=590$.
}
\label{tab:sims1v2}
\end{table}

Wall-bounded turbulent flows are often characterised by very long structures on the order of tens of channel half heights \citep{Kim99,Lozanoduran14}.
This becomes very expensive to simulate for conventional full-span channel simulations, so  shorter domain lengths of $L_x/h=2\pi$ are often used \citep{Chin10,Munters16}.
These seem to capture the majority of the outer-layer dynamics despite their relatively short length.
However, these lengths are only necessary due to the large outer-layer structures which are present in the full-span simulations, which have streamwise lengths on the order of the channel half-height, $h$. 

The minimal-span channel only captures the inner-layer flow which is not dependent on $h$, suggesting a shorter domain length can be used.
 The near-wall cycle of the buffer layer produces streaky structures with streamwise lengths of 1000 viscous units \citep{Kline67}. However, these low- and high-speed streaks are nearly two-dimensional, which suggests they could be represented in the infinite ($k_x=0$) mode and hence the domain length doesn't have to be this long. These streaks are accompanied by quasi-streamwise vortices with streamwise lengths of 200--300 viscous units \citep{Jeong97}. The narrowest minimal-span channels with $L_y^+\approx100$ therefore require $L_x^+\approx200$--300 to capture the near-wall cycle in the buffer layer \citep{Jimenez91}. The inertial logarithmic layer also have similar vortex clusters \citep{Delalamo06}, which scale as 2--3 times their spanwise width \citep{Flores10,Hwang15}.  Statistically stationary and homogenous shear turbulence, which shows similarities to the logarithmic layer  in wall-bounded turbulence, also suggest a scaling of 2 times the spanwise length \citep{Sekimoto16}.
    Simulations with larger spanwise domain widths would produce vortices with longer streamwise lengths, so that the largest captured eddy would have  spanwise and streamwise lengths of \mbox{$\lambda_y = L_y$ and $\lambda_x=2$--$3L_y$}.

Two different spanwise domain widths are investigated for varying the streamwise length (table \ref{tab:sims1v2}). Firstly, the smallest width of $L_y^+\approx120$ would have the largest eddies having streamwise lengths of $\lambda_x^+=240$--350. Various streamwise domain lengths of $L_x^+ = (190,300,460,930,1850)$ are chosen to see how this affects the largest eddies. The second spanwise domain width of $L_y^+\approx 350$ would be able to capture larger eddies, the largest of which would have a  streamwise length of approximately $\lambda_x^+=710$--$1100$. The same set of streamwise domain lengths are simulated for this wider domain. A standard-height channel is used, with no outer-layer damping, $K_i=0$.

\setlength{\unitlength}{1cm}
\begin{figure}
\centering
 \captionsetup[subfigure]{labelformat=empty}
	\subfloat[]{
		\includegraphics[width=0.49\textwidth,trim=0 11 0 0, clip = true]{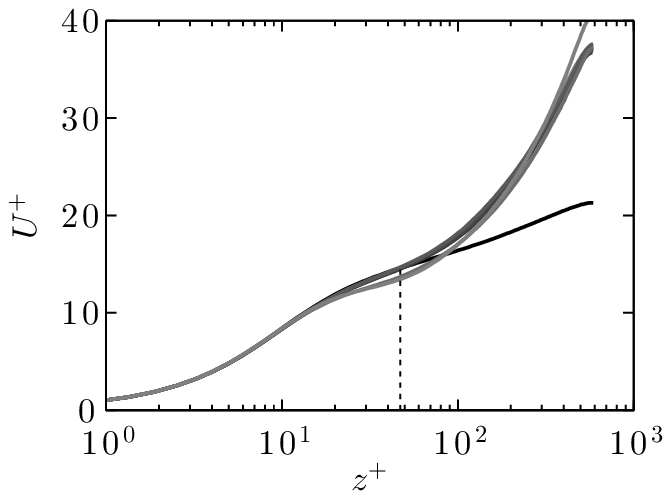}
		\label{fig:uxLengthSm}
	}
	\subfloat[]{
		\includegraphics[width=0.49\textwidth,trim=0 11 0 0, clip = true]{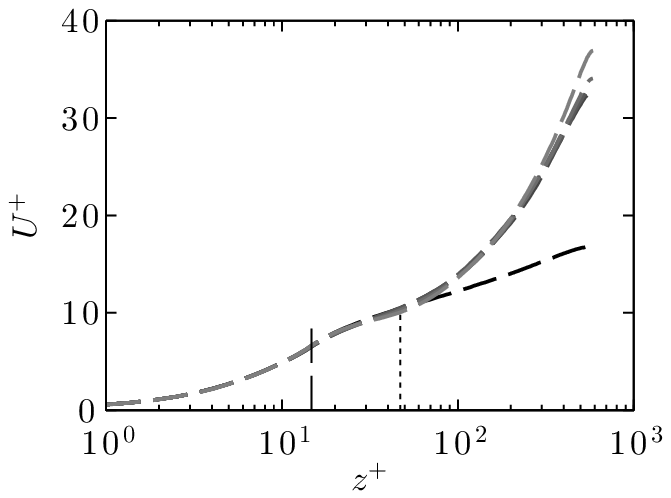}
		\label{fig:uxLengthRo}
	}
	\put(-13.5,4.25){(\emph{a})}
	\put(-6.7,4.25){(\emph{b})}
	\put(-9.5,3.9){Smooth,}
	\put(-9.5,3.5){$L_y^+$$\approx$$120$}
	\put(-9.3,1.3){$L_x^+$$\approx$$190,300$}
	\put(-9.85,1.80){\line(-2,1){0.57}}
	\put(-11.8,2.0){$L_x^+\ge 460$}
	\put(-9.65,1.75){\line(1,-1){0.36}}
	\put(-2.5,3.9){Rough,}
	\put(-2.5,3.5){$L_y^+$$\approx$$120$}
	\put(-5.6,1.5){\includegraphics{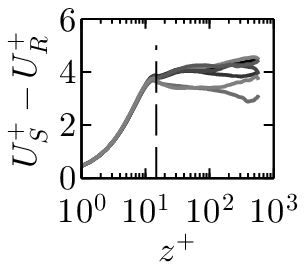}}
	\put(-3.65,2.7){$L_x^+$$\approx$$190,300$}
	\put(-3.65,2.85){\line(-1,4){0.12}}
	\vspace{-2.5\baselineskip}
	\\
	\subfloat[]{
		\includegraphics[width=0.49\textwidth]{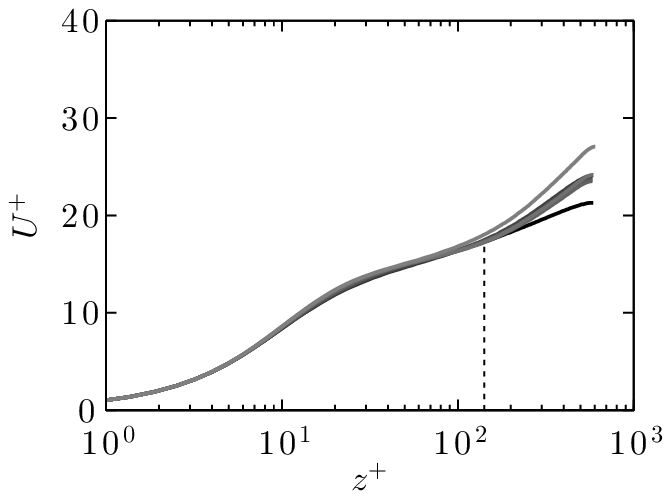}
		\label{fig:uWiderlengthSm}
	}
	\subfloat[]{
		\includegraphics[width=0.49\textwidth]{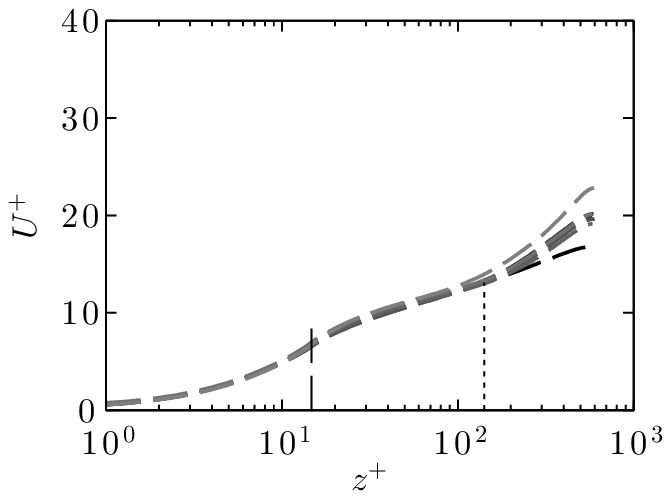}
		\label{fig:uWiderlengthRo}
	}
	\put(-13.5,4.6){(\emph{c})}
	\put(-6.7,4.6){(\emph{d})}
	\put(-9.6,4.2){Smooth,}
	\put(-9.5,3.8){$L_y^+$$\approx$$350$}
	\put(-9.8,3.3){$L_x^+$$\approx$$190$}
	\put(-8.56,3.38){\line(1,-1){0.36}}
	\put(-2.6,4.2){Rough,}
	\put(-2.5,3.8){$L_y^+$$\approx$$350$}
	\put(-2.68,3.22){$L_x^+$$\approx$$190$}
	\put(-1.45,3.3){\line(1,-1){0.36}}
	\put(-5.6,1.8){\includegraphics{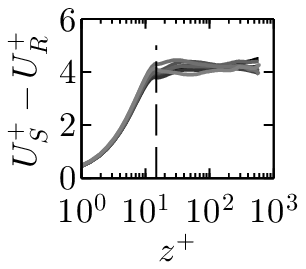}}
	\vspace{-1.8\baselineskip}
	\caption{Streamwise velocity profile for varying streamwise length for (\emph{a},\emph{c}) smooth- and (\emph{b},\emph{d}) rough-wall standard-height channels.
	Spanwise domain width is (\emph{a},\emph{c}) $L_y^+\approx120$ and (\emph{b},\emph{d}) $L_y^+\approx350$.
	Line styles are:
	black, full-span channel;
	grey, minimal-span channel with varying streamwise length (dark grey is longer), refer table \ref{tab:sims1v2};
	vertical dashed line indicates the roughness height $k = h/40$;
	vertical dotted line indicates critical height $z_c=0.4L_y$.
	Insets of (\emph{b},\emph{d}) shows difference in smooth and rough wall flows.
	}
	\label{fig:uLength}
\end{figure}

Figure \ref{fig:uLength} shows the mean velocity profile for the two spanwise widths, for all the streamwise lengths tested. 
The minimal channels (grey lines) tend to agree with the full-span channel (black line) up until the critical wall-normal location $z_c=0.4L_y$ (vertical dotted line), at which point the velocity profile of the minimal channels increases. For the smaller domain width (figure \ref{fig:uLength}\emph{a},\emph{b}), the shortest streamwise lengths of $L_x^+\approx190$ and $300$ result in a reduction in the smooth-wall velocity above $z^+\gtrsim 20$. There is also a slight increase in the centreline velocity. As a result, the difference in smooth- and rough-wall velocity (inset of figure \ref{fig:uLength}\emph{b}) decreases relative to the full-span and longer minimal-span cases. 
This effect is not due to the differences in grid resolution of the two shortest channels (table \ref{tab:sims1v2}), as the wider channel has the same differences in grid resolution but will be seen to not have this effect.
 Little discernible difference can be seen between the longer streamwise lengths of $L_x^+\ge460$, in agreement with the scalings discussed above. This also indicates that the low- and high-speed streaks, with lengths of 1000 viscous units, do not need to be explicitly captured as they are aliased into the $k_x=0$ mode. The case of $L_x^+\approx300=2.5L_y^+$ produces an incorrect profile and this suggests that a minimum streamwise length of $L_x^+ = 3L_y^+$ is required, especially for narrow channels in which $z_c$ is closer to the buffer layer than log layer of the flow.

The smallest streamwise length of $L_x^+\approx190$ was unable to sustain turbulence for a prolonged period. It would decay to a laminar state after $T^+=T U_\tau^2/\nu \approx 14.8\times10^{4}$ in the smooth-wall channel, and after $T^+\approx9.7\times10^{4}$ in the rough-wall channel. The data shown in the previous figures for this streamwise length is averaged only when the flow is turbulent. This behaviour is similar to what was observed in \cite{Jimenez91}, who showed that  turbulence could not be maintained in channels with streamwise lengths of around 200 viscous units. As discussed above, this is because it is shorter than the quasi-streamwise vortices which have streamwise lengths of $\lambda_x^+\approx200$--$300$.

The wider spanwise domain width of $L_y^+\approx354$ (figure \ref{fig:uLength}\emph{c},\emph{d}) results in a centreline velocity that is closer to the full-span channel, as these wider channels capture a wider range of turbulent motions. However, the case with the shortest streamwise length of $L_x^+\approx190\approx0.54L_y^+$ has a larger centreline velocity, with the critical wall-normal height appearing lower than $z_c=0.4L_y$. This somewhat resembles the channels with a narrower spanwise width of $L_y^+\approx120$ (figure \ref{fig:uLength}\emph{a},\emph{b}), suggesting that the very short domain length restricts the size of the largest eddies. Their spanwise width would now be smaller than the width of the domain, i.e. the streamwise length is now the limiting length scale.  
  Interestingly, the streamwise length of $L_x^+\approx 300=0.84L_y^+$ appears to agree quite well with the cases with longer lengths. This is in contrast to the narrow domain (figure \ref{fig:uLength}\emph{a}), which shows a clear difference for $L_x^+\approx 300\approx2.5L_y^+$. Even the case with $L_x^+\approx 460\approx1.3L_y^+$ is producing a velocity profile and roughness function comparable to that of channels with longer streamwise lengths, despite not having the requisite scaling of $L_x^+\gtrsim 2$--$3L_y^+$ discussed above. This is similar to the results of \cite{Toh05} who looked at wide spanwise channels with very short streamwise lengths. The velocity profiles they obtained look similar to a full-span, full-length channel, with no apparent increase in streamwise velocity in the outer layer that is characteristic of minimal-span channels. A possible explanation for the results seen here is that the wall-normal critical location $z_c^+\approx 140=0.24 h^+$ is outside the log layer, which is generally believed to end at $z\approx 0.15h$ \citep{Marusic13}. As such, the expected streamwise length scale of 2--3$L_y$ is no longer appropriate in the outer layer. In this case, the largest eddy at the edge of the logarithmic layer would have a streamwise length of approximately $\lambda_x^+\approx $440--660, which could explain why the simulation with $L_x^+\approx 460$ produces a reasonable velocity profile.

\setlength{\unitlength}{1cm}
\begin{figure}
\centering
 \captionsetup[subfigure]{labelformat=empty}
	
 	\vspace{0.2\baselineskip}
 	\subfloat[]{
 		\includegraphics[width=0cm,height=4cm,trim= 0 0 0 0,clip = true]{white.eps} 
 	}
	\subfloat[]{
		\includegraphics[width=0.45\textwidth,trim=0 15 7 0,clip = true]{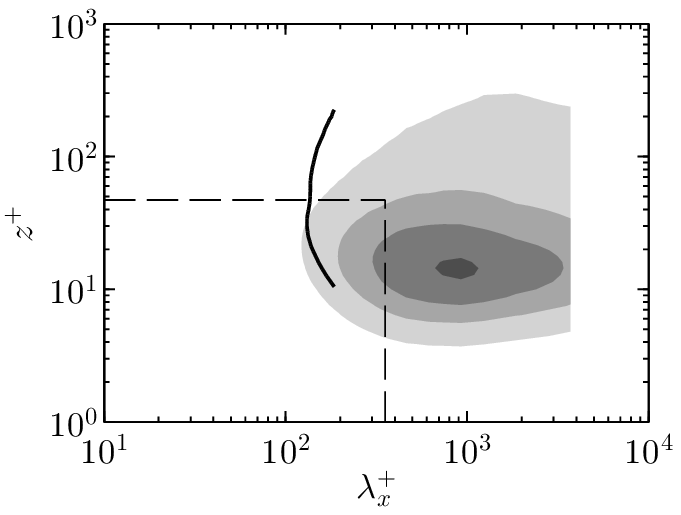} 
		\label{fig:specx1}
	}
	\subfloat[]{ 
		\includegraphics[width=0.45\textwidth,trim=0 15 7 0,clip = true]{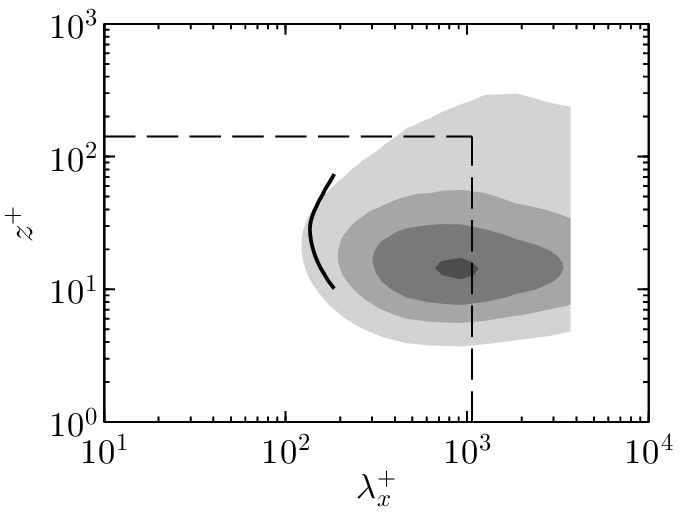} 
		\label{fig:specx5}
	}
	\put(-8.0,4.3){Increasing width $L_y$}
	\put(-5.0,4.4){\vector(1,0){0.5}}
	\put(-8.7,3.55){(\emph{a}) $L_y^+\approx 120$}
	\put(-2.4,3.55){(\emph{b}) $L_y^+\approx 350$}
	\put(-8.85,0.6){$\lambda_x^+= 3L_y^+$}
	\put(-1.85,0.6){$\lambda_x^+=3L_y^+$}
	\put(-11.25,2.15){\includegraphics[scale=0.8]{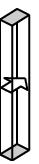}}
	\put(-5.0,2.15){\includegraphics[scale=0.8]{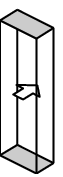}}
	\vspace{-2.5\baselineskip}
	\subfloat[]{
 		\includegraphics[width=0cm,height=4cm,trim= 0 0 0 0,clip = true]{white.eps} 
 	}
	\subfloat[]{
		\includegraphics[width=0.45\textwidth,trim=0 15 7 0,clip = true]{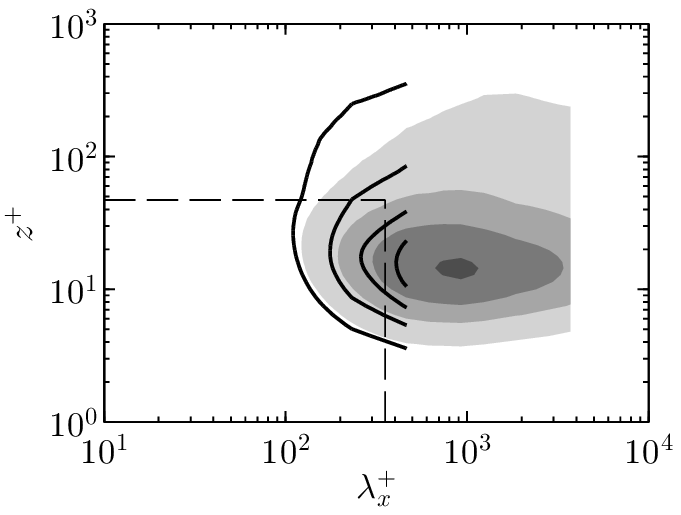} 
		\label{fig:specx2}
	}
	\subfloat[]{
		\includegraphics[width=0.45\textwidth,trim=0 15 7 0,clip = true]{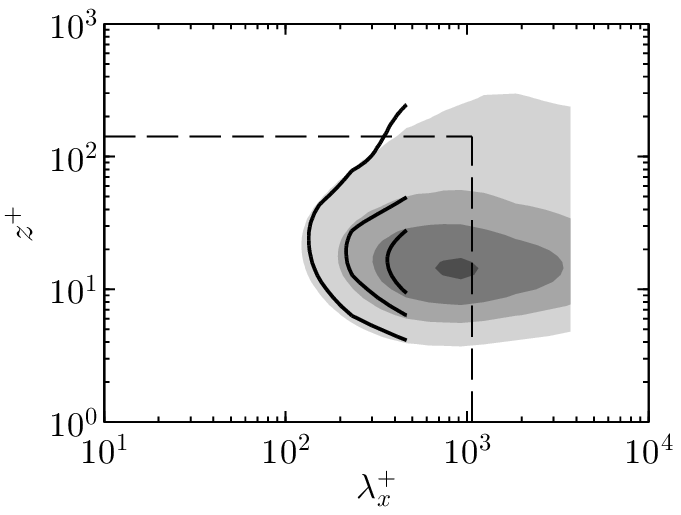} 
		\label{fig:specx6}
	}
	\put(-12.9,-1){\rotatebox{90}{Increasing length $L_x$}}
	\put(-8.7,3.55){(\emph{c}) $L_y^+\approx 120$}
	\put(-2.4,3.55){(\emph{d}) $L_y^+\approx 350$}
	\put(-8.85,0.6){$\lambda_x^+= 3L_y^+$}
	\put(-1.85,0.6){$\lambda_x^+=3L_y^+$}
	\put(-11.25,2.15){\includegraphics[scale=0.8]{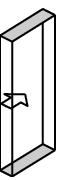}}
	\put(-5.0,2.15){\includegraphics[scale=0.8]{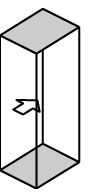}}
	\vspace{-3.8\baselineskip}
	\subfloat[]{
 		\includegraphics[width=0cm,height=4cm,trim= 0 0 0 0,clip = true]{white.eps} 
 	}
	\subfloat[]{
		\includegraphics[width=0.45\textwidth,trim=0 15 7 0,clip = true]{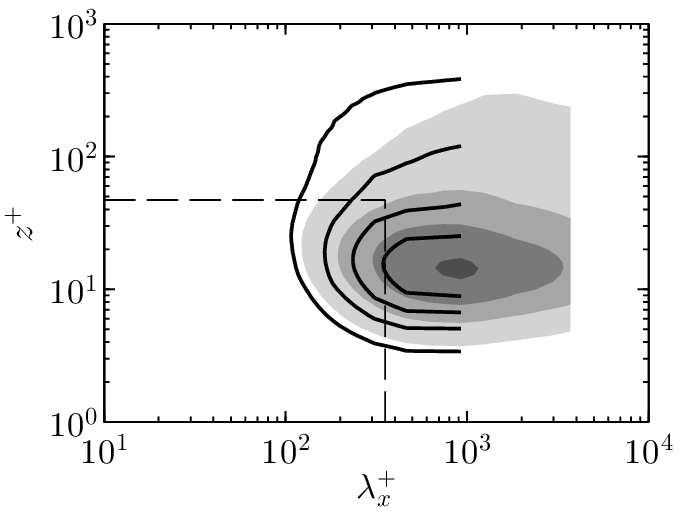} 
		\label{fig:specx3}
	}
	\subfloat[]{
		\includegraphics[width=0.45\textwidth,trim=0 15 7 0,clip = true]{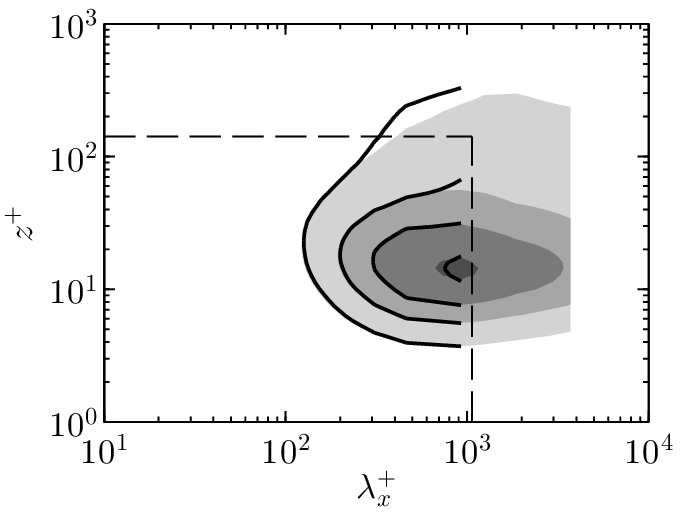} 
		\label{fig:specx7}
	}
	\put(-12.75,3.3){\vector(0,-1){0.75}}
	\put(-8.7,3.55){(\emph{e}) $L_y^+\approx 120$}
	\put(-2.4,3.55){(\emph{f}) $L_y^+\approx 350$}
	\put(-8.85,0.6){$\lambda_x^+= 3L_y^+$}
	\put(-1.85,0.6){$\lambda_x^+=3L_y^+$}
	\put(-11.25,2.15){\includegraphics[scale=0.8]{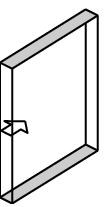}}
	\put(-5.0,2.15){\includegraphics[scale=0.8]{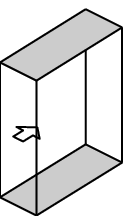}}
	\vspace{-2.5\baselineskip}
	\subfloat[]{
 		\includegraphics[width=0cm,height=4cm,trim= 0 0 0 0,clip = true]{white.eps} 
 	}
	\subfloat[]{
		\includegraphics[width=0.45\textwidth,trim=0  0 7 0,clip = true]{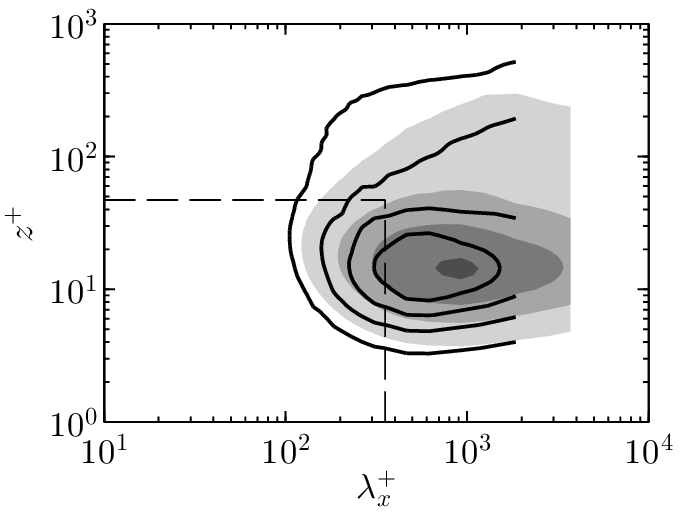} 
		\label{fig:specx4}
	}
	\subfloat[]{
		\includegraphics[width=0.45\textwidth,trim=0 0 7 0,clip = true]{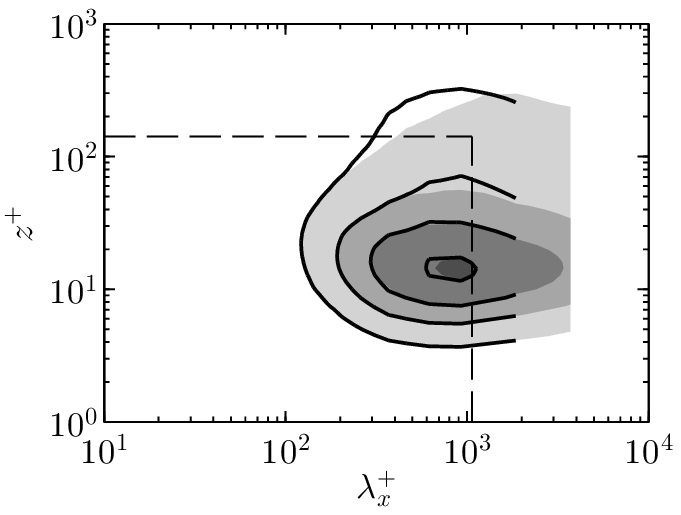} 
		\label{fig:specx8}
	}
	\put(-8.7,4.05){(\emph{g}) $L_y^+\approx 120$}
	\put(-2.4,4.05){(\emph{h}) $L_y^+\approx 350$}
	\put(-8.85,1.1){$\lambda_x^+= 3L_y^+$}
	\put(-1.85,1.1){$\lambda_x^+=3L_y^+$}
	\put(-11.25,2.3){\includegraphics[scale=0.8]{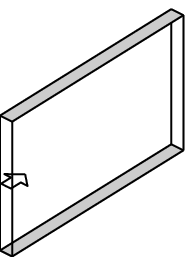}}
	\put(-5.0,2.2){\includegraphics[scale=0.8]{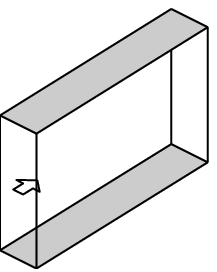}}
	\vspace{-1.7\baselineskip}
	\caption{Smooth-wall premultiplied one-dimensional streamwise energy spectra of streamwise velocity, $k_x E_{u u}^+$, for varying streamwise lengths of
	(\emph{a},\emph{b}) $L_x^+\approx 190$, 
	(\emph{c},\emph{d}) $L_x^+\approx 460$, 
	(\emph{e},\emph{f}) $L_x^+\approx 930$, 
	and 
	(\emph{g},\emph{h}) $L_x^+\approx 1850$. 
	Spanwise domain width (\emph{a},\emph{c},\emph{e},\emph{g}) $L_y^+\approx 120$, and (\emph{b},\emph{d},\emph{f},\emph{h}) $L_y^+\approx 350$.
	Shaded contour is smooth-wall full-span channel (same for all figures).  Contour levels $k_x E_{u u}^+ = 0.5$, 1.0, 1.5, 2.0.  Inset box gives channel scale, arrow shows direction of streamwise flow.
	Vertical dashed line shows $\lambda_x^+=3L_y^+$, horizontal dashed line shows $z_c^+=0.4 L_y^+$. Square dashed box of wider channel (\emph{b},\emph{d},\emph{f},\emph{h}) shows the additional captured region.
	}
	\label{fig:specx}
\end{figure}

\setlength{\unitlength}{1cm}
\begin{figure}
\centering
 \captionsetup[subfigure]{labelformat=empty}
	
 	\vspace{0.2\baselineskip}
 	\subfloat[]{
 		\includegraphics[width=0cm,height=4cm,trim= 0 0 0 0,clip = true]{white.eps} 
 	}
	\subfloat[]{
		\includegraphics[width=0.45\textwidth,trim=0 15 7 0,clip = true]{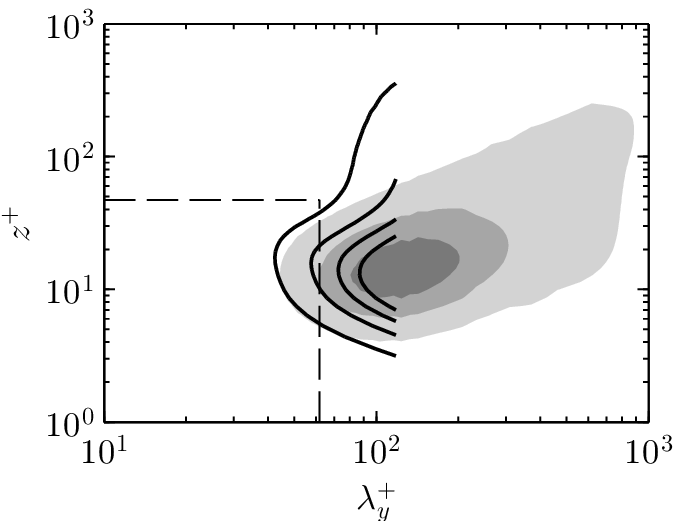} 
		\label{fig:specy1}
	}
	\subfloat[]{ 
		\includegraphics[width=0.45\textwidth,trim=0 15 7 0,clip = true]{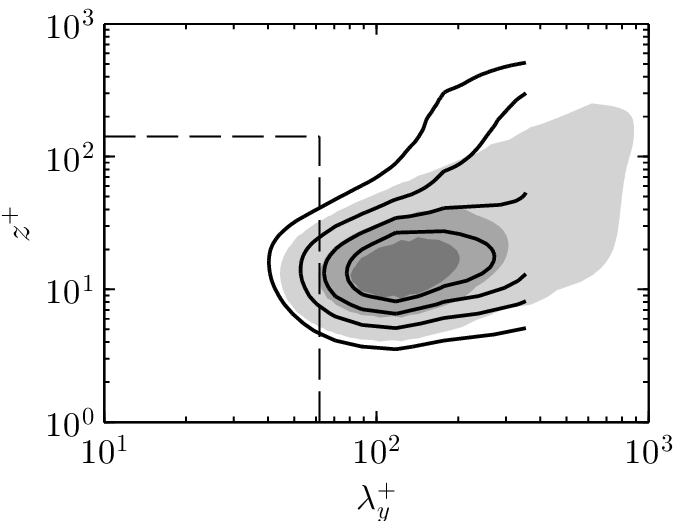} 
		\label{fig:specy5}
	}
	\put(-8.0,4.3){Increasing width $L_y$}
	\put(-5.0,4.4){\vector(1,0){0.5}}
	\put(-8.7,3.7){(\emph{a}) $L_x^+\approx 190$}
	\put(-2.4,3.7){(\emph{b}) $L_x^+\approx 190$}
	\put(-9.45,0.6){$\lambda_y^+= L_x^+/3$}
	\put(-3.2,0.6){$\lambda_y^+=L_x^+/3$}
	\put(-11.25,2.2){\includegraphics[scale=0.8]{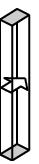}}
	\put(-5.0,2.2){\includegraphics[scale=0.8]{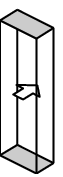}}
	\vspace{-2.5\baselineskip}
	\subfloat[]{
 		\includegraphics[width=0cm,height=4cm,trim= 0 0 0 0,clip = true]{white.eps} 
 	}
	\subfloat[]{
		\includegraphics[width=0.45\textwidth,trim=0 15 7 0,clip = true]{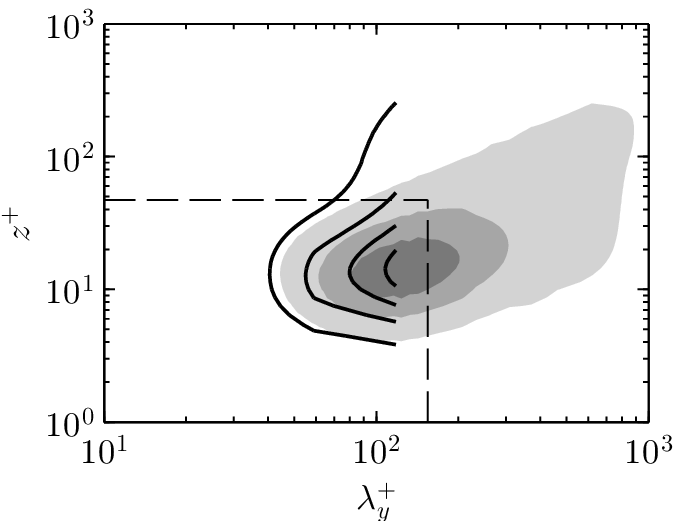} 
		\label{fig:specy2}
	}
	\subfloat[]{
		\includegraphics[width=0.45\textwidth,trim=0 15 7 0,clip = true]{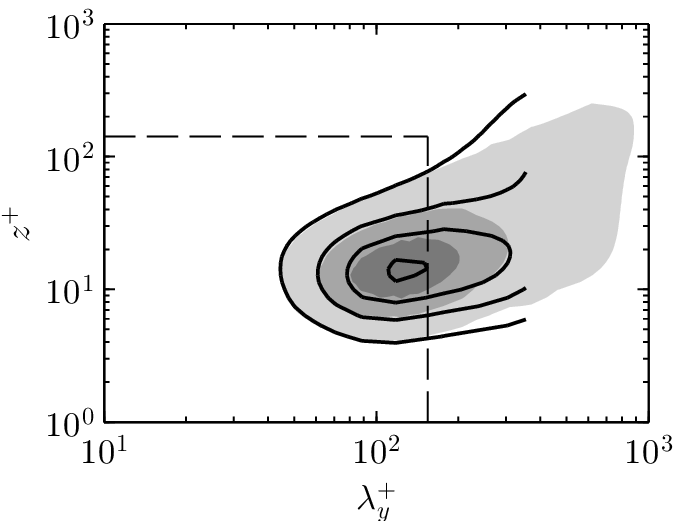} 
		\label{fig:specy6}
	}
	\put(-12.9,-1){\rotatebox{90}{Increasing length $L_x$}}
	\put(-8.7,3.65){(\emph{c}) $L_x^+\approx 460$}
	\put(-2.4,3.65){(\emph{d}) $L_x^+\approx 460$}
	\put(-8.5,0.6){$\lambda_y^+=L_x^+/3$}
	\put(-2.2,0.6){$\lambda_y^+=L_x^+/3$}
	\put(-11.25,2.2){\includegraphics[scale=0.8]{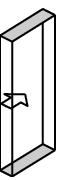}}
	\put(-5.0,2.2){\includegraphics[scale=0.8]{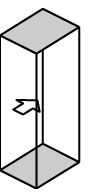}}
	\vspace{-3.8\baselineskip}
	\subfloat[]{
 		\includegraphics[width=0cm,height=4cm,trim= 0 0 0 0,clip = true]{white.eps} 
 	}
	\subfloat[]{
		\includegraphics[width=0.45\textwidth,trim=0 15 7 0,clip = true]{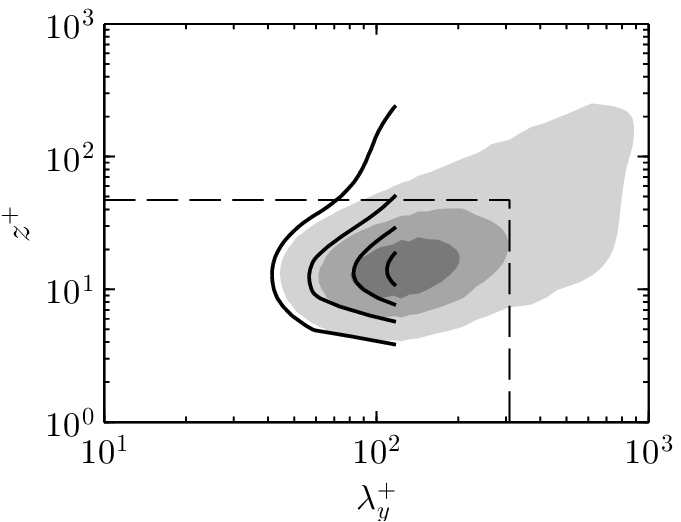} 
		\label{fig:specy3}
	}
	\subfloat[]{
		\includegraphics[width=0.45\textwidth,trim=0 15 7 0,clip = true]{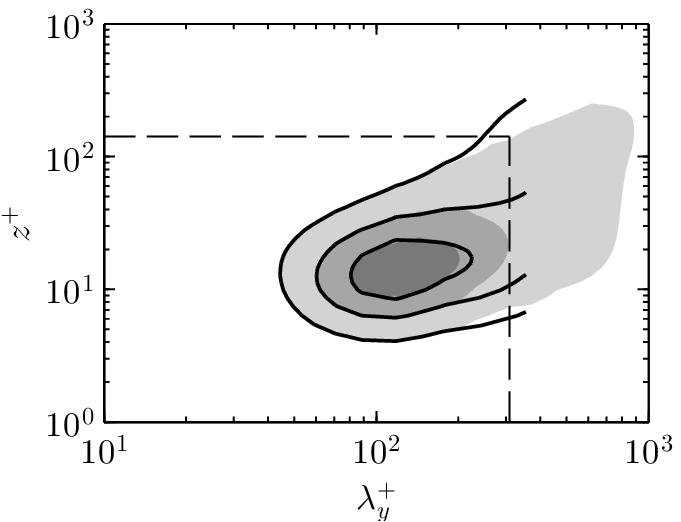} 
		\label{fig:specy7}
	}
	\put(-12.75,3.3){\vector(0,-1){0.75}}
	\put(-8.7,3.65){(\emph{e}) $L_x^+\approx 930$}
	\put(-2.4,3.65){(\emph{f}) $L_x^+\approx 930$}
	\put(-9.5,0.6){$\lambda_y^+= L_x^+/3$}
	\put(-3.25,0.6){$\lambda_y^+=L_x^+/3$}
	\put(-11.25,2.2){\includegraphics[scale=0.8]{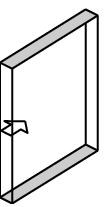}}
	\put(-5.0,2.2){\includegraphics[scale=0.8]{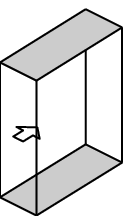}}
	\vspace{-2.5\baselineskip}
	\subfloat[]{
 		\includegraphics[width=0cm,height=4cm,trim= 0 0 0 0,clip = true]{white.eps} 
 	}
	\subfloat[]{
		\includegraphics[width=0.45\textwidth,trim=0  0 7 0,clip = true]{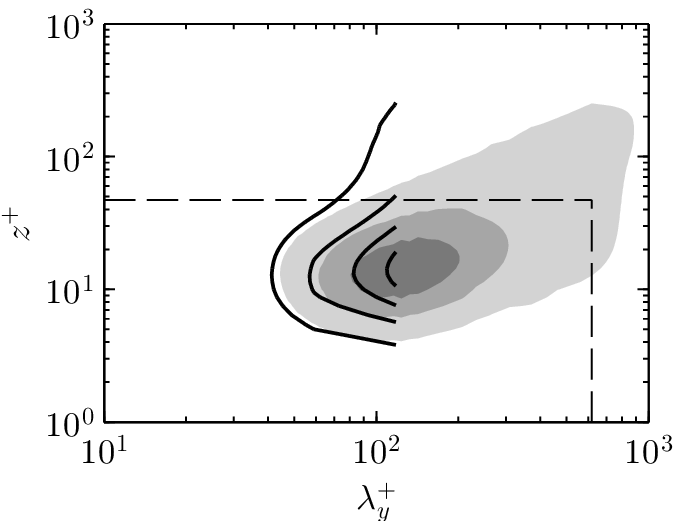} 
		\label{fig:specy4}
	}
	\subfloat[]{
		\includegraphics[width=0.45\textwidth,trim=0 0 7 0,clip = true]{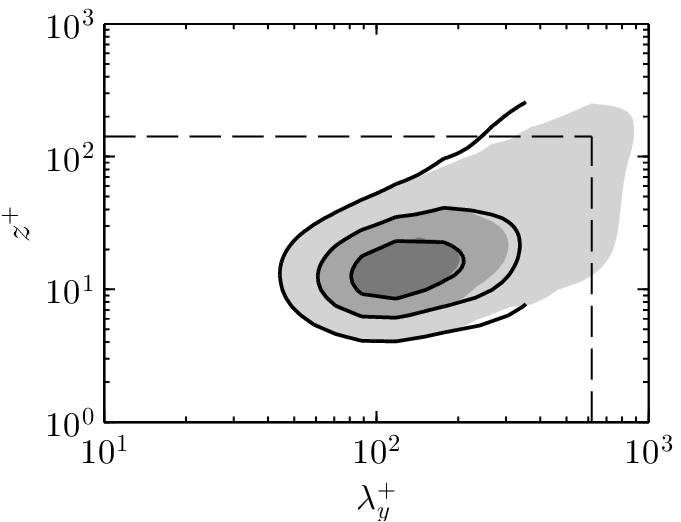} 
		\label{fig:specy8}
	}
	\put(-8.7,4.1){(\emph{g}) $L_x^+\approx 1850$}
	\put(-2.4,4.1){(\emph{h}) $L_x^+\approx 1850$}
	\put(-8.8,1.1){$\lambda_y^+= L_x^+/3$}
	\put(-2.5,1.1){$\lambda_y^+=L_x^+/3$}
	\put(-11.25,2.4){\includegraphics[scale=0.8]{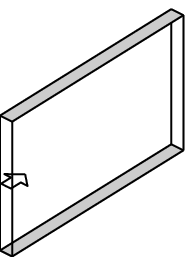}}
	\put(-5.0,2.2){\includegraphics[scale=0.8]{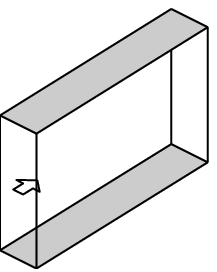}}
	\vspace{-1.7\baselineskip}
	\caption{Smooth-wall premultiplied one-dimensional spanwise energy spectra of streamwise velocity, $k_y E_{u u}^+$, for varying streamwise lengths of 
	(\emph{a},\emph{b}) $L_x^+\approx 190$, 
	(\emph{c},\emph{d}) $L_x^+\approx 460$, 
	(\emph{e},\emph{f}) $L_x^+\approx 930$, 
	and 
	(\emph{g},\emph{h}) $L_x^+\approx 1850$. 
	Spanwise domain width (\emph{a},\emph{c},\emph{e},\emph{g}) $L_y^+\approx 120$, and (\emph{b},\emph{d},\emph{f},\emph{h}) $L_y^+\approx 350$.
	Shaded contour is smooth-wall full-span channel (same for all figures). Contour levels $k_y E_{u u}^+=1.0$, 2.0, 3.0, 4.0.  Inset box gives channel scale, arrow shows direction of streamwise flow.
	Vertical dashed line shows $\lambda_y^+= L_x^+/3$, horizontal dashed line shows $z_c^+=0.4 L_y^+$.}
	\label{fig:specy}
\end{figure}

Premultiplied one-dimensional streamwise energy spectra of streamwise velocity, $k_x E_{uu}^+$, are shown in figure \ref{fig:specx}, where ${{u}_{rms}^{\prime2}}=\int_0^\infty k_x  E_{uu}\id\log(k_x)$.  Only the smooth-wall flow is shown as  the rough-wall data exhibits the same qualitative trends as the smooth wall and so is omitted. 
This occurs because the rough-wall flow is in the transitionally rough regime, which leads to a weakening of the near-wall cycle \citep{MacDonald16}. As it hasn't been destroyed, the buffer-layer streaks and quasi-streamwise vortices are active and so the same trends as the smooth-wall flow are observed. Note that in the fully rough regime, this cycle is believed to be destroyed with the inertial logarithmic region started at the roughness crests \citep{Jimenez04}. However, the scaling of $\lambda_x\approx2$--$3\lambda_y$ would likely still hold as this is generated by a self-sustaining process in the logarithmic layer \citep{Flores10,Hwang15}.
The vertical dashed lines correspond to the longest streamwise scale based on the log-layer scaling of $3L_y^+$ discussed above, while the termination of the line contours of the minimal channels show the channel streamwise length. The dashed square in the wider channels (figure \ref{fig:specx}\emph{b},\emph{d},\emph{f},\emph{h}) shows the extra captured region due to having a larger $L_y^+$, compared to the narrower minimal channels (figure \ref{fig:specx}\emph{a},\emph{c},\emph{e},\emph{g}).

It is clear that for the narrowest channel with $L_y^+\approx120$, the shortest streamwise length of $L_x^+\approx190=1.6L_y^+$ (figure \ref{fig:specx}\emph{a}) is too short. This is seen to result in an increase in turbulent energy above $z_c^+$ (horizontal dashed line), relative to the full-span channel (shaded contour).
  A similar effect is observed for $L_x^+\approx 300=2.5L_y^+$ (not shown). 
This increase in energy at smaller streamwise length scales above $z_c^+$  is in agreement with previous minimal-channel simulations \citep{Hwang13}.
When the streamwise domain length exceeds the $3L_y$ scaling with $L_x^+\approx 460=3.9L_y^+$ (figure \ref{fig:specx}\emph{c}) then reasonable agreement with the full-span channel is observed, despite the narrow range of scales captured by the minimal channel.  Further increases to the streamwise length for this domain width (figure \ref{fig:specx}\emph{e},\emph{g}) do not improve the collapse with the full-span channel, especially near $z_c^+$, as no additional turbulent motions are captured according to the $3L_y^+$ scaling discussed above. The increased length is however able to capture the near-wall cycle with peak at $z^+\approx 15$ and $\lambda_x^+\approx1000$ (figure \ref{fig:specx}\emph{g}).

A similar picture emerges for the wider minimal channel with $L_y^+\approx354$. The shortest domain length tested, $L_x^+\approx190\approx0.54L_y^+$ (figure \ref{fig:specx}\emph{b}), is unable to capture much of the turbulent motions, similar to that observed in the narrower minimal channel of figure \ref{fig:specx}(\emph{a}). 
A streamwise domain length of $L_x^+\approx463\approx1.3L_y^+$ (figure \ref{fig:specx}\emph{d}) has better agreement with the full-span channel, although there is still some discrepancy. The $3L_y$ scaling is almost reached in figure \ref{fig:specx}(\emph{f}) with $L_x+\approx927\approx2.6L_y$, and excellent agreement is observed with the full-span channel, with only a slight increase in energy above $z_c^+$. Further increasing $L_x^+$ above $3L_y$ (figure \ref{fig:specx}\emph{h}) does not provide any improvement other than capturing more of the near-wall cycle, further supporting this scaling argument.

The premultiplied spanwise energy spectra of streamwise velocity are shown in figure \ref{fig:specy}. Here, the vertical dashed line now shows the widest spanwise length scale that can exist based on the streamwise domain length using the $\lambda_y=3L_x$ scaling. For the narrower channel with $L_y^+\approx 120$, the scaling is reached when $L_x^+\approx460=3.9L_y^+$ (figure \ref{fig:specy}\emph{c}), and further increases to $L_x^+$ (figure \ref{fig:specy}\emph{e},\emph{g}) do not result in any change to the turbulent energy. Similarly for the wider channel with $L_y^+\approx 350$, the scaling is approximately reached in figure \ref{fig:specy}(\emph{f}) with $L_x^+\approx 930=2.6L_y^+$ and increasing $L_x^+$ to 1850 viscous units (figure \ref{fig:specy}\emph{h}) does not result in any substantial change to the spectra. The cases with $L_x\lesssim2$--$3L_y$ (figure \ref{fig:specy}\emph{a},\emph{b},\emph{d}) result in enhanced turbulence energy, particularly in the near-wall peak at $z^+\approx 15$ and $\lambda_y^+\approx 100$.

Given the above results, it is possible to restrict the streamwise length to $L_x=3L_y$. This is due to quasi-streamwise vortices that exist in the logarithmic layer (where neither viscosity or the channel half height are relevant length scales), suggesting this scaling is independent of $Re_\tau$.
 Smaller streamwise lengths than this scaling lead to poor agreement between the minimal and full-span channel, as seen in figures \ref{fig:specx} and \ref{fig:specy}.
However, for very narrow channels this can result in a streamwise domain length less than 1000 viscous units. While it appears the buffer-layer streaks do not need to be captured in the domain, having such a small streamwise domain exacerbates the bursty nature of the minimal-span channel \citep{Jimenez15}. Only one or maybe two quasi-streamwise vortices are present in the domain, which makes obtaining converged statistics difficult as the simulation needs to be run for a significantly long time, an issue discussed in detail in \S4. We believe that a minimum streamwise length of approximately 1000 viscous units seems a reasonable length in these cases, so that several of the smallest quasi-streamwise vortices are present. 
 As such, we recommend that $L_x\gtrsim\max(\hphantom{.}3L_y,1000\nu/U_\tau,\lambda_{r,x})$, where the last requirement pertains to the streamwise length scale, $\lambda_{r,x}$, of the roughness which is absent in the current roughness forcing model.

%
%
\subsection{Half-height channel (table \ref{tab:sims2v2})}
\label{sect:half}
\begin{table}
\centering
\begin{tabular}{ c c  c c c c   c c c  c c c c  c  c}
ID	& $Re_\tau$	& $\tfrac{L_x}{h}$	&  $L_x^+$	& $L_y^+$	& $\tfrac{L_z}{h}$	& $N_x$	& $N_y$	& $N_z$	& $\Delta x^+$	& $\Delta y^+$	& $\Delta z_w^+$	& $\Delta z_h^+$	& $\Delta t^+_{S}$	& $\Delta t^+_{R}$\\[0.5em] 
  \multicolumn{15}{c}{Full-span channel} \\[0.5em] 
FS	& 590		& $2\pi$	& 3707	& 1854	&2	& 384		& 384		& 256		& 9.7		& 4.8		& 0.04	& 7.2		& 0.26	& 0.29	\\
FSH	& 590		& $2\pi$	& 3707	& 1854	&1	& 384		& 384		& 128		& 9.7		& 4.8		& 0.04	& 7.2		& 0.29	& 0.32	\\[0.5em] 
  \multicolumn{15}{c}{Minimal-span channel} \\[0.5em] 
MS6	& 590		& $2\pi$	& 3707	& 118		&2	& 384		& 24		& 256		& 9.7		& 4.9		& 0.04	&7.2		& 0.35	& 0.40	\\
MSH6	& 590		& $2\pi$	& 3707	& 118 		&1	& 384		& 24		& 128		& 9.7		& 4.9		& 0.04	& 7.2		& 0.36	& 0.41	\\
\end{tabular}
\caption{Full-span and minimal-span channel simulations, for standard-height (two no-slip walls, $L_z/h=2$) and half-height (one no-slip and one slip wall, $L_z/h=1$) channels.
Entries are same as table \ref{tab:sims1v2}.
}
\label{tab:sims2v2}
\end{table}

\setlength{\unitlength}{1cm}
\begin{figure}
\centering
 \captionsetup[subfigure]{labelformat=empty}
	\subfloat[]{
		\includegraphics[width=0.49\textwidth]{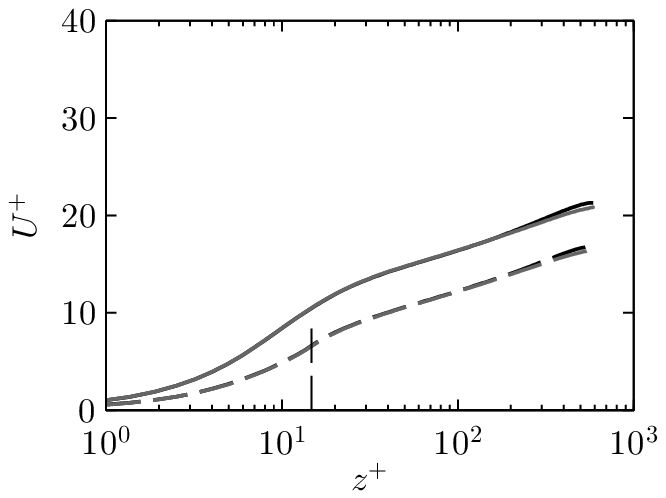}
		\label{fig:UxValid1}
	}
	\subfloat[]{
		\includegraphics[width=0.49\textwidth]{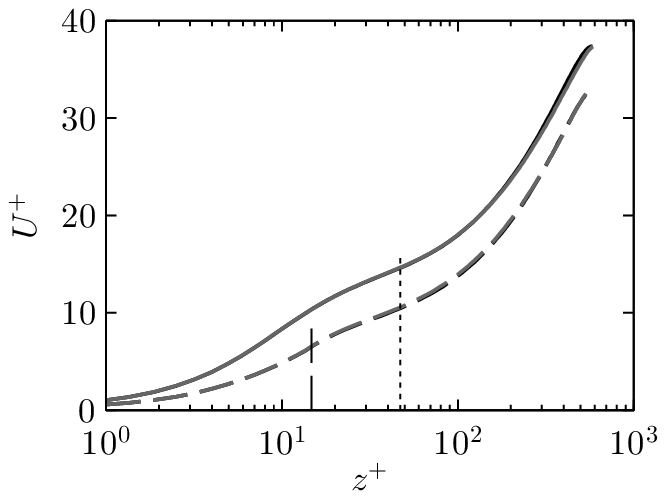}
		\label{fig:UxValid2}
	}
	\put(-12.45,1.9){\includegraphics{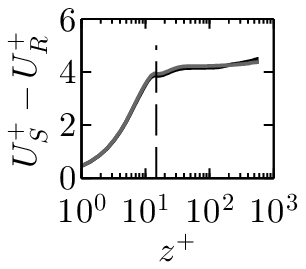}}
	\put(-13.55,4.6){(\emph{a})}
	\put(-6.7,4.6){(\emph{b})}
	\put(-9.3,4.2){Full-span}
	\put(-5.6,1.9){\includegraphics{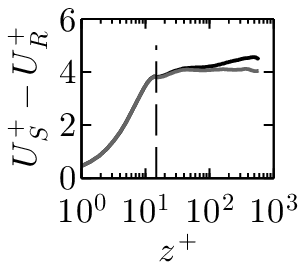}}
	\put(-2.8,4.2){ Minimal-}
	\put(-2.5,3.9){span}
	\vspace{-1.8\baselineskip}
	\caption{Mean streamwise velocity against wall-normal position for (\emph{a}) full-span channel  and (\emph{b}) minimal-span channel. Line styles are:
	black, standard-height channel;
	grey, half-height channel;
	solid, smooth wall;
	dashed, rough-wall model.
	Vertical dashed line shows roughness crest, $k=h/40$,
	vertical dotted line shows critical height, $z_c=0.4L_y$.
	Inset shows difference in smooth- and rough-wall velocity, $U_S^+-U_R^+$.}
	\label{fig:velProfHalf}
\end{figure}

The previous section looked at the effect of the streamwise domain. Now we will consider the effect of the wall-normal domain, particularly in terms of the outer-layer flow which is unphysical in minimal channels. First, we will consider a half-height (open) channel which consists of a slip wall positioned at the channel centreline. Intuitively, changing the boundary condition at $z=h$ is unlikely to effect the flow at $z=z_c$, given the distance between these two heights ($z_c/h\approx0.1$ in this study). For efficient roughness simulations, this has the benefit of  reducing the size of the grid by a half when compared to a conventional standard-height channel with two no-slip walls.
The simulations are detailed in table \ref{tab:sims2v2} and both full-span and minimal-span channels are simulated. Here, the streamwise length is fixed at $L_x^+\approx3700$ and there is no outer-layer damping, $K_i = 0$. 

Figure \ref{fig:velProfHalf} shows the effect of the half-height channel in terms of the mean velocity profile, when compared to the standard-height channel. This is done for both full-span (figure \ref{fig:velProfHalf}\emph{a}) and minimal-span (figure \ref{fig:velProfHalf}\emph{b}) channels, for smooth and rough walls. For clarity, the full-span and minimal-span channels are shown in different figures, however the near-wall behaviour is identical up until $z_c= 0.4L_y\Rightarrow z_c^+\approx50$, as observed previously (figure \ref{fig:uLength}). Both sets of figures show that the use of the half-height channel with slip wall has a negligible effect on the flow. The main difference is seen in the wake region, where the half-height channel restricts the outer-layer structures, resulting in a slight decrease in the mean velocity. Crucially, this difference is the same for both smooth and rough walls, meaning that the difference between them, $U_S^+-U_R^+$, is the same. Moreover, the change in the half-height channel occurs above the critical wall-normal location, where the minimal-span flow is already altered compared to the full-span flow.

\setlength{\unitlength}{1cm}
\begin{figure}
\centering
 \captionsetup[subfigure]{labelformat=empty}
	\subfloat[]{
		\includegraphics[width=0.49\textwidth]{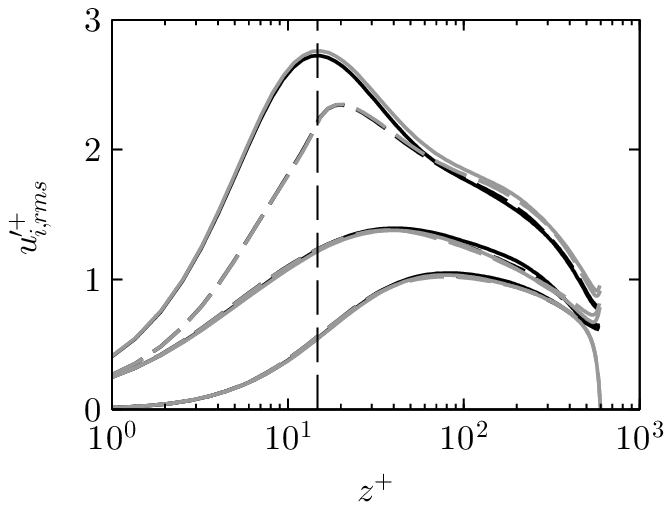}
		\label{fig:uflucValid1}
	}
	\subfloat[]{
		\includegraphics[width=0.49\textwidth]{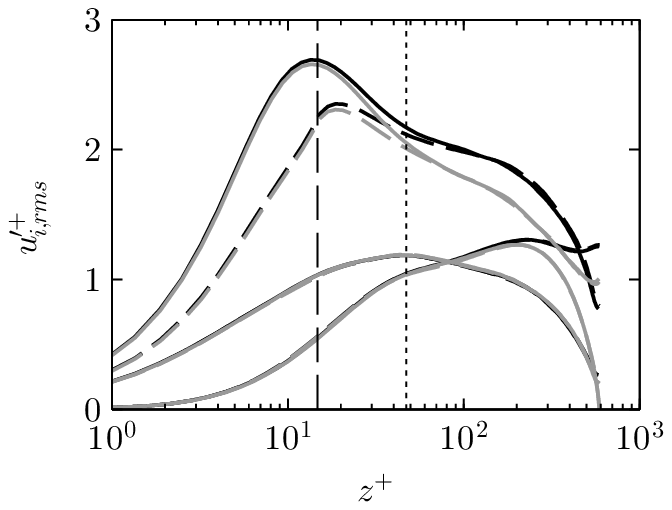}
		\label{fig:uflucValid2}
	}
	\put(-13.35,4.65){(\emph{a})}
	\put(-6.5,4.65){(\emph{b})}
	\put(-9.0,4.2){Full-span}
	\put(-12.2,3.5){$u_{x,rms}'^+$}
	\put(-10.3,2.9){$u_{y,rms}'^+$}
	\put(-10.2,1.65){$u_{z,rms}'^+$}
	\put(-2.7,4.2){Minimal-span}
	\put(-5.4,3.5){$u_{x,rms}'^+$}
	\put(-3.5,2.6){$u_{y,rms}'^+$}
	\put(-3.35,1.65){$u_{z,rms}'^+$}
	\vspace{-1.8\baselineskip}
	\caption{Streamwise, spanwise, and wall-normal root-mean-squared velocity fluctuations against wall-normal position for (\emph{a}) full-span channel  and (\emph{b}) minimal-span channel. Line styles are same as figure \ref{fig:velProfHalf}.}
	\label{fig:turbIntHalf}
\end{figure}

The streamwise, spanwise, and wall-normal root-mean-squared velocity fluctuations are shown in figure \ref{fig:turbIntHalf} for both full-span and minimal-span channels, comparing standard-height and half-height channels. For the full-span channel (figure \ref{fig:turbIntHalf}\emph{a}), these velocity fluctuations show very good agreement between the standard-height (black lines) and half-height (grey lines) channels in the near-wall region, in agreement with previous full-span half-height channel studies \citep{Handler99,Leonardi10}. The half-height channel has zero wall-normal velocity fluctuations at the channel centre due to the impermeability constraint of the slip wall. The streamwise velocity fluctuations of the half-height channels are slightly enhanced above $z^+\gtrsim40$ relative to the standard-height channels, however the difference is relatively small. Moreover, outer-layer similarity is still maintained between the smooth- and rough-wall channels, suggesting the effect on the difference between the two flows will be minor.

For the minimal-span channel (figure \ref{fig:turbIntHalf}\emph{b}), a slightly different picture emerges. Firstly, it should be noted that a standard-height minimal-span channel (black lines of figure \ref{fig:turbIntHalf}\emph{b}) have enhanced streamwise and wall-normal velocity fluctuations compared to its full-span counterpart (black lines of figure \ref{fig:turbIntHalf}\emph{a}) in the outer layer. 
This has been noted in other minimal-span channel studies (e.g. \citealt{Hwang13,MacDonald16}), so will not be discussed in depth here.
However, when a half-height channel is used in a minimal-span channel, the streamwise velocity fluctuations (grey lines of figure \ref{fig:turbIntHalf}\emph{b}) are damped from $z^+\gtrsim20$ compared to the standard-height minimal-span channel. This is the opposite of what occurred in a full-span channel.
Interestingly, \cite{Hwang13} found that when the spanwise uniform eddies (that is, the $k_y=0$ mode) were filtered out of a minimal-span channel, a similar behaviour was observed in that the wall-normal and streamwise velocity fluctuations were damped compared to the unfiltered channel. It is unclear whether the half-height channel is performing a similar operation to this $k_y=0$ filtering, as this would imply the impermeability constraint is limiting the infinite spanwise motions in some way. However, it is clear that the near-wall flow is preserved, despite the imposition of an outer-layer boundary condition.
 This shows that a half-height channel flow can be simulated without significant near-wall detriment. 

\setlength{\unitlength}{1cm}
\begin{figure}
\centering
 \captionsetup[subfigure]{labelformat=empty}
	\subfloat[]{
		\includegraphics[width=0.49\textwidth]{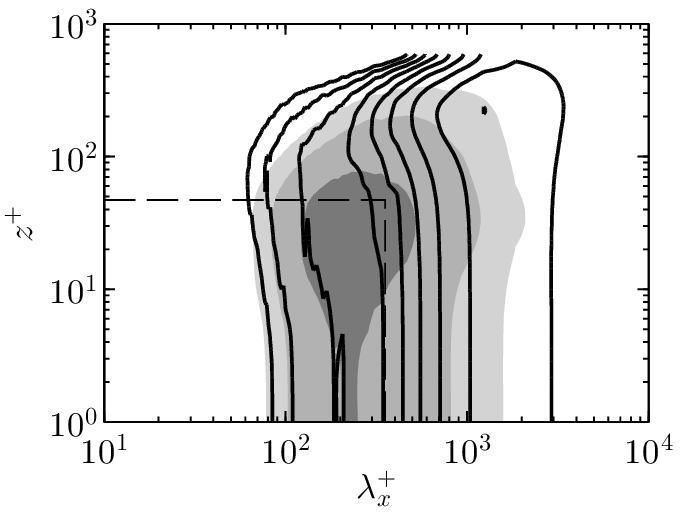}
		\label{fig:pspectra}
	}
	\subfloat[]{
		\includegraphics[width=0.49\textwidth]{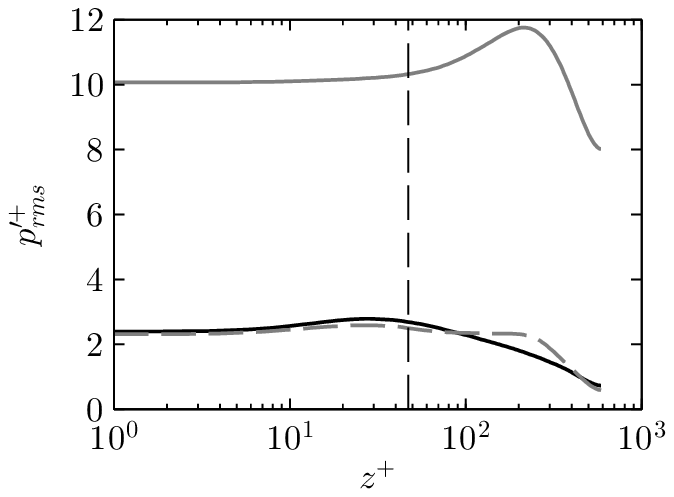}
		\label{fig:prmsCorrect}
	}
	\put(-13.35,4.65){(\emph{a})}
	\put(-6.5,4.65){(\emph{b})}
	\vspace{-1.8\baselineskip}
	\caption{
	(\emph{a}) Smooth-wall premultiplied one-dimensional streamwise energy spectrum of pressure, $k_x E_{pp}^+$.
	Shaded contour is full-span channel, line contour is minimal-span half-height channel.
	Horizontal dashed line shows $z_c^+=0.4L_y^+$, vertical dashed line shows $3L_y^+$ (\S \ref{sect:slength}).
	Contour levels are logarithmically spaced over 8 intervals between 0.75 and 66.
(\emph{b}) 
Pressure root-mean-squared fluctuations against wall-normal position. Line styles:
	black, full-span channel;
	grey solid, minimal-span half-height channel;
	grey dashed, minimal-span channel with altered velocity field $u^a$ (see text).
	Vertical dashed line shows $z_c^+$.
	}
	\label{fig:pressCorrect}
\end{figure}

The pressure fluctuations, meanwhile, show a substantial increase in the minimal-span channel compared to the full-span channel. The premultiplied streamwise spectra of pressure is shown in figure \ref{fig:pressCorrect}(\emph{a}), and it is clear that the fluctuations are larger at longer wavelengths ($\lambda_x^+\approx1000$--2000) and at heights above $z_c^+$. The fluctuations at wavelengths shorter than $3L_y^+$ and below $z_c^+$ are, however, in reasonable agreement with the full-span channel. Despite this, because the root-mean-squared pressure fluctuations are the integral of this spectrum ($p_{rms}^{\prime2}=\int_0^\infty k_x E_{p p}d\log(k_x)$), then the profile is nearly an order of magnitude larger in the minimal-span channel (grey line, figure \ref{fig:pressCorrect}\emph{b}). Note that the mean pressure is still correct, as this must obey the averaged wall-normal momentum equation, $\overline{w^2}+\overline{p}=C$, where $C$ is a constant.

A possible solution to this behaviour is to consider decomposing the pressure into its rapid and slow parts \citep{Kim89}. Here, the rapid pressure emerges due to the mean streamwise velocity gradient while the slow pressure is due to the velocity fluctuation gradients. The mean velocity gradient $dU/dz$ in the minimal-span channel is enhanced above $z_c$, which means the rapid pressure will also be larger. If we instead define an altered streamwise velocity field whereby $u^a(x,y,z)=u(x,y,z)-(U(z)-U(z_c))$ for $z>z_c$ where $U(z_c)$ is the mean streamwise velocity at $z=z_c$, this will then set the mean $dU^a/dz=0$. If the other two velocity components are unchanged and $u^a=u$ for $z\le z_c$, then the rapid pressure would be zero above $z_c$. The pressure field arising from this altered velocity field, $\nabla^2 p^a=-u^a_{i,j}u^a_{j,i}$ can then be computed and is seen to be in much better agreement with the full-span channel (figure \ref{fig:pressCorrect}\emph{b}). There is a slight difference in the near-wall peak near $z^+\approx30$, however this may be due to the discontinuity in $dU^a/dz$ at $z=z_c$.
 This correction to pressure is an additional step that needs to be performed whenever pressure statistics are to be outputted. The unaltered (discrete) pressure is still necessary at each time step to ensure the flow remains divergence free.

%
%
\subsection{Outer-layer damping (table \ref{tab:sims3v2})}
\label{sect:outerForcing}
\begin{table}
\centering
\begin{tabular}{ c c  c c c c   c c c  c c c c  c  c c}
ID	& $Re_\tau$	& $\tfrac{L_x}{h}$	&  $L_x^+$	& $L_y^+$	& $\tfrac{L_z}{h}$	& $N_x$	& $N_y$	& $N_z$	& $\Delta x^+$	& $\Delta y^+$	& $\Delta z_w^+$	& $\Delta z_h^+$	& $z_d^+$ 	& $\Delta t^+_{S}$	& $\Delta t^+_{R}$\\[0.5em] 

  \multicolumn{16}{c}{Minimal-span, half-height channel with outer-layer damping} \\[0.5em] 

MSHD1	& 590		& $2\pi$	& 3707	& 118		&1		& 384		& 24		& 128		& 9.7		& 4.9		&0.04		&7.2		& 100 		& 0.48	& 0.59	\\
MSHD2	& 590		& $2\pi$	& 3707	& 118		&1		& 384		& 24		& 128		& 9.7		& 4.9		& 0.04	&7.2		& 150 		& 0.46	& 0.54	\\	
MSHD3	& 590		& $2\pi$	& 3707	& 118		&1		& 384		& 24		& 128		& 9.7		& 4.9		& 0.04	& 7.2		& 200 		& 0.44	& 0.51	\\
MSHD4	& 590		& $2\pi$	& 3707	& 118		&1		& 384		& 24		& 128		& 9.7		& 4.9		& 0.04	&7.2		& 300 		& 0.41	& 0.47	\\
MSH6	&590		& $2\pi$	& 3707	& 118 		&1		& 384		& 24		& 128		& 9.7		& 4.9		& 0.04	& 7.2		& - 		& 0.36	& 0.41	\\[0.5em] 

  \multicolumn{16}{c}{Minimal-span, half-height channel with outer-layer damping (wider span)} \\[0.5em] 
MSWHD1	& 590		& $2\pi$	& 3707	& 236		&1		& 384		& 48		& 128		& 9.7		& 4.9		& 0.04	&7.2		& 200 		& 0.43	& 0.48	\\
MSWHD2	& 590		& $2\pi$	& 3707	& 236		&1		& 384		& 48		& 128		& 9.7		& 4.9		&0.04		& 7.2		& 250		& 0.43	& 0.47	\\
MSWHD3	& 590		& $2\pi$	& 3707	& 236		&1		& 384		& 48		& 128		& 9.7		& 4.9		& 0.04	&7.2		& 300 		& 0.42	& 0.46	\\
MSWH6	&590		& $2\pi$	& 3707	& 236		&1		& 384		& 48		& 128		& 9.7		& 4.9		& 0.04	& 7.2		& -		& 0.40	& 0.43	\\[0.5em] 

  \multicolumn{16}{c}{Minimal-span, half-height channel with outer-layer damping (increased $Re_\tau$)} \\[0.5em] 
MSHRD1	&  2000		& \Retwo	& 3707	& 300		&1		& 384		& 60		& 384		& 9.7		& 5.0		& 0.02	& 8.2		& 300 		& 0.25	& 0.37	\\
MSHRD2	&  2000		& \Retwo	& 3707	& 300		&1		& 384		& 60		& 384		& 9.7		& 5.0		& 0.02	& 8.2		& 400 		& 0.24	& 0.35	\\
MSHR6	&2000		& \Retwo	& 3707	& 300		&1		& 384		& 60		& 384		& 9.7		& 5.0		& 0.02	& 8.2		& -		& 0.24	& 0.31	\\

\end{tabular}
\caption{Minimal-span simulations with outer-layer damping. Refer to table \ref{tab:sims1v2} for definitions. $z_d^+$ indicates location where damping starts (figure \ref{fig:forcingZones}), cases with no damping indicated by a hyphen.
}
\label{tab:sims3v2}
\end{table}

The previous section showed that a half-height (open) channel did not significantly alter the healthy near-wall flow. A more aggressive alteration is now attempted in which the outer-layer flow is damped through use of the $K_i$ forcing term (\ref{eqn:outerForcing}) in a half-height channel.
This will reduce the large centreline velocity (figure \ref{fig:velProfHalf}\emph{b}) of the minimal-span channel, which places a restriction on the time step due to the $\textit{CFL}$ number.
The streamwise domain length is fixed at $L_x^+\approx3700$, and other relevant parameters are detailed in table \ref{tab:sims3v2}.
The height where the damping begins, $z_d$, is varied in channels of two different spanwise widths; $L_y^+\approx120$ and $L_y^+\approx240$. These minimal channels will have a healthy turbulence region up to $z_c^+\approx47$ and $z_c^+\approx94$, respectively. 

\setlength{\unitlength}{1cm}
\begin{figure}
\centering
 \captionsetup[subfigure]{labelformat=empty}
	\subfloat[]{
		\includegraphics[width=0.49\textwidth,trim=0 10 0 0,clip=true]{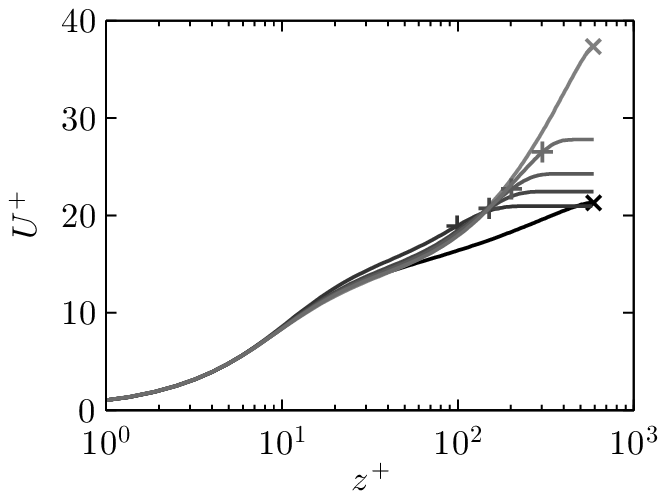}
		\label{fig:uOuterForceSm}
	}
	\subfloat[]{
		\includegraphics[width=0.49\textwidth,trim=0 10 0 0,clip=true]{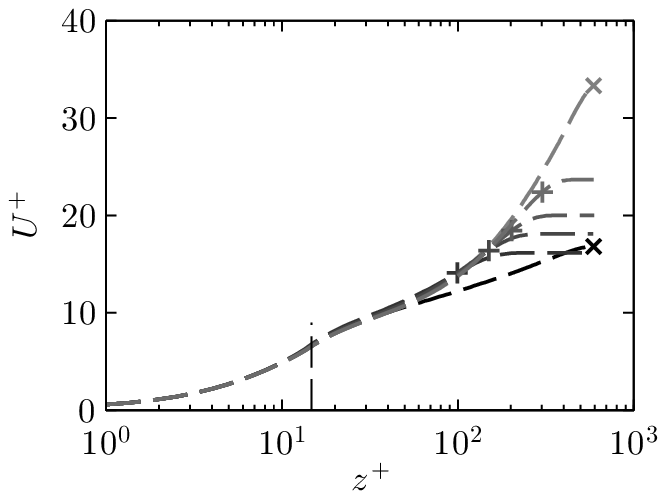}
		\label{fig:uOuterForceRo}
	}
	\put(-13.55,4.25){(\emph{a})}
	\put(-6.7,4.25){(\emph{b})}
	\put(-9.5,3.85){Smooth,}
	\put(-9.5,3.5){$L_y^+$$\approx$$120$}
	\put(-10.5,2.2){Inc.\ $z_d$}
	\put(-10.0,2.1){\vector(1,-2){0.25}}
	\put(-8.5,2.35){\vector(1,2){0.5}}
	\put(-2.5,3.85){Rough,}
	\put(-2.5,3.5){$L_y^+$$\approx$$120$}
	\put(-1.8,1.9){\vector(3,4){0.8}}
	\put(-5.5,1.5){\includegraphics{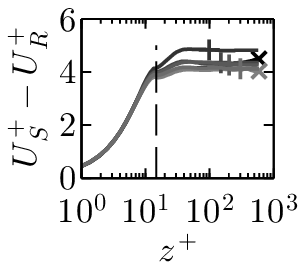}}
	\put(-3.85,3.7){\vector(1,-2){0.25}}
	\vspace{-2.75\baselineskip}
	\\
	\subfloat[]{
		\includegraphics[width=0.49\textwidth]{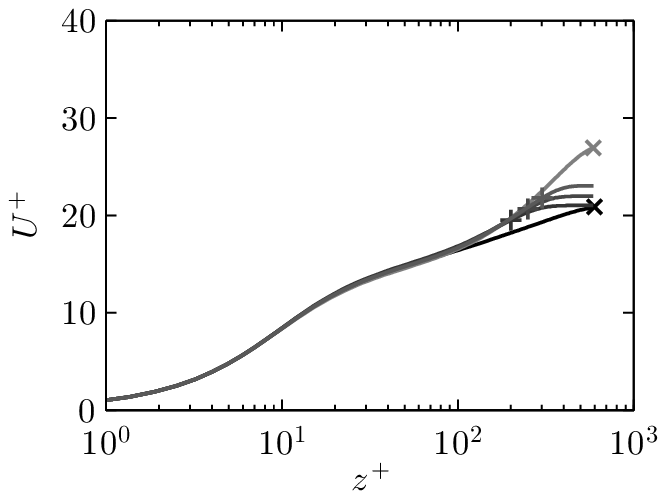}
		\label{fig:uOuterForceSmWider}
	}
	\subfloat[]{
		\includegraphics[width=0.49\textwidth]{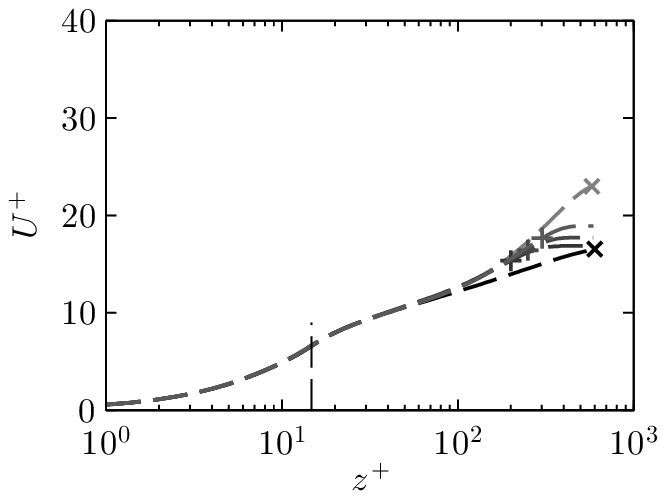}
		\label{fig:uOuterForceRoWider}
	}
	\put(-13.5,4.6){(\emph{c})}
	\put(-6.7,4.6){(\emph{d})}
	\put(-9.5,4.2){Smooth,}
	\put(-9.5,3.85){$L_y^+$$\approx$$240$}
	\put(-8.25,2.73){\vector(1,1){0.5}}
	\put(-2.5,4.2){Rough,}
	\put(-2.5,3.85){$L_y^+$$\approx$$240$}
	\put(-1.5,2.3){\vector(1,1){0.5}}
	\put(-5.5,1.85){\includegraphics{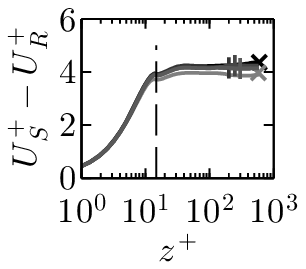}}
	\vspace{-1.8\baselineskip}
	\caption{Streamwise velocity profile for outer-layer damping for (\emph{a},\emph{c}) smooth- and (\emph{b},\emph{d}) rough-wall half-height channels.
	Spanwise domain width is (\emph{a},\emph{b}) $L_y^+\approx120$ and (\emph{c},\emph{d}) $L_y^+\approx240$ (table \ref{tab:sims3v2}).
	Line styles are:
	black, full-span channel;
	grey, minimal-span channel with outer-layer damping (\ref{eqn:outerForcing}) for varying damping heights of (\emph{a},\emph{b}) $z_d^+ = (100,150,200,300)$  and (\emph{c},\emph{d}) $z_d^+ = (200,250,300)$, shown by \protect\raisebox{0.4ex}{\protect\scalebox{1.0}{$\boldsymbol{\pmb{+}}$}};
	light-grey, minimal-span with no damping.
	Vertical dashed line indicates the roughness height $k = h/40$.
	Inset of (\emph{b},\emph{d}) shows the difference in smooth- and rough-wall velocity.
	Arrows indicates trend of increasing $z_d$.
	}
	\label{fig:uOuterForce}
\end{figure}

Figure \ref{fig:uOuterForce} shows the mean streamwise velocity profile for smooth- and rough-wall minimal-span channels, for both of the spanwise widths at $Re_\tau=590$. The smallest spanwise width, $L_y^+\approx120\Rightarrow z_c^+\approx47$ (figure \ref{fig:uOuterForce}\emph{a},\emph{b}) has four different positions for $z_d$ examined, with values of $z_d^+ = (100,150,200,250)$. In all cases, the damping tends to limit the streamwise velocity above $z>z_d$ to the value of $U(z=z_d)$. It is clear in the smooth-wall flow (figure \ref{fig:uOuterForce}\emph{a}) that the two smallest values of $z_d^+=100$ and 150 are too close to the wall, contaminating the healthy near-wall flow and resulting in an increase in the streamwise velocity below $z_d$. In particular, $z_d^+=100$ results in an obvious overestimation of $\Delta U^+$, as seen in the inset of figure \ref{fig:uOuterForce}(\emph{b}). A value of $z_d^+=150$ has a similar, although not as strong effect. This is clearly not desirable and suggests that values of $z_d^+\ge 200$ are necessary to avoid contamination of the healthy near-wall flow.

To investigate how the location of $z_d$ is related to $z_c$ will require cases with a larger spanwise domain width, as $z_c\approx0.4L_y$. This is done in figure \ref{fig:uOuterForce}(\emph{c},\emph{d}), where the spanwise width is now $L_y^+\approx240\Rightarrow z_c^+\approx94$.
Here, three different positions for $z_d$ are analysed, with values of $z_d^+=(200,250,300)$. There is little difference in the near-wall profiles of the three different $z_d$ positions. This suggests that, even though $z_c$ has doubled, we still require the damping to begin at least 200 viscous-units away from the wall.
 Having the outer-layer damping begin closer to the wall (i.e. reducing $z_d$) leads to its effects permeating the entire near-wall region, increasing the viscous stress in the log-layer and even buffer layer. This results in an overestimation of the mean streamwise velocity. The strong effect of the damping when $z_d$ is small could also be due to the strength of the damping, $\gamma h/U_\tau\approx 1$ in (\ref{eqn:outerForcing}). Reducing the strength would likely reduce the strong near-wall effects that the damping has, although this is not pursued here.

\begin{figure}
\centering
 \captionsetup[subfigure]{labelformat=empty}
	\subfloat[]{
		\includegraphics[width=0.49\textwidth]{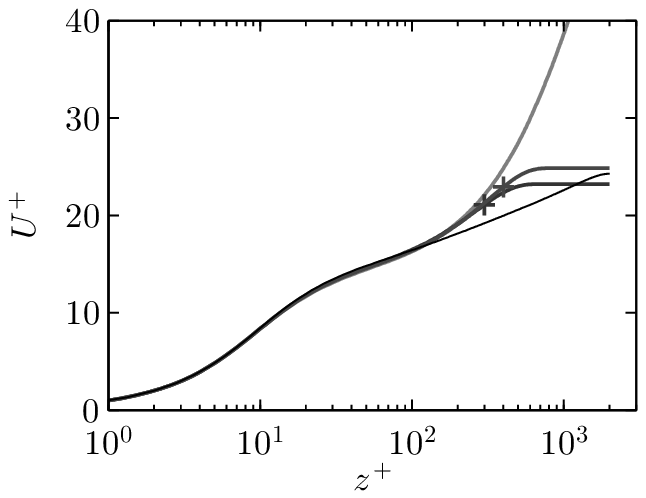}
		\label{fig:uOuterForceSmReHigh}
	}
	\subfloat[]{
		\includegraphics[width=0.49\textwidth]{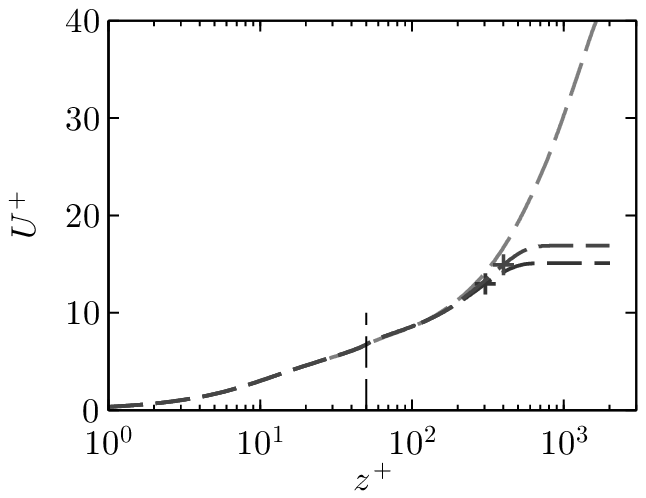}
		\label{fig:uOuterForceRoReHigh}
	}
	\put(-13.5,4.8){(\emph{a})}
	\put(-6.6,4.8){(\emph{b})}
	\put(-9.5,4.4){Smooth}
	\put(-8.1,3.60){Inc.\ $z_d$}
	\put(-8.3,3.0){\vector(1,1){0.5}}
	\put(-2.3,4.4){Rough}
	\put(-1.3,2.25){\vector(1,1){0.5}}
	\put(-5.5,1.7){\includegraphics{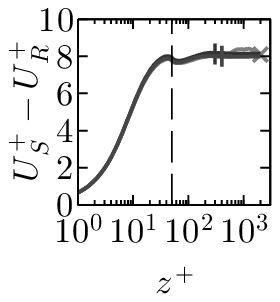}}
	\vspace{-1.8\baselineskip}
	\caption{Streamwise velocity profile for (\emph{a}) smooth- and (\emph{b}) rough-wall half-height channels at $Re_\tau=2000$ (table \ref{tab:sims3v2}).
	Line styles are same as figure \ref{fig:uOuterForce}, however full-span smooth-wall channel data taken from \cite{Hoyas06}
	and damping heights are $z_d^+ = (200,300)$, shown by \protect\raisebox{0.4ex}{\protect\scalebox{1.0}{$\boldsymbol{\pmb{\times}}$}}.
	}
	\label{fig:uOuterForceReHigh}
\end{figure}

Next, the friction Reynolds number is increased to $Re_\tau=2000$. Since the roughness height is fixed as a ratio of the channel half-height, $k=h/40$, then the roughness Reynolds number increases to $k^+ = 50$. The spanwise width also needs to be increased to submerge the roughness sublayer in healthy turbulence \citep{Chung15}; here it takes a value of $L_y^+=300$ so that the critical wall-normal location is $z_c^+ \approx 120=2.4k$. Figure \ref{fig:uOuterForceReHigh} shows the mean velocity profile for these higher $Re_\tau$ simulations in which $z_d^+=(200,300)$. Only smooth-wall full-span channel data is available from \cite{Hoyas06} to compare with the minimal-span channel, and good agreement is seen up until $z^+\approx120=0.4L_y^+$ (figure \ref{fig:uOuterForceReHigh}\emph{a}). Here, both positions of $z_d$ produce velocity profiles which compare well with the minimal-span channel with no outer-layer damping, up to the forcing location $z_d$.

\setlength{\unitlength}{1cm}

\setlength{\unitlength}{1cm}
\begin{figure}
\centering
 \captionsetup[subfigure]{labelformat=empty}
	\subfloat[]{
		\includegraphics[width=0.49\textwidth]{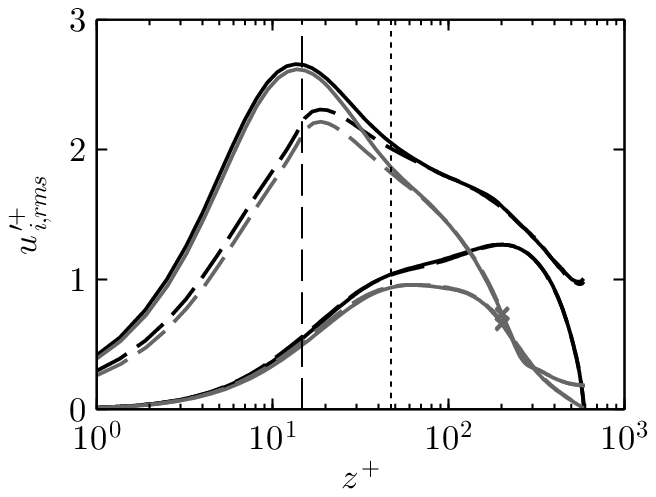}
		\label{fig:turbIntOuterForcing}
	}
	\subfloat[]{
		\includegraphics[width=0.49\textwidth]{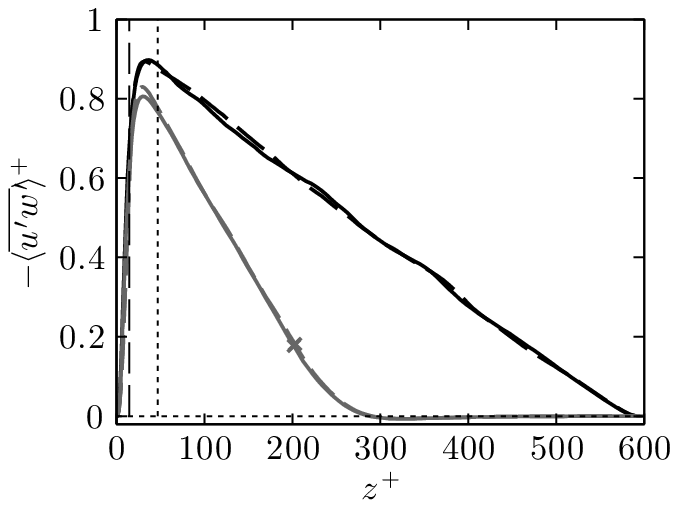}
		\label{fig:reStressOuterForcing}
	}
	\put(-13.5,4.6){(\emph{a})}
	\put(-6.5,4.6){(\emph{b})}
	\put(-12.3,3.5){$u_{x,rms}'^+$}
	\put(-10.1,1.6){$u_{z,rms}'^+$}
	\vspace{-1.6\baselineskip}
	\caption{Effect of the outer-layer damping on (\emph{a}) the root mean squared velocity fluctuations in the streamwise and wall-normal directions, and (\emph{b}) the Reynolds stress.
Line styles:
black, minimal-span, half-height channel (no outer-layer damping);
grey, minimal-span, half-height channel with outer-layer damping starting at $z_d^+=200$ shown by \protect\raisebox{0.4ex}{\protect\scalebox{1.0}{$\boldsymbol{\pmb{\times}}$}};
solid, smooth wall;
dashed, rough wall.
Vertical dashed line shows roughness crest, vertical dotted line shows critical height $z_c=0.4L_y$.
}
	\label{fig:turbIntOuterForce}
\end{figure}

It is also of interest to see what effect the outer-layer damping  has on second-order turbulence statistics. This is shown in figure \ref{fig:turbIntOuterForce} for the streamwise and wall-normal velocity fluctuations (figure \ref{fig:turbIntOuterForce}\emph{a}) and the Reynolds shear stress (figure \ref{fig:turbIntOuterForce}\emph{b}), for the case of outer forcing with $z_d^+=200$ and $L_y^+=120$. This is compared to the minimal-span channel with the same spanwise width, but with no outer-layer damping. The turbulence intensity up until $z^+\approx 30$ agrees reasonably well  between the case with forcing and without, however because the forcing term scales as $(u_i - \langle u_i\rangle)^2$, fluctuations are almost zero above $z>z_d$. The fact that they are not exactly zero is likely because only the spatially averaged velocity $\langle u_i\rangle$ is used in the forcing term, and so will vary slightly with time given the small spatial averaging domain of the minimal channel.

The Reynolds shear stress (figure \ref{fig:turbIntOuterForce}\emph{b}) is almost zero above $z^+>300$, meaning the damping is not dissimilar to having a lower friction Reynolds number while keeping the roughness Reynolds number the same. However, flow structures can enter into the damping region and remain coherent for a period of time which likely depends on the damping strength $\gamma$. This is different to a half-height channel at a lower friction Reynolds number, in which the impermeability constraint confines the structures at the slip wall.
An advantage of higher $Re_\tau$ with damping compared to lower $Re_\tau$ without damping is that
the higher friction Reynolds number of the flow with damping will reduce the effect of low friction Reynolds number flows.
\cite{Chan15} showed that the roughness function $\Delta U^+$ was overestimated for $Re_\tau\approx180$ compared to $Re_\tau\approx 360$ and $Re_\tau\approx540$ for matched roughness Reynolds numbers. This is due to there being a non-negligible pressure gradient effect, which as discussed in \S\ref{sect:sweep} is $\varDelta_p=-1/Re_\tau$.
It would therefore be preferable to conduct higher friction Reynolds number simulations with damping, rather than a lower friction Reynolds number simulations. While not investigated here, it also seems reasonable that the grid spacings above $z_d$ could be relaxed without altering the near-wall flow.

This damping is similar to the masks, or sponge regions, employed in \cite{Jimenez99}. The authors explicitly filtered out disturbances in the outer layer, resulting in laminar-like flow in the outer layer but still maintaining a healthy near-wall cycle. The present damping is similar, although some fluctuations are still present in the outer layer (figure \ref{fig:turbIntOuterForce}\emph{a}) and the velocity is fixed rather than forming a laminar velocity profile. A benefit of this damping is that it reduces the computational time step, which is limited by the $\textit{CFL}$ number,
$\textit{CFL} = \max_{1\le i \le 3}\left(|u_i|\Delta t/\Delta x_i\right)$. Here $|u_i|$ is the maximum instantaneous velocity in the $i$th direction and $\Delta x_i$ is the corresponding grid spacing at that location. Note that the viscous time step restriction does not become significant in any of the simulations performed here.
For minimal-span channels  at $Re_\tau=590$ with no outer-layer damping, the maximum \textit{CFL} number occurs in the outer layer due to the large streamwise velocity. We see the time step improves by approximately 20--24\% with the use of the outer-layer damping (table \ref{tab:sims3v2}, $L_y^+\approx118$) than without. This means the simulations can be completed quicker and hence use less computational resources. However, for the present simulations at higher friction Reynolds numbers of $Re_\tau=2000$, the time step for the smooth wall with outer-layer damping is unaltered compared to the case without damping. 
This is due to the use of the cosine mapping to define the wall-normal grid, which produces an excessively fine grid near the wall especially for high Reynolds number cases. 
As a result of such a small $\Delta z$, the wall-normal $\textit{CFL}$ number is now maximum and this occurs in the near-wall region, which is independent of the outer-layer damping. 
This clustering of points near the wall is a known issue of the cosine mapping for high $Re_\tau$ simulations \citep{Lenaers14}, but is used here to ensure a systematic, controlled study. If a more appropriate mapping was used for these high friction Reynolds numbers (e.g. that in \citealt{Lee15}), we would expect a similar or better improvement in time step as at $Re_\tau=590$.

The most appropriate position for $z_d$ is difficult to establish from the above data. A minimum of $z_d^+\approx200$ seems necessary to ensure that the damping does not interact with the near-wall flow. It also seems reasonable that $z_d$ should scale in some way on $z_c$, for larger domain widths. The three sets of simulations suggest that $z_d$ should be approximately two times $z_c$, so a tentative rule of thumb would be $z_d^+ \gtrsim \max(200,\hphantom{.}2z_c^+)$. For an appropriate wall-normal mapping, this allows a 20--24\% improvement in the computational time step.

%
%
\subsection{Temporal sweep (table \ref{tab:sims4v2})}
\label{sect:sweepres}
\begin{table}
\centering
\begin{tabular}{c c  c c c c   c c c  c c c c   c c}
ID	& $Re_\tau$	& $\tfrac{L_x}{h}$	&  $L_x^+$	& $L_y^+$	& $\tfrac{L_z}{h}$	& $N_x$	& $N_y$	& $N_z$	& $\Delta x^+$	& $\Delta y^+$	& $\Delta z_w^+$	& $\Delta z_h^+$	& $\Delta t^+_{S}$	& $\Delta t^+_{R}$\\[0.5em] 

  \multicolumn{15}{c}{Temporal sweep (at initial conditions)} \\[0.5em] 

MS6	& 590		& $2\pi$	& 3707	& 354		&2		& 384		& 80		& 256		& 9.6		& 4.4		& 0.04	& 7.2				& 0.38	& 0.43	\\

\end{tabular}
\caption{Initial conditions for temporal sweep in a minimal-span channel.
 Refer to table \ref{tab:sims1v2} for definitions.
}
\label{tab:sims4v2}
\end{table}

In this section, the temporal sweep (\S \ref{sect:sweep}) is performed in a standard-height channel, with no outer-layer damping ($K_i=0$) and a streamwise length of $L_x/h=2\pi$ (table \ref{tab:sims4v2}).
The initial friction Reynolds number is $Re_\tau=590$ which reduces to $Re_\tau=180$ via an adverse pressure gradient with different rates investigated. The roughness height is fixed at $k=h/40$, so the roughness Reynolds number varies as $5\lesssim k^+\lesssim15$.
Three different sweeps were conducted in which the pressure gradient parameter at the end of the sweep is $\varDelta_{p,Re_\tau=180}=(0.03,0.07,0.15)$.

\begin{figure}
\centering
 \captionsetup[subfigure]{labelformat=empty}
	\subfloat[]{
		\includegraphics[width=0.49\textwidth,trim = 0 22 0 0,clip = true]{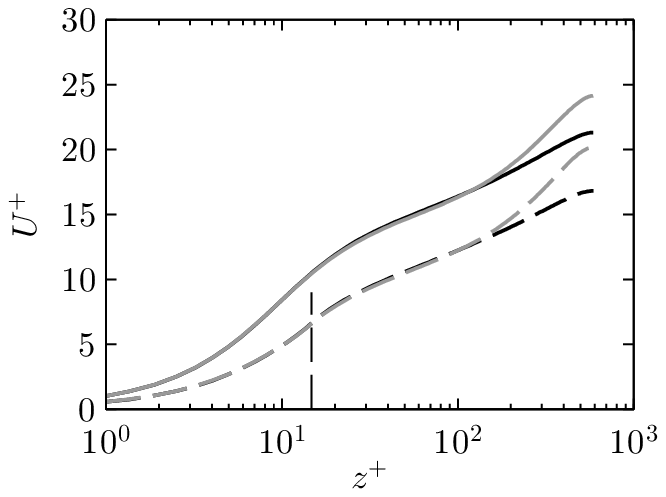}
		\label{fig:uxSweep1}
	}
	\subfloat[]{
		\includegraphics[width=0.49\textwidth,trim = 0 22 0 0,clip = true]{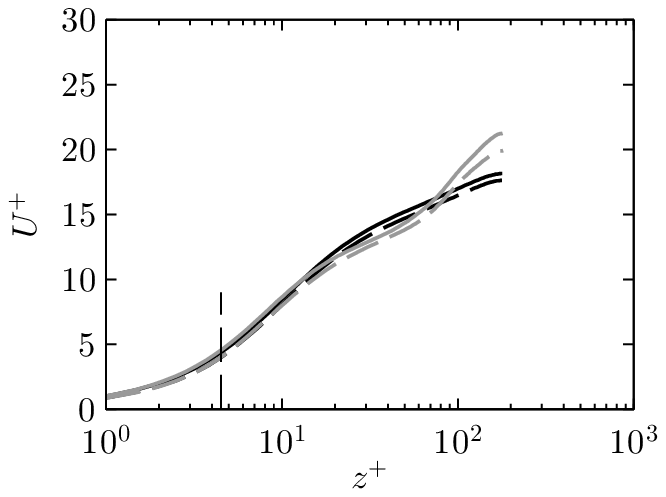}
		\label{fig:uxSweep2}
	}
	\put(-13.5,3.8){(\emph{a})}
	\put(-6.7,3.8){(\emph{b})}
	\put(-9.9,3.5){$\varDelta_{p,Re_\tau=180} $=$ 0.03$}
	\put(-12.5,1.5){\includegraphics[scale=0.88]{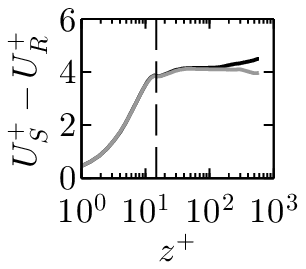}}
	\put(-3.0,3.5){$\varDelta_{p,Re_\tau=180} $=$ 0.03$}
	\put(-5.6,1.5){\includegraphics[scale=0.88]{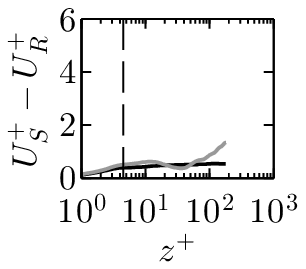}}
	\vspace{-2.2\baselineskip}
	\\
	\subfloat[]{
		\includegraphics[width=0.49\textwidth]{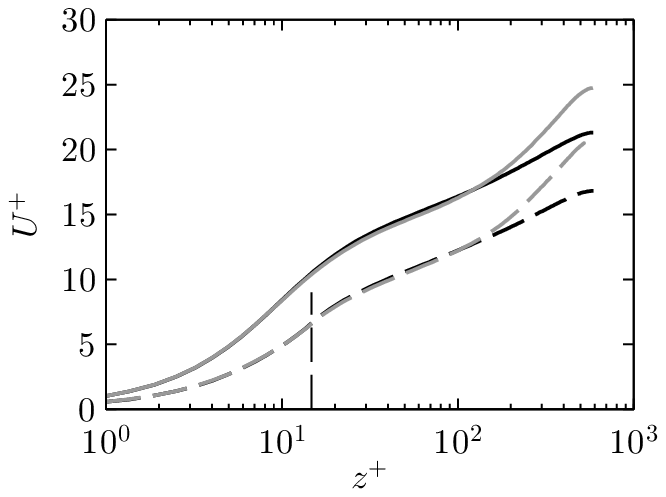}
		\label{fig:uxSweep3}
	}
	\subfloat[]{
		\includegraphics[width=0.49\textwidth]{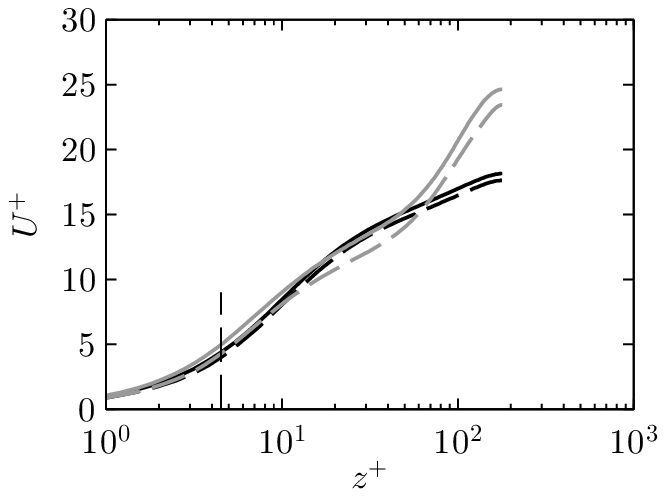}
		\label{fig:uxSweep4}
	}
	\put(-13.5,4.6){(\emph{c})}
	\put(-6.7,4.6){(\emph{d})}
	\put(-9.9,4.2){$\varDelta_{p,Re_\tau=180} $=$ 0.15$}
	\put(-12.5,2.25){\includegraphics[scale=0.88]{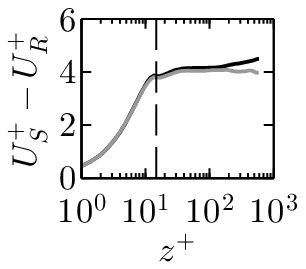}}
	\put(-3.0,4.2){$\varDelta_{p,Re_\tau=180} $=$ 0.15$}
	\put(-5.6,2.25){\includegraphics[scale=0.88]{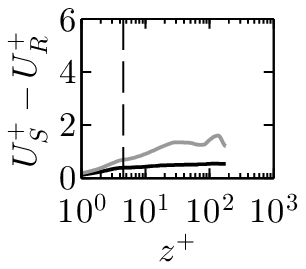}}
	\vspace{-1.8\baselineskip}
	\caption{Mean streamwise velocity profile for (\emph{a},\emph{c}) $Re_\tau\approx590$ and (\emph{b},\emph{d}) $Re_\tau\approx180$, from 
	full-span steady flow (black line) and 
	temporal sweep (grey line).
	(\emph{a},\emph{b}) slowest sweep ($\varDelta_{p,Re_\tau=180} = 0.03$) and
	(\emph{c},\emph{d}) fastest sweep ($\varDelta_{p,Re_\tau=180} = 0.15$),  \S \ref{sect:sweep}.
	Line styles:
	Solid line, smooth wall;
	dashed line, rough wall;
	vertical dashed line, roughness crest, $k = h/40$.
	Inset shows difference in smooth- and rough-wall velocity.
	}
	\label{fig:uxSweep}
\end{figure}

Figure \ref{fig:uxSweep} shows the mean velocity profiles for the fastest and slowest sweeps, comparing the sweep data with full-span steady-flow data at the highest and lowest friction Reynolds numbers tested. Figure \ref{fig:uxSweep}(\emph{a},\emph{b}) shows the mean velocity for the slowest sweep, when the pressure gradient parameter (\ref{eqn:deltap}) at $Re_\tau\approx180$ was $\varDelta_{p,Re_\tau=180} = 0.03$. Figure \ref{fig:uxSweep}(\emph{c},\emph{d}) shows the mean velocity for the fastest sweep, with $\varDelta_{p,Re_\tau=180} = 0.15$. To obtain statistics for a desired $Re_\tau$, time-averaging is performed over a window of $\pm10\%$ of that value.
At the initial friction Reynolds number of the simulation, $Re_\tau=590$, both the sweeps (figure \ref{fig:uxSweep}\emph{a},\emph{c}) show good agreement with the full-span channel in the near-wall region. As the spanwise width is $L_y/h=0.6$, then the critical wall-normal location is approximately $z_c^+\approx0.4L_y^+=142$ which agrees well with the figure. As $\varDelta_p$ is not significant yet (figure \ref{fig:sweept}\emph{b}) then there is little difference between the two sweeps. 

Differences start to emerge at the end of the sweep when the friction Reynolds number is close to $Re_\tau=180$ and the $\varDelta_p$ is maximum. The slowest sweep (figure \ref{fig:uxSweep}\emph{b}) has reasonable agreement very close to the wall below $z^+\lesssim15$, but the pressure gradient has reduced the streamwise velocity compared to the full-span steady flow. However, the difference between the smooth and rough-wall sweep flows (inset of figure \ref{fig:uxSweep}\emph{b}) agrees quite well with the full-span steady data. Taking the difference between two flows with the same weak pressure gradient ostensibly still produces a correct estimate of $\Delta U^+$. However, when the pressure gradient is increased to to $\varDelta_{p,Re_\tau=180} = 0.15$ (figure \ref{fig:uxSweep}\emph{d}) then larger differences are seen. The near-wall region has been noticeably changed, and the difference in smooth and rough wall velocity (inset) is seen to overestimate the full-span data. Note that the data have been ensemble averaged over four runs, so that the amount of time the flow is averaged over is the same as in the slowest sweep. This therefore indicates that the overestimation of $\Delta U^+$ is a direct result of the stronger (adverse) pressure gradient. 

\setlength{\unitlength}{1cm}

\begin{figure}
\begin{center}
\includegraphics[scale=0.9]{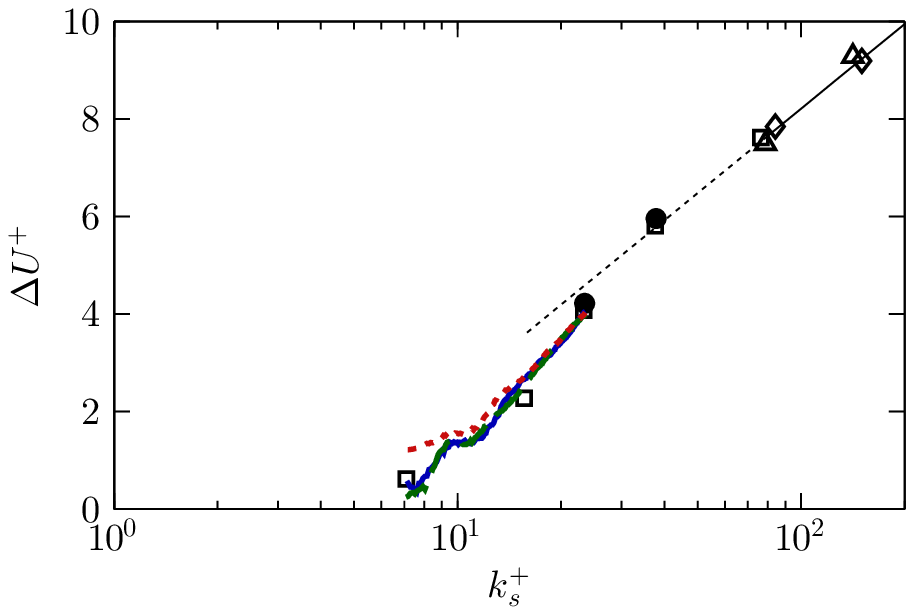}
\end{center}
\vspace{-1.0\baselineskip}
\caption{(Colour online) Hama roughness function for the temporal sweep with modelled roughness at fixed $k=h/40$, where $k_s^+\approx1.6k^+$.
Symbols:
black symbols, steady flow data for the same modelled roughness \citep{Chung15};
\protect\raisebox{0.8ex}{\color{myblue}\linethickness{0.5mm}\line(1,0){0.6}}, sweep data for $\varDelta_{p,Re_\tau=180} = 0.03$ (slowest sweep);
\protect\raisebox{0.8ex}{\color{mygreen}\linethickness{0.5mm}\line(1,0){0.3}\hspace{0.15cm}\line(1,0){0.3}}, sweep data for $\varDelta_{p,Re_\tau=180} = 0.07$;
\protect\raisebox{0.8ex}{\color{myred}\linethickness{0.5mm}\line(1,0){0.1}\hspace{0.15cm}\line(1,0){0.1}\hspace{0.15cm}\line(1,0){0.1}}, sweep data for $\varDelta_{p,Re_\tau=180} = 0.15$ (fastest sweep).
}
\label{fig:DUsweep}
\end{figure}

The roughness function $\Delta U^+$ is now computed from the difference in smooth and rough-wall flows at matched $U_\tau$ values.
 As seen in the insets of figure \ref{fig:uxSweep}, the smooth- and rough-wall velocity difference is not constant with $z^+$ as in the steady data. This indicates that the flow has not been averaged for long enough at the desired $Re_\tau$ value. As such, multiple sweeps from different initial conditions would need to be conducted to obtain a converged outer region. However, for the purposes of this investigatory study, the velocity difference is averaged from the crest of the roughness to the channel centre to obtain an estimate of $\Delta U^+$. This is done for a range of friction Reynolds numbers, with the resulting data plotted in figure \ref{fig:DUsweep} against the equivalent sandgrain roughness, $k_s^+$. Ideally, the temporal sweep data (lines) should agree with the steady flow data (black symbols). Indeed, the temporal sweep does show promise in this regard, especially for the weaker pressure gradient cases. 
   If we interpolate the steady roughness function data from \cite{Chung15}, then an overall relative error term  can be approximated as $\varepsilon = \int|1-\Delta U^+_{steady}/\Delta U^+_{sweep}|\id k_s^+$. The two slowest sweeps, with $\varDelta_p=0.03$ and 0.07 have similar relative errors of $\varepsilon\approx12\%$ and 13\%, respectively, while the fastest sweep is noticeably larger with an  error of $\varepsilon\approx26\%$.

It appears that the sweep approach could work for determining the hydraulic behaviour of roughness. A pressure gradient parameter of $\varDelta_p\le0.03$--0.07 is required, which is not too far from the recommendation in \cite{Patel65} that zero pressure gradient conditions can be assumed when $\varDelta_p\le0.01$.
 A value of $\varDelta_p \gtrsim0.15$ results in pressure gradient effects that lead to inaccuracies in $\Delta U^+$. 
Future studies could prescribe a constant value of $\varDelta_p$ rather than linearly varying the bulk velocity, as this would ensure pressure gradient effects remain negligible, while efficiently traversing the range of roughness Reynolds numbers.

%
%
\section{Cost and Convergence}
\label{sect:cost}

Quantifying the cost of the simulations is extremely important when it comes to evaluating the benefit of minimal-span channels. There are two important costs to consider in these simulations;
the cost of memory (or the number of processors required) and the cost of simulation time. The memory cost can be readily described by the size of the grid: the number of spatial degrees of freedom for full-span simulations scale as $Re_\tau^{9/4}$ \citep[\S 9.1.2]{Pope00} which for $Re_\tau\gtrsim10^3$ requires national-level high-performance computing facilities. The minimal channel, meanwhile, scales as $k_s^{+9/4}$ \citep{Chung15}, making it feasible for smaller facilities at the university and industry level, or potentially high-end desktop computers. 

The cost of the simulation runtime is conventionally given in terms of CPU hours, however an obvious question emerges; does a minimal channel with a domain volume that is, say, 10 times smaller than a full-span channel require the simulation runtime to be 10 times longer? This would imply a similar cost in terms of CPU hours, but to answer this question definitively we need to know how long it takes to obtain converged statistics, particularly in the first-order statistics necessary to obtain the roughness function. A benefit of this analysis is that it will enable us to predict how many CPU hours are required for a minimal-span simulation before actually performing the simulation.

Generally, full-scale turbulence simulations are run for approximately 10 large-eddy turnover times, where the largest eddy captured in the domain are of characteristic size $h$ and velocity $U_\tau$. This implies a simulation run time of $T_{sim}U_\tau/h\approx10$ (see, for example, \citealt{Hoyas06,Lozanoduran14,Lee15}) and this long runtime is why DNS of full-span channels are so expensive. However, the largest captured eddies in the minimal-span channel will be of characteristic size $z_c$, implying we need $T_{sim}U_\tau/z_c\approx10$ eddy turnover times of this $z_c$-sized eddy. In this study, these are generally of the size $h/z_c\approx10$, implying that $T_{sim}U_\tau/z_c\approx10$ would be only 1 turnover time of the $h$-sized eddy (if it existed), representing an order of magnitude saving in time.

This argument, however, does not make reference to the wall-parallel size of the domain.
\cite{Lozanoduran14} suggested that channels with smaller domain sizes need to be run for proportionately longer, and showed that the statistical uncertainties
in $u'(z=h$)
of full-span channels decrease in a manner inversely proportional to the square root of the wall area. However, this result was for two channels that, from the point of view of the current study, were full-span channels. 
The largest characteristic eddies  in both channels were of size $h$, which explains the simple relationship between channel domain size and simulation runtime. 
In contrast, the largest characteristic eddies in minimal channels are of size $z_c$ which varies depending on the width of the channel. This means that such a simple relationship between channel domain size and run time is unlikely to hold.
In the context of these $z_c$-sized eddies, a full-span simulation could potentially have hundreds of eddies distributed throughout the domain, while a minimal-span simulation will only have a few. 
 For this minimal-span channel to obtain the same level of converged statistics  as a full-span channel would presumably require the same number of these $z_c$-sized eddies to pass through the domain. 
 It therefore becomes useful to count the number of $z_c$-sized eddies present in the domain. 
 These eddies have a spanwise width of $\lambda_{y,z_c}=L_y\approx z_c/0.4=2.5 z_c$ and, as discussed in the previous section, a streamwise length of approximately three times the span, $\lambda_{x,z_c}\approx 3 \lambda_{y,z_c}=7.5z_c$. Hence, the number of $z_c$-sized eddies at any instant is
 \begin{equation}
\label{eqn:Ninstant}
N_{\text{instant}} = \f{L_x}{\lambda_{x,z_c}}\f{L_y}{\lambda_{y,z_c}} \cdot \f{L_z}{h} =\f{L_x}{7.5z_c}\f{L_y}{2.5z_c} \cdot  \f{L_z}{h}.
\end{equation}
The first factor is an approximation of the number of $z_c$-sized eddies present on one wall of the channel.
The other factor, $L_z/h$, is present as a half-height channel ($L_z/h=1$) will have half as many $z_c$-sized eddies  as a standard-height channel ($L_z/h=2$).

\setlength{\unitlength}{1cm}
\begin{figure}
\centering
 \captionsetup[subfigure]{labelformat=empty}
	\subfloat[]{
		\includegraphics[width=0.47\textwidth]{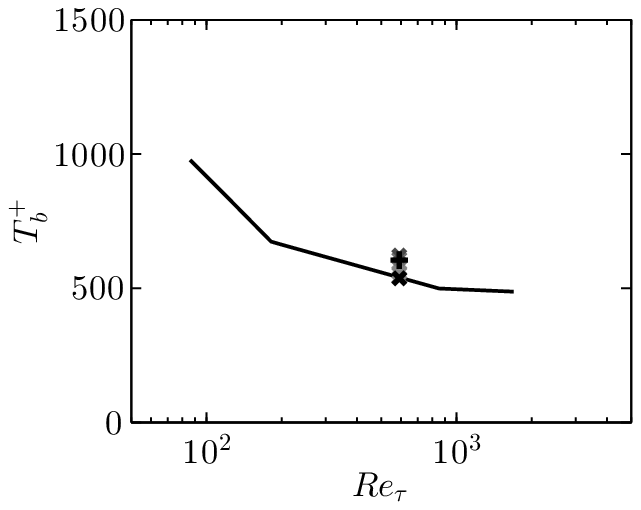}
	}
	\subfloat[]{
		\includegraphics[width=0.49\textwidth]{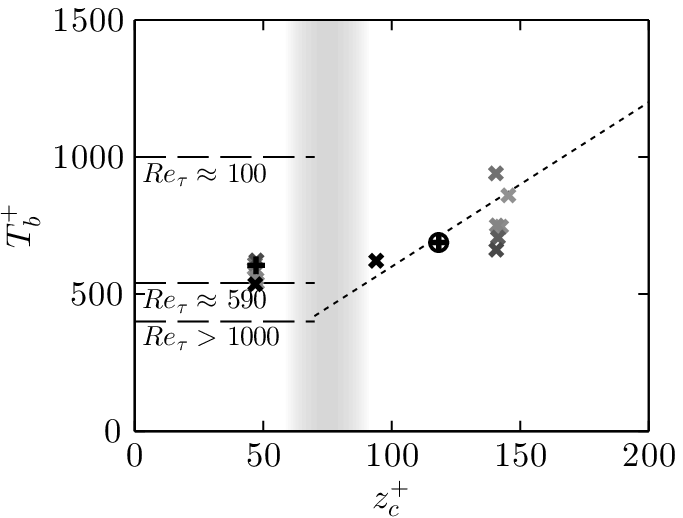}
	}
	\put(-13.4,4.7){(\emph{a})}
	\put(-6.8,4.7){(\emph{b})}
	\put(-5.35,4.35){$T_\text{buffer}^+$}
	\put(-3.00,4.35){$T_\text{log}^+$}
	\vspace{-2.3\baselineskip}
	\caption{(\emph{a}) Buffer-layer bursting period $T_b^+$ against friction Reynolds number.
	Solid line denotes data of \cite{Jimenez05}, symbols are current data for buffer-layer minimal channels.
	(\emph{b}) Bursting period $T_b^+$ against wall-normal critical height $z_c^+$.
	Symbols:
	\protect\raisebox{0.4ex}{\protect\scalebox{1.0}{$\boldsymbol{\pmb{\times}}$}}, $Re_\tau=590$ standard-height channel;
	\protect\raisebox{0.4ex}{\protect\scalebox{1.0}{$\boldsymbol{\pmb{+}}$}},  $Re_\tau=590$ half-height channel;
	\protect\raisebox{0.4ex}{\protect\scalebox{1.0}{$\boldsymbol{\pmb{\oplus}}$}}, $Re_\tau=2000$ half-height channel.
	Lighter grey symbols denote shorter streamwise channel lengths.
	Line styles:
\protect\raisebox{0.8ex}{\linethickness{0.5mm}\line(1,0){0.25}\hspace{0.15cm}\line(1,0){0.25}}, buffer-layer bursting period at different $Re_\tau$;
\protect\raisebox{0.8ex}{\linethickness{0.5mm}\line(1,0){0.1}\hspace{0.15cm}\line(1,0){0.1}\hspace{0.15cm}\line(1,0){0.1}}, log-layer bursting period $T_b U_\tau/z_c = 6$ \citep{Flores10,Hwang16}.
}
	\label{fig:burstingperiod}
\end{figure}

 Over the course of the simulation, these $z_c$-sized eddies will grow and decay in a process termed `bursting', so it becomes necessary to consider the  time scale of these eddies.
 The bursting process in the buffer layer was investigated by \cite{Jimenez05}, who showed that the time scale depended on the friction Reynolds number. Low Reynolds number flows have a long bursting period of $T_b^+\approx1000$ for $Re_\tau\approx100$, however this saturates to $T_b^+\approx 400$  for $Re_\tau\gtrsim1000$. 
 This was determined for minimal channels with small spanwise widths such that they only captured the buffer layer, i.e. $L_y^+\lesssim 125$--$200\Rightarrow z_c^+\lesssim 50$--$80$. 
Figure \ref{fig:burstingperiod}(\emph{a}) shows this buffer-layer bursting period, along with the current data for buffer-layer minimal channels ($L_y^+\lesssim 125$--$200$) at $Re_\tau=590$. Good agreement is seen, with the bursting period at this friction Reynolds number being approximately $T_{b}^+\approx 540$.
 The bursting period was determined in the same manner as \cite{Jimenez05}, defined as the weighted logarithmic average of the frequency spectra of the wall-shear stress.

In the logarithmic layer, \cite{Flores10} determined that the bursting period scales with distance from the wall according to $T_b U_\tau/z_c =6$, which was supported by \cite{Hwang16}.  The bursting period of the largest eddies in the minimal channel, $T_b^+$, therefore depends on both $z_c$ and $Re_\tau$.  Narrow minimal channels which only capture the buffer layer ($z_c^+\lesssim 50$--$80$) have an $Re_\tau$-dependence on the bursting period (figure \ref{fig:burstingperiod}\emph{a}), while wider minimal channels which capture part of the logarithmic layer ($z_c^+\gtrsim50$--$80$) have a bursting period that depends on $z_c$. This is summarised in figure \ref{fig:burstingperiod}(\emph{b}), which also shows the present data. 
The behaviour for intermediate $z_c^+$ values between buffer and logarithmic flows (grey band) is unknown and likely depends on both $z_c^+$ and $Re_\tau$. Here, the inertial motions of the logarithmic layer are scarce and are likely competing with the $Re_\tau$-dependent motions of the buffer layer, with neither completely dominating. This distinction becomes irrelevant for higher $Re_\tau$ channels as the buffer-layer bursting period saturates to $T_b^+\approx 400$.

The product of the bursting timescale, $T_b$, and the instantaneous number of $z_c$-sized eddies, $N_\text{instant}$, gives the total number of $z_c$-sized eddies that exist over the simulation runtime $T_{sim}$,
\begin{equation}
\label{eqn:Cstar}
C^\star = \f{T_{sim}}{T_{b}}N_{\text{instant}} = \f{T_{sim}}{T_b} \cdot\f{L_z}{h}\cdot \f{L_x}{7.5z_c}\f{L_y}{2.5z_c}.
\end{equation}
$C^\star$ is simply a running count of how many $z_c$-sized eddies have been sampled, and increases with simulation runtime.
To investigate whether this is the correct measure for statistical convergence, the statistical uncertainty of streamwise velocity in different minimal channels is analysed.
There are various ways of quantifying the statistical uncertainty; here we will estimate it using the
standard error of the mean.
Instantaneous snapshots of $u(x,y,z,t_i)$ at time $t_i\le T_{sim}$ are spatially averaged in the wall-parallel plane to get $\langle u(z,t_i)\rangle$.
The temporal average of these $N$ snapshots gives the overall mean,
\begin{equation}
\label{eqn:umean}
{U}(z,T_{sim}) = \frac{1}{N}\sum_{i=0}^{N-1}\langle u(z,t_i)\rangle
\end{equation}
The standard error of this mean for statistically independent samples can be computed as $\epsilon = s/\sqrt{N}$, where $s$ is the unbiased standard deviation of the signal $\langle u(z,t_i)\rangle$. The quantity $\epsilon$ has the same units as $u$, so can be non-dimensionalised to $\epsilon^+$.  A 95\% confidence interval can then be given in which we are 95\% certain the true mean falls within $U^+\pm 1.96\epsilon^+$ \citep{Benedict96}. The roughness function, being the difference in two estimated velocities, therefore has a 95\% probability of falling within the range $\Delta U^+ \pm 1.96\sqrt{\epsilon_s^{+2}+\epsilon_r^{+2}}$, where subscripts $s$ and $r$ refers to the uncertainties in the smooth-wall and rough-wall velocities.

In turbulence simulations the samples may not necessarily be statistically independent, so the assumption that $\epsilon=s/\sqrt{N}$ would not hold. Following \cite{Trenberth84} and \cite{Oliver14}, an effective number of statistically independent samples $N_{eff}=N/T_0$ can be defined, where
 \begin{equation}
 \label{eqn:T0}
 T_0 = 1+2\sum_{k=1}^{N-1}\left(1-\frac{k}{N}\right){\rho}(k)
 \end{equation}
 is the decorrelation separation distance, ${\rho}$ being the autocorrelation function of $\langle u(z,t)\rangle$
 normalised on its variance.
  The standard error can then be estimated as $\epsilon = s/\sqrt{N_{eff}}$, where
 \begin{equation}
 \label{eqn:s}
s^2 = \frac{1}{N-T_0}\sum_{i=0}^{N-1}\left(\langle u(z,t_i)\rangle  - U(z,T_{sim})\right)^2.
 \end{equation} 
 If the samples were truly independent, then the autocorrelation $\rho$ should be near zero, resulting in $T_0=1\Rightarrow N_{eff}=N$ and the familiar unbiased standard deviation for $s$ is obtained (with divisor $N-1$) . The autocorrelation function can be noisy, especially for small $N$, prompting \cite{Oliver14} to develop an open-source code to fit an autoregressive time series model to a given signal. However we have found that for minimal channel simulations $N$ is sufficiently large such that the autocorrelation definition appears to be adequate, with $T_0$ usually being $O(1)$.
The coarse-graining approach for estimating the standard error for correlated samples, detailed in the appendix of \cite{Hoyas08}, was also attempted although no appreciable differences were detected, again because the samples are nearly independent.

\begin{figure}
\centering
 \captionsetup[subfigure]{labelformat=empty}
	\subfloat[]{
		\includegraphics[width=0.48\textwidth]{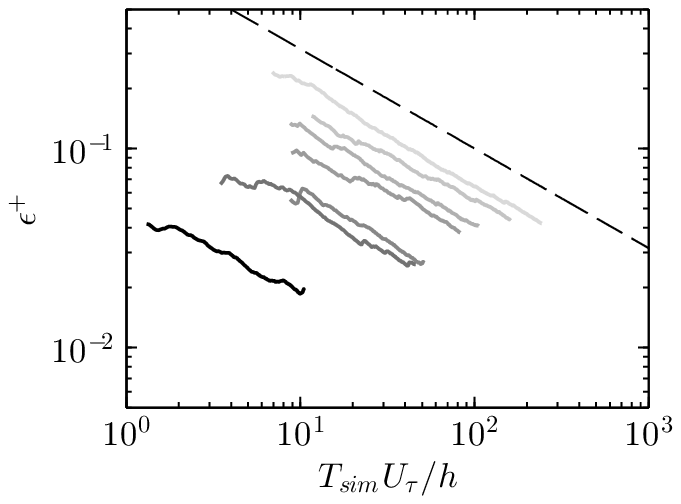}
	}
	\subfloat[]{
		\includegraphics[width=0.49\textwidth]{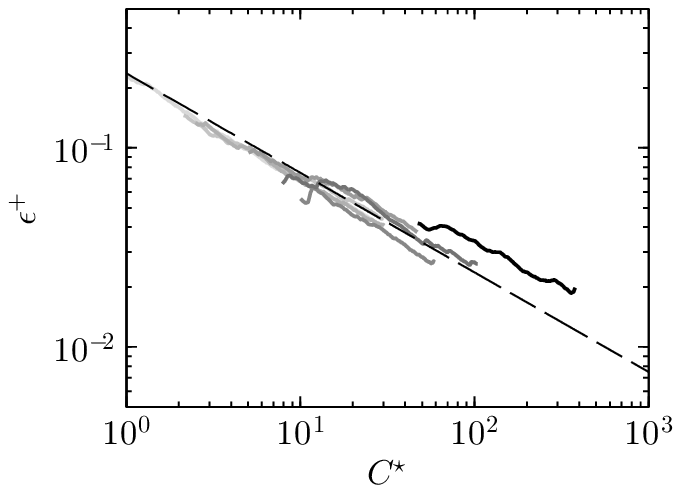}
	}
	\put(-13.4,4.6){(\emph{a})}
	\put(- 6.65,4.6){(\emph{b})}
	\put(-9.4,3.4){\vector(-1,-1){1.2}}
	\put(-10.5,1.9){Inc. $L_x^+$}
	\vspace{-2.0\baselineskip}
\vspace{-0.5\baselineskip}
\caption{
Standard error of the velocity $\langle u\rangle$ at $z^+\approx47$ as a function of (\emph{a}) large-eddy turnover time $T_{sim} U_\tau/h$ and (\emph{b}) number of $z_c$ eddies $C^\star$ (\ref{eqn:Cstar}).
Line styles:
black, full-span channel;
grey, minimal-span channel with $z_c^+\approx47$ where darker grey line indicates increasing length $L_x^+$ (table \ref{tab:sims1v2}).
Dashed line in (\emph{a}) shows $(T_{sim} U_\tau/h)^{-1/2}$, the expected trend in standard error,
and dashed line in (\emph{b}) shows $0.75(C^\star)^{-1/2}$
}
\label{fig:convCost}
\end{figure}

  Increasing the simulation runtime $T_{sim}$ reduces the statistical uncertainties as evidenced in figure \ref{fig:convCost}(\emph{a}), which shows the standard error of the velocity at a representative wall location $z^+=47$. The very short minimal-span channels (light grey lines) only have one or two $z_c$-sized eddies present, and the bursting nature of only a few eddies greatly increases the statistical uncertainties, requiring a very long runtime. The dashed line shows $(T_{sim}U_\tau/h)^{-1/2}$, indicating that the uncertainties decrease inversely proportional to the square root of time. This is essentially the same result as \cite{Hoyas08} and \cite{Lozanoduran14} who argued the decrease was inversely proportional to the square-root of wall area, the difference here  being that  we have considered the snapshots as a sampling of time rather than multiples of wall area.
  
When the standard error is instead considered as a function of $C^\star$ (figure \ref{fig:convCost}\emph{b}) we obtain a better collapse, indicating that counting $z_c$-sized eddies as in the $C^\star$ formulation is indeed an appropriate measure to use. 
The data show a collapse with $\epsilon^+=K(C^\star)^{-1/2}$, where here it was determined $K=0.75$ for $z_c^+\approx 47$. This relationship shows that obtaining a desired error tolerance requires capturing a certain number of $z_c$-sized eddies ($C^\star$). Increasing the streamwise domain length does not affect this measure, other than to simulate more eddies at once.

\begin{figure}
\centering
 \captionsetup[subfigure]{labelformat=empty}
	\subfloat[]{
		\includegraphics[width=0.49\textwidth]{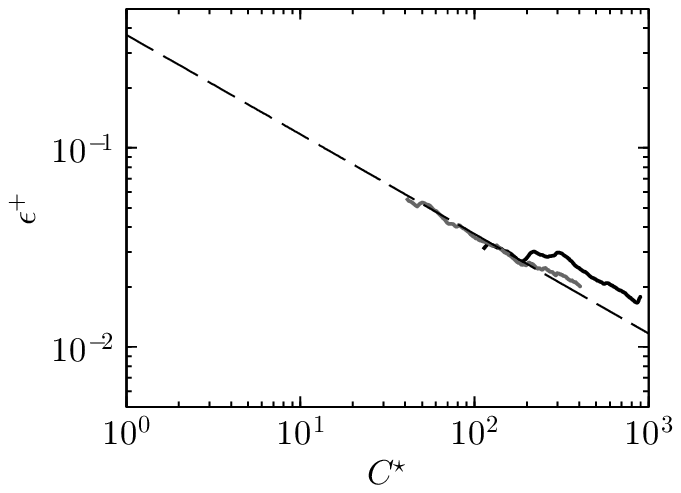}
	}
	\subfloat[]{
		\includegraphics[width=0.49\textwidth]{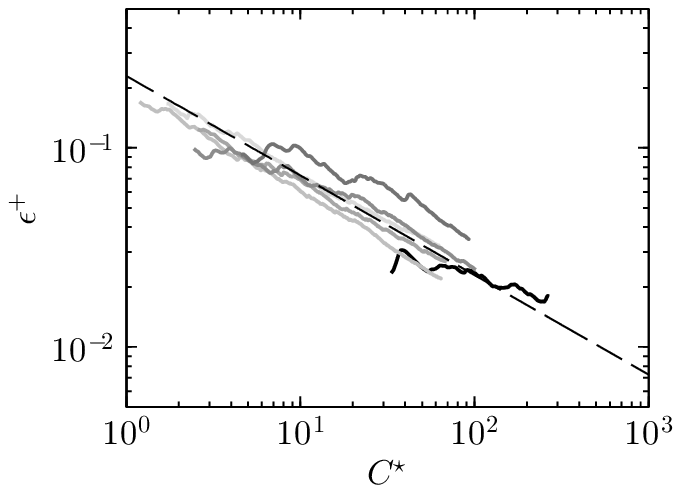}
	}
	\put(-13.4,4.6){(\emph{a})}
	\put(-10.4,4.25){$z_c^+\approx94$, $Re_\tau=590$}
	\put(-3.7,4.25){$z_c^+\approx142$, $Re_\tau=590$}
	\put(- 6.65,4.6){(\emph{b})}
	\vspace{-2.0\baselineskip}
	\\
	\subfloat[]{
		\includegraphics[width=0.49\textwidth]{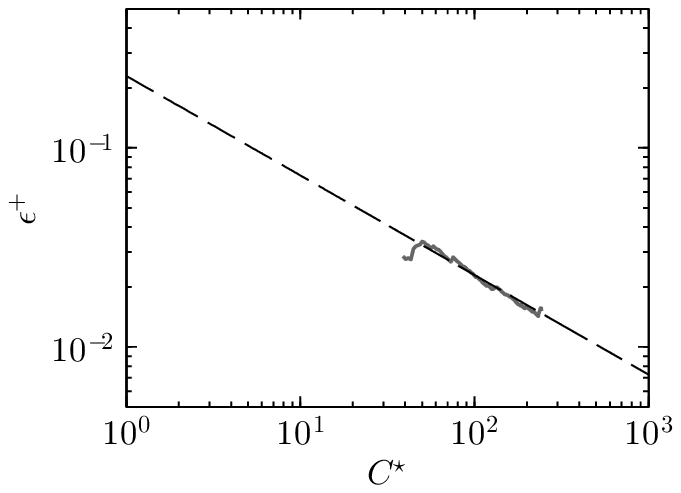}
	}
	\subfloat[]{
		\includegraphics[width=0.475\textwidth]{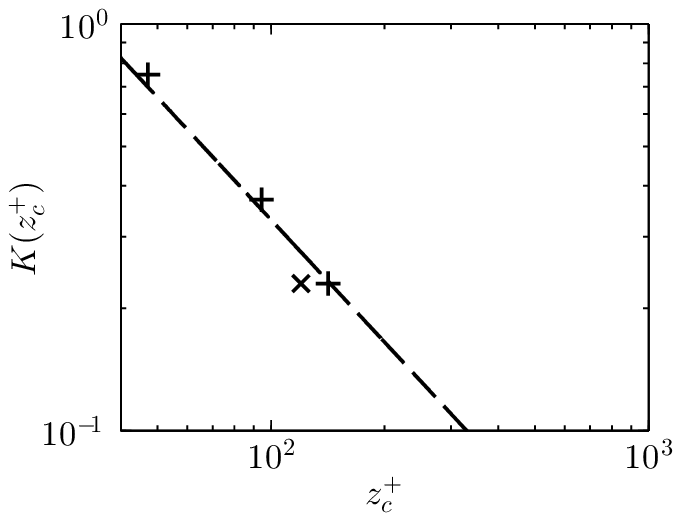}
	}
	\put(-13.4,4.6){(\emph{c})}
	\put(-10.5,4.25){$z_c^+\approx120$, $Re_\tau=2000$}
	\put(- 6.65,4.6){(\emph{d})}
	\vspace{-2.0\baselineskip}
\vspace{-0.5\baselineskip}
\caption{
Standard error of the velocity  as a function of the number of $z_c$-sized eddies eddies $C^\star$ (\ref{eqn:Cstar}) for
(\emph{a}) $z_c^+\approx 94$, $Re_\tau=590$ (table \ref{tab:sims3v2}),
(\emph{b}) $z_c^+\approx 142$, $Re_\tau=590$ (table \ref{tab:sims1v2}),
(\emph{c}) $z_c^+\approx 120$, $Re_\tau=2000$ (table \ref{tab:sims3v2}).
Line styles are same as figure \ref{fig:convCost}.
Dashed line shows $\epsilon^+=K(C^\star)^{-1/2}$, where (\emph{a}) $K = 0.37$, (\emph{b}) $K = 0.23$, and (\emph{c}) $K = 0.23$.
Coefficient $K$ is shown in (\emph{d}) as a function of $z_c^+$.
Symbols:
\protect\raisebox{0.4ex}{\protect\scalebox{1.0}{$\boldsymbol{\pmb{+}}$}}, $Re_\tau=590$;
\protect\raisebox{0.4ex}{\protect\scalebox{1.0}{$\boldsymbol{\pmb{\times}}$}},  $Re_\tau=2000$.
Dashed line shows $K=33/z_c^+$.
}
\label{fig:convzc}
\end{figure}

Figure \ref{fig:convzc} shows that there is not a universal number of $z_c$-sized eddies that need to be simulated to obtain a desired $\epsilon^+$, that is, the coefficient $K$ in $\epsilon^+=K (C^\star)^{-1/2}$ is a function of $z_c^+$.
 Figure \ref{fig:convzc}(\emph{a}) shows that for $z_c^+\approx94$, $K=0.37$, 
 while figure \ref{fig:convzc}(\emph{b}) with $z_c^+\approx142$ has $K=0.23$. 
 Figure \ref{fig:convzc}(\emph{d}) shows that this coefficient decreases with $z_c^+$,
which implies that a wider minimal channel will have a smaller statistical uncertainty than a narrower one, for the same $C^\star$ value. This is  possibly due to the way in which eddies aggregate in wider channels. As the channel widens and $z_c^+$ increases, the largest $z_c$-sized eddy increases, however there remains a hierarchy of smaller eddies below this largest eddy. The uncertainty $\epsilon^+$ of the mean velocity
would have contributions from all of these eddies and so would depend on $z_c^+$.  
The reason this effect is not already captured in $C^\star$ is that this quantity only considers the largest $z_c$-sized eddy, whereas $\epsilon^+$ ostensibly considers all eddies up to and including those of $z_c$ size. The data in figure \ref{fig:convzc}(\emph{d}) show what appears to be a $-1$ power law, in which case a fit gives $K(z_c)=33/z_c^+$.

Recall that the 95\% confidence interval of the roughness function is $\Delta U^+\pm 1.96\sqrt{\epsilon_s^{+2}+\epsilon_r^{+2}}$. If both smooth- and rough-wall simulations were to have the same uncertainty, then this interval can be rewritten as 
\begin{equation}
\label{eqn:duerr}
\Delta U^+\pm2.77\epsilon^+ \approx \Delta U^+\pm 2.77 K(C^\star)^{-1/2}\approx \Delta U^+\pm91.4(C^\star)^{-1/2}/z_c^+.
\end{equation}
The advantage of this formulation is that setting a desired error tolerance in $\Delta U^+$ enables the simulation run time (and hence CPU hours) to be determined \emph{a priori}. To see this, the spanwise channel width, $L_y\gtrsim\max(100\nu/U_\tau,k/0.4,\lambda_{r,y})$ \citep{Chung15}, and streamwise channel length, $L_x\gtrsim\max(3L_y,1000\nu/U_\tau,\lambda_{r,x})$ (\S\ref{sect:slength}), must first be determined, which sets $z_c=0.4L_y$. The user must then determine a desired error tolerance, $\zeta$, for $\Delta U^+\pm\zeta$. This error tolerance will depend on the application; riblets have a roughness function $-\Delta U^+\lesssim1$ \citep{Spalart11}, which would require a smaller level of statistical uncertainty, than say an engineering roughness problem where $\Delta U^+\approx 10$. Once set however, this can be related to $C^\star$ through $\Delta U^+\pm91.4(C^\star)^{-1/2}/z_c^+$.
 From here, (\ref{eqn:Cstar}) can be solved for the simulation time $T_{sim}$ using the minimal channel dimensions and bursting period $T_b$. 
 The number of CPU hours is then
\begin{equation}
\label{eqn:cpu}
\text{CPU hours} = \frac{T_{sim}^+}{\Delta t^+} \mathcal{T}_1 N_{CPU},
\end{equation}
where $\Delta t^+$ is the computational time step, so $T^+/\Delta t^+$ is the number of time steps performed during the simulation. $\mathcal{T}_1$ is the average wall clock time taken to perform a single step and $N_{CPU}$ is the number of processors employed in the computation. The time step $\Delta t^+$ can be estimated if the streamwise CFL number at the channel centre is unity, i.e. $U_h\Delta t/\Delta x=CFL=1$, which gives $\Delta t^+=\Delta t U_\tau^2/\nu=\Delta x^+ U_\tau/U_h$. If $\Delta x^+\approx 10$ and, for a full-span channel, the centreline velocity is $U_h/U_\tau\approx25$ then we have $\Delta t^+\approx0.4$, which agrees with that recommended in \cite{Choi94}. Body-fitted roughness grids may have a smaller time step due to having a finer grid to resolve the roughness geometry. $\mathcal{T}_1$ and $N_{CPU}$ will depend on the code being used. 
The full-span channel used in this study, run for $T U_\tau/h=10$ large-eddy turnover times with $\Delta t^+\approx0.18$, $\mathcal{T}_1\approx26.4$ seconds and $N_{cpu}=64$ requires approximately 15800 CPU hours. For a  minimal channel with $h/z_c\approx10$ and 95\% confidence interval of $\Delta U^+\pm0.1$, the above analysis leads to a requirement of only 1420 CPU hours (for $N_{CPU}=64$ and $\mathcal{T}_1=2.4$ seconds, based on case MS6 of table \ref{tab:sims2v2}), which is over ten times less than the full-span channel.

%
%

\section{Minimal channel applied to pyramids}
\label{sect:pyramids}
\subsection{Computational setup}
\label{sect:compsetup}

\begin{table}
\centering
\begin{tabular}{ l c  c c c  c c  c c c  c c c c  c c }
Case & 	$Re_\tau$	& $k_t^+$ & $\lambda^+$ & $\alpha$ &  $L_x^+$	& $L_y^+$	& $N_x$	& $N_y$	& $N_z$	& $\Delta x^+$	& $\Delta y^+$	& $\Delta z_w^+$	& $\Delta z_h^+$ & $C_{f,mod}$ & $\Delta U^+$\\[0.5em]

Sm\_198 	& 834 		& - 	& - 	& - 	& 1180 & 197 & 288 & 48 & 320 & 4.1 & 4.1 & 0.14 & 6.2 & 0.0053 & -\\
Sm\_313 	& 850 		& - 	& - 	& - 	& 1201 & 316 & 288 & 72 & 320 & 4.2 & 4.4 & 0.14 & 6.3 & 0.0051 & -\\
Sm\_396  	& 851		& - 	& - 	& - 	& 1204 & 401 & 288 & 96 & 320 & 4.2 & 4.2 & 0.14 & 6.3 & 0.0050 & -\\
40\_198 	& 848 		& 40.4 & 200 & 22 	& 1200 & 200 & 288 & 48 & 320 & 4.2 & 4.2 & 0.14 & 6.3 & 0.015 & 8.0\\
40\_198\_f & 837 	& 39.2 & 197 & 22 	& 1184 & 197 & 504 & 84 & 560 & 2.3 & 2.3 & 0.24 & 3.6 & 0.015 & 7.9\\
40\_198\_2 & 879 	& 41.7 & 207 & 22 	& 1243 & 414 & 288 & 96 & 320 & 4.3 & 4.3 & 0.15 & 6.5 & 0.015  & 8.3\\
63\_313 	& 883 		& 66.3 & 329 & 22 	& 1249 & 329 & 288 & 72 & 320 & 4.3 & 4.6 & 0.24 & 6.5 & 0.021 & 10.2\\
80\_396	 & 868 	& 82.4 & 409 & 22 	& 1227 & 409 & 288 & 96 & 320 & 4.3 & 4.3 & 0.24 & 6.4 & 0.024 & 11.0\\[1.0em]
SF09			& 4020& 39.9	& 198	& 16	& -		& -	& -	& -	& -		& -		& -		& -		& -	&-	& 6.4 	\\
SF09			& 3900& 62.7	& 310	& 16	& -		& -	& -	& -	& -		& -		& -		& -		& -	&-	& 7.0 	\\
SF09			& 4290& 87.5	& 433	& 16	& -		& -	& -	& -	& -		& -		& -		& -		& -	&-	& 8.7 	\\
DCO16		& 670	& 39		& 147	& 28	& -		& -	& -	& -	& -		& -		& -		& -		& -	&-	& 8.9 	\\
HKS11			& 3520& 63.3	& 442	& 16	& -		& -	& -	& -	& -		& -		& -		& -		& -	&-	& 8.2\\

\end{tabular}
\caption{Description of the simulations performed with the finite volume code for the square-based pyramids.
$k_t^+$ is the trough-to-peak pyramid height,
$\lambda^+$ the pyramid wavelength,
$\alpha$ the pyramid edge angle,
and
$C_{f,mod}=2U_\tau^2/U_{b,full}^2$ the predicted full-span skin friction coefficient (see text).
Other entries are same as table \ref{tab:sims1v2}.
All simulations conducted using a half-height channel with no outer-layer damping.
SF09 refers to \cite{Schultz09}, HKS11 to \cite{Hong11}, and DCO16 to \cite{DiCicca16}.
}
\label{tab:simsB}
\end{table}

Following the insights discussed in this paper, a finite-volume code is used to determine the roughness function of square-based pyramids. This is the same code as in \cite{Chan15} and \cite{Chung15}, based on the work of \cite{Mahesh04} and \cite{Ham04}. No outer-layer damping will be used ($\mathbf{K}=\mathbf{0}$) and since a body-fitted grid is used, no roughness forcing is required ($\mathbf{F}=\mathbf{0}$). Three pyramids are simulated with different heights, but all with nominal $Re_\tau=840$
and angles between the pyramid edge and the horizontal of $\alpha=22^\circ$. Each pyramid will be referred to by its nominal trough-to-peak height, $k_t^+$ and wavelength $\lambda^+$, as a unique identifier given by the notation $k_t^+$\underline{\hspace{0.2cm}}$\lambda^+$. Hence the case 40\_198 has a height $k_t^+\approx40$ and wavelength $\lambda^+\approx198$. The other two cases are 63\_313 and 80\_396. 
These three cases were designed to match the pyramids studied in \cite{Schultz09}, herein referred to as SF09. The angle of SF09 was reported as the edge angle, however this should be the angle between the pyramid face and the horizontal (M. Schultz, personal communication). The correct edge angles of $16^\circ$ are given in table \ref{tab:simsB}. This makes the present pyramids steeper than SF09, however their study included pyramids with $\alpha=35^\circ$ and observed no appreciable difference in $\Delta U^+$, suggesting this discrepancy is insignificant. The smallest pyramid also has a similar $k_t^+$ to \cite{DiCicca16}, herein referred to as DCO16, who had $k_t^+=39$. Their pyramid slope angle was $\alpha=28^\circ$ and their pyramids were separated from each other by a small gap of 32 viscous units. However, given that the roughness height is matched, these difference shouldn't result in a substantially different roughness function.
Additionally, the case 63\_313 matches the roughness height of \cite{Hong11} (herein HKS11). The cases that are simulated here, and those available in the literature, are summarised in table \ref{tab:simsB}. The expected full-span skin-friction coefficient, $C_{f,mod}=2U_\tau^2/U_{b,full}^2$, is also given in this table. Due to the altered outer-layer velocity profile of the minimal-span channel, the composite profile of \cite{Nagib08} is fitted to estimate the full-span velocity profile for $z>z_c$.  This therefore allows for an estimate of the full-span bulk velocity $U_{b,full}$ to be obtained. The estimated smooth-wall coefficient value of $C_{f,mod}\approx0.005$ is in good agreement with the empirical fit of \cite{Dean78}, in which $C_{f}=0.073 (Re_b)^{-1/4}=0.0054$ where $Re_b=2hU_b/\nu$ is the bulk Reynolds number. This shows that the minimal-span channel framework can still be used to estimate the full-span skin-friction coefficient, which is commonly used by engineers.

Now that the dimensions of the roughness are known, the channel domain size can be determined.
Following the guidelines of \cite{Chung15}, the spanwise domain width must satisfy $L_y\gtrsim\max(100\nu/U_\tau,k_t/0.4,\lambda_{r,y})$. 
For the smallest pyramids (case 40\_198), we have $L_y^+\gtrsim \max(100, 100,198)$ so it is the large wavelength which necessitates $L_y=\lambda$.  An additional case with the same channel dimensions as 40\_198 is simulated (case 40\_198\_f), but with a finer computational grid to ensure that all the turbulent scales are resolved. This spanwise width of $L_y=\lambda$ means that the critical wall-normal location $z_c=0.4L_y$ for these pyramids is set at $z_c^+=80\approx 2k_t^+$. The roughness sublayer may be larger than $2k_t$, so a channel with a larger width of $L_y=2\lambda$ is also simulated (case 40\_198\_2), so that the critical location is now $z_c=4k_t$. For the larger pyramids (cases 63\_313 and 80\_396), the pyramid wavelength is again the limiting constraint, so $L_y=\lambda$.
 The streamwise domain length was investigated in \S\ref{sect:slength}, where the recommendation was $L_x\gtrsim\max(3L_y,1000\nu/U_\tau,\lambda_{r,x})$. 
 For the smaller pyramids (40\_198), we have $L_x^+\gtrsim(594,1000,198)$ so it is the second constraint that is the limiting one. To ensure we have complete a number of pyramids in the domain, a streamwise length of $L_x = 6\lambda$ is selected which in viscous units is $L_x^+=1188$. Case 63\_313 is also limited by the second constraint.
 For the largest pyramids ($L_y^+=396$), the first constraint ($L_x\gtrsim3L_y$) requires a streamwise length of $L_x=3L_y\Rightarrow L_x^+=1188$.

\setlength{\unitlength}{1cm}
\begin{figure}
\centering
 \captionsetup[subfigure]{labelformat=empty}
	\subfloat[]{
		\includegraphics[width=0.47\textwidth]{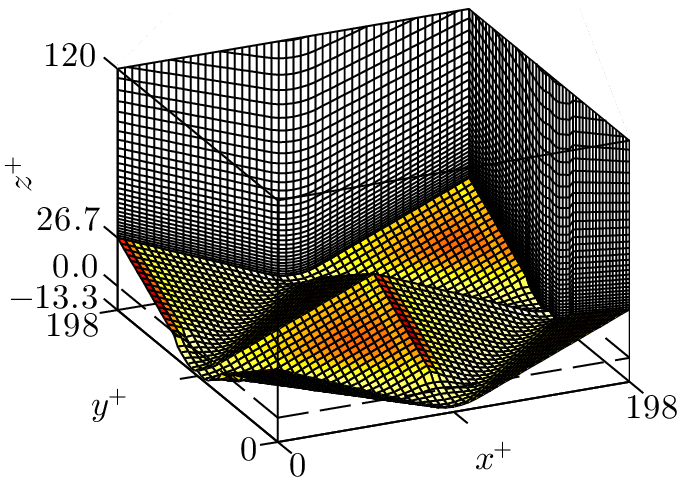}
	}
	\subfloat[]{
		\includegraphics[width=5.75cm,trim = 20 0 0 0]{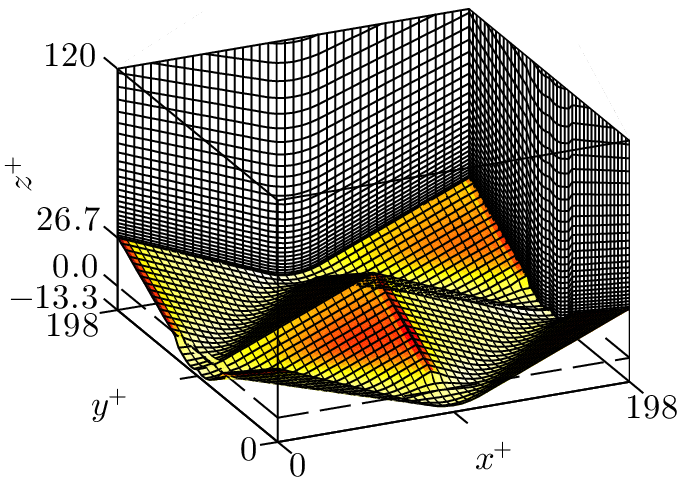}
	}
	\subfloat[]{
		\includegraphics[width=0.98cm,trim = 10 10 0 0]{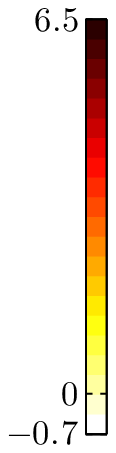}
	}
	\put(-13.7,4.2){(\emph{a})}
	\put(-7.7,4.2){(\emph{b})}
	\vspace{-2.3\baselineskip}
	\caption{(Colour online) Pyramid geometry and body-fitted mesh for pyramid with $k_t^+=40$, $\lambda^+=198$ at $Re_\tau=840$
		for (\emph{a}) normal mesh 40\_198 and (\emph{b}) finer mesh 40\_198\_f.
	Only every (\emph{a}) fifth or (\emph{b}) tenth wall-normal node is shown, and only every second spanwise and streamwise node in (\emph{b}).
Contour shows time-averaged total drag (viscous + pressure) force per unit area.
}
	\label{fig:mesh}
\end{figure}

The length of time to achieve a converged result can now be estimated since the spanwise width $L_y$ and hence critical wall-normal location $z_c=0.4L_y$ have been set. 
For the smallest pyramid case (40\_198), the spanwise width is $L_y^+=198$ so $z_c^+=79$.
If we wish to be 95\% confident that the roughness function is in the range $\Delta U^+\pm 0.1$, then from (\ref{eqn:duerr}) the number of $z_c$-sized eddies that must be observed are
$C^\star = (91.4/(0.1z_c^+))^2 \approx 134$.
We want to use this to determine the simulation runtime $T_{sim}$ through (\ref{eqn:Cstar}), which requires estimating the bursting period $T_b$.
Since the channel is relatively wide and $z_c^+=79$, it's likely to have the beginnings of a logarithmic layer (figure \ref{fig:burstingperiod}\emph{b}), which implies $T_bU_\tau/z_c=6\Rightarrow T_b^+=6z_c^+\approx 475$. 
All that remains is to substitute this value and the channel dimensions into (\ref{eqn:Cstar}) to determine that $T_{sim}^+\approx 3.1\times10^4$. As the nominal streamwise grid spacing is known to be $\Delta x^+\approx 4.2$, and if we assume a centreline velocity of $U_h^+\approx30$, then the time step $\Delta t^+=\Delta x^+/U_h^+\approx 0.14$ for a CFL number of unity at the centreline. Finally, using (\ref{eqn:cpu}) with an assumed wall clock time per step of $\mathcal{T}_1=6.4$ seconds and $N_{CPU}=128$, we expect to use around 52,000 CPU hours. A similar computation for cases 63\_313 ($z_c^+\approx 125$, $C^\star\approx 53$) and 80\_395 ($z_c^+\approx158$, $C^\star\approx33$) predicts 53,000 and 100,000 CPU hours, respectively. For comparison, a single full-span channel ($L_x=2\pi$, $L_y=\pi$, $L_z=2h$ and $T_{sim}U_\tau/h=10$) with pyramids would likely require at least $2\times 10^6$ CPU hours using the current code, more than a full order of magnitude greater than the minimal channel.

The pyramids are aligned so that the trough between neighbouring pyramids is 45$^\circ$ to the mean flow direction. However, because hexahedral cells are used which are aligned with the mean flow direction, the trough line falls across opposite vertices of the cell. This would mean the cell needs to be split into two pentahedrals to fit the roughness geometry. To mitigate this issue, the trough of the pyramids are rounded off and faceted.  The resulting trough has a radius of curvature of approximately half the pyramid crest-to-peak height, $R = 0.53k_t$. The crest of the pyramid retains its sharp edge, as evidenced in figure \ref{fig:mesh}. The total volume of fluid in the channel with pyramids matches that of a smooth-wall channel, which would ensure a collapse of flow statistics in the outer layer, if it existed \citep{Chan15}. In other words, the hydraulic half height (the channel equivalent to the hydraulic radius of a pipe) matches the smooth wall, which implies the virtual origin is at one third the height of the pyramid.

\subsection{Pyramid results}

\setlength{\unitlength}{1cm}
\begin{figure}
\centering
 \captionsetup[subfigure]{labelformat=empty}
	\subfloat[]{
		\includegraphics[width=0.49\textwidth]{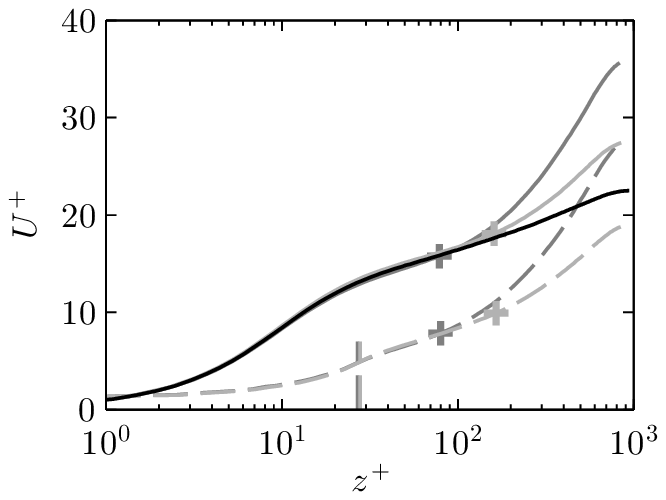}
	}
	\subfloat[]{
		\includegraphics[width=0.49\textwidth]{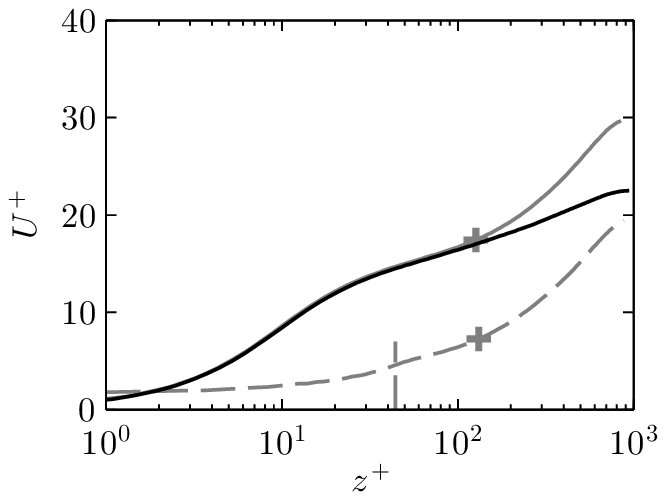}
	}
	\put(-13.5,4.65){(\emph{a})}
	\put(-6.7,4.65){(\emph{b})}
	\put(-9.0,4.2){$k_t^+\approx40$}
	\put(-2.0,4.2){$k_t^+\approx63$}
	\put(-12.45,2.0){\includegraphics{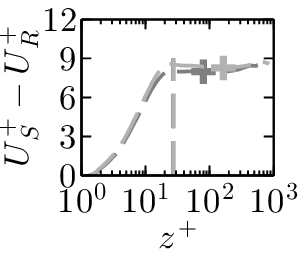}}
	\put(-5.6,2.0){\includegraphics{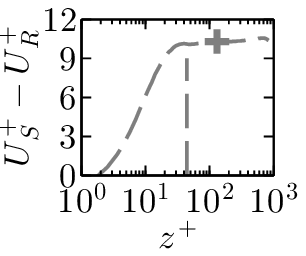}}
	\vspace{-2.6\baselineskip}
	\\
	\subfloat[]{
		\includegraphics[width=0.49\textwidth]{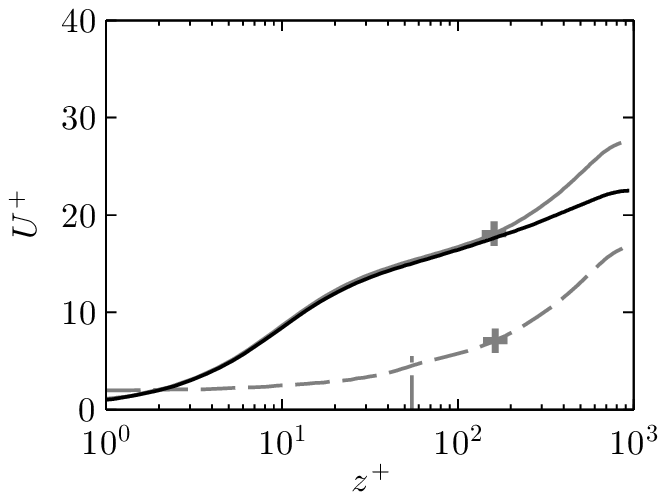}
	}
	\subfloat[]{
		\includegraphics[width=0.49\textwidth]{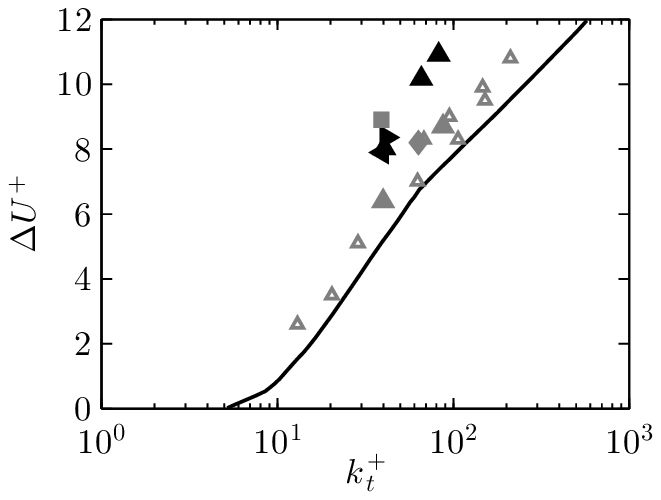}
	}
	\put(-13.5,4.65){(\emph{c})}
	\put(-9.0,4.2){$k_t^+\approx80$}
	\put(-6.7,4.65){(\emph{d})}
	\put(-12.45,2.0){\includegraphics{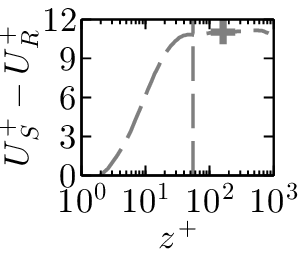}}
	\vspace{-2.1\baselineskip}
	\caption{(\emph{a}--\emph{c}) Mean velocity profile for pyramid roughness. Line styles:
	\protect\raisebox{0.8ex}{\color{myblack}\linethickness{0.5mm}\line(1,0){0.5}} full-span channel, $Re_\tau=950$ \citep{Chung15};
	\protect\raisebox{0.8ex}{\color{LGrey}\linethickness{0.5mm}\line(1,0){0.5}}, minimal span.
	Solid line, smooth wall;
	dashed line, rough wall with pyramid height (\emph{a}) $k_t^+\approx 40$,  (\emph{b}) $k_t^+\approx 63$, (\emph{c}) $k_t^+\approx 80$.
	\protect\raisebox{0.4ex}{\protect\scalebox{1.0}{$\boldsymbol{\pmb{+}}$}} indicate wall-normal critical location $z_c=0.4L_y$.
	Inset shows difference in smooth-wall and pyramid-roughness velocity.
	(\emph{d}) Roughness function against trough-to-peak pyramid height, $k_t^+$.
	Symbols:
	$\color{LGrey}\blacktriangle$, SF09 data in table \ref{tab:simsB};
	$\color{LGrey}\pmb\triangle$, other SF09 data for $\alpha=16^\circ$ pyramids;
	$\color{LGrey}\blacksquare$, DCO16 data;
	$\color{LGrey}\blacklozenge$, HKS11 data;
	\protect\raisebox{0.8ex}{\linethickness{0.5mm}\line(1,0){0.6}}, \cite{Nikuradse33} sandgrain data;
	$\blacktriangle$, current data (40\_198, 63\_313, 80\_396);
	$\blacktriangleright$, 40\_198\_2;
	$\blacktriangleleft$, 40\_198\_f.
	}
	\label{fig:pyramidUx}
\end{figure}

Figure \ref{fig:pyramidUx}(\emph{a}--\emph{c}) shows the mean velocity profile for the pyramid roughness data. The insets show the difference in velocity between smooth-wall and pyramid-roughness channels, and it can be seen to be relatively constant from the pyramid crest to the channel centre. This indicates that the velocity profile for the smooth-wall and rough-wall are matched from the crest (apart from the offset $\Delta U^+$), or that the roughness sublayer for pyramids is small. The roughness function is obtained by averaging the difference in smooth- and rough-wall velocities over $k_t^++5<z<z_c^+$.

 For the smaller pyramids ($k_t^+=40$) in a narrow channel, we obtained a roughness function of $\Delta U^+=8.0$ which agrees well with the refined mesh ($\Delta U^+=7.9$). This indicates that both meshes are resolving all the relevant scales in the roughness sublayer. When the channel is widened to include two repeating pyramids in the spanwise direction (40\_198\_2), we achieve a similar result of  $\Delta U^+=8.3$. While slightly larger than the previous two values, this difference is predominantly due to the increase in $Re_\tau$ for this channel. The roughness Reynolds number, $k_t^+\approx41.8$, is almost 5\% over the target value of $k_t^+=40$, which means the roughness function would also be larger. The otherwise good agreement between this case and the narrower case 40\_198 supports the conclusion that the roughness sublayer is less than $2k_t$ in height above the pyramid crest. If the roughness sublayer was larger than $2k_t$, then the narrower minimal channel with $z_c=2k_t$ could not fully capture the roughness effects and hence would have a different roughness function to the wider minimal channel.  Moreover, this agreement with case 40\_198, when $L_y=\lambda_y$, suggests that the interaction of flow structures generated between neighbouring roughness elements is small, as this case only has a single repeating element in the spanwise direction.  This agrees with our previous study \citep{MacDonald16}, which compared a three-dimensional sinusoidal roughness in a full-span channel and a minimal-span channel
  in which there is only one repeating roughness element in the spanwise direction. It would only be when the spanwise roughness wavelength was very large, or infinite, as in the case of two-dimensional spanwise bar roughness and $d$-type surfaces that these spanwise effects would become significant. However, this would require further study and is outside the scope of this paper.
 The above three values of the roughness function can be compared with values of $6.4\pm10\%$ from SF09 and $8.9$ from DCO16. The current data therefore falls between these two sources (figure \ref{fig:pyramidUx}\emph{d}).

For the case with $k_t^+\approx63$, the roughness function was found to be $\Delta U^+=10.2$, which is larger than $7.0\pm10\%$ from SF09 and 8.2 from HKS11.
For the largest pyramids, where $k_t^+\approx80$, the roughness function was found to be  $\Delta U^+=11.0$, which is again larger than the value of $8.7\pm10\%$ from SF09. As a result, the equivalent sandgrain roughness appears to be $k_s\approx 3.5 k_t$, compared to $k_s\approx1.5 k_t$ from SF09. Note that DCO11 would predict an even larger equivalent sandgrain roughness than $k_s\approx3.5k_t$, given their larger $\Delta U^+$ at $k_t^+\approx40$.

\setlength{\unitlength}{1cm}
\begin{figure}
\centering
 \captionsetup[subfigure]{labelformat=empty}
	\subfloat[]{
		\includegraphics[width=0.49\textwidth]{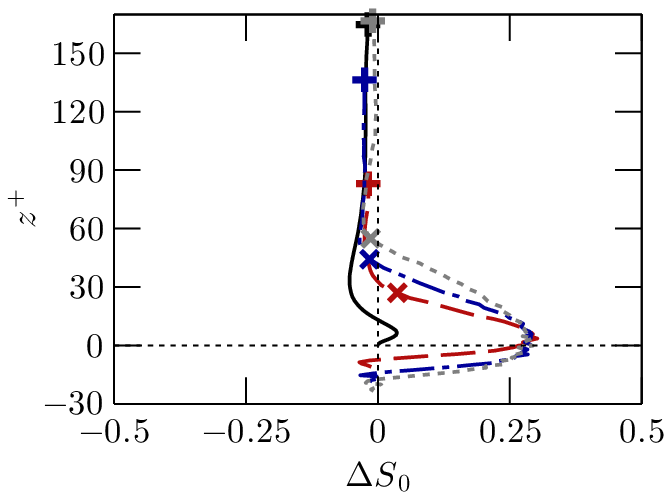}
	}
	\subfloat[]{
		\includegraphics[width=0.49\textwidth]{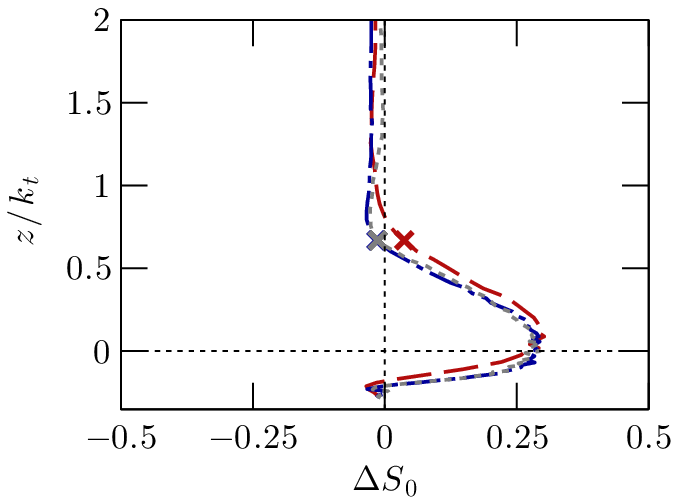}
	}
	\put(-13.6,4.65){(\emph{a})}
	\put(-6.7,4.65){(\emph{b})}
	\put(-9.3,1.7){\vector(0,1){0.75}}
	\put(-9.7,2.5){Inc. $k_t^+$}
	\vspace{-1.5\baselineskip}
	\caption{(Colour online) Difference $\Delta S_0$ between sweep and ejection stress contributions \citep{Raupach81} as a function of (\emph{a}) $z^+$ and (\emph{b}) $z/k$.
	Line styles:
	\protect\raisebox{0.8ex}{\color{myblack}\linethickness{0.5mm}\line(1,0){0.6}},  smooth wall (sm\_396);
	\protect\raisebox{0.8ex}{\color{myred}\linethickness{0.5mm}\line(1,0){0.3}\hspace{0.15cm}\line(1,0){0.3}},  $k_t^+\approx40$ (40\_198);
	\protect\raisebox{0.8ex}{\color{myblue}\linethickness{0.5mm}\line(1,0){0.08}\hspace{0.15cm}\line(1,0){0.3}\hspace{0.15cm}\line(1,0){0.08}}, $k_t^+\approx63$;
	\protect\raisebox{0.8ex}{\color{LGrey}\linethickness{0.5mm}\line(1,0){0.1}\hspace{0.15cm}\line(1,0){0.1}\hspace{0.15cm}\line(1,0){0.1}}, $k_t^+\approx80$.
	Symbols:
	\protect\raisebox{0.4ex}{\protect\scalebox{1.0}{$\boldsymbol{\pmb{+}}$}}, $z_c^+$;
	\protect\raisebox{0.4ex}{\protect\scalebox{1.0}{$\boldsymbol{\pmb{\times}}$}},  pyramid crest.
	}
	\label{fig:sweepEjections}
\end{figure}

To show some of the higher order statistics that can be obtained from these simulations, the difference between sweep and ejection Reynolds stress contributions are shown in figure \ref{fig:sweepEjections}. This difference, $\Delta S_0=(\{u'w'\}_{4,0} - \{u'w'\}_{2,0})/\langle \overline{u'w'}\rangle$, was developed in \cite{Raupach81} in which the Reynolds shear stress is conditionally averaged on sweeps ($u'>0$, $w'<0$, quadrant 4) and ejections ($u'<0$, $w'>0$, quadrant 2), where $\{\cdot\}$ denotes the conditional average. The hyperbolic hole region $H=0$. From this statistic, it can be seen in figure \ref{fig:sweepEjections}(\emph{a}) that sweeps  are the dominant means of momentum transfer within the roughness canopy, in agreement with \cite{Raupach81}. These sweeps are much more dominant than in a smooth wall, which has a slight preference for sweeps very close to the the wall ($z^+\lesssim12$), with ejections dominant above. Above the roughness crest, both rough-wall and smooth-wall flows collapse with ejections being slightly larger than sweeps. The outer-layer is associated with ejections being dominant, with $\Delta S_0$ tending to $-1$ at $z=h$. However, in the minimal-span channel this statistic stays approximately constant above $z>z_c$, with $\Delta S_0$ being close to zero. If instead the wall-normal coordinate is normalised on $k_t$ (figure \ref{fig:sweepEjections}\emph{b}) then we see that all three rough-wall flows nearly collapse, although  the case with the smallest height ($k_t^+\approx40$, red dashed line) there is a slight difference, with  sweeps remaining dominant at the roughness crest.

\setlength{\unitlength}{1cm}
\begin{figure}
\centering
 \captionsetup[subfigure]{labelformat=empty}
	\subfloat[]{
		\includegraphics[width=0.486\textwidth]{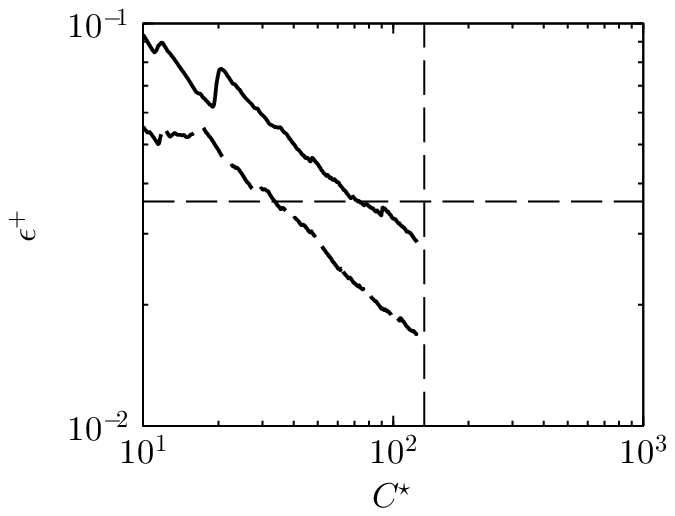}
		\label{fig:cdpcstar1}
	}
	\subfloat[]{
		\includegraphics[width=0.486\textwidth]{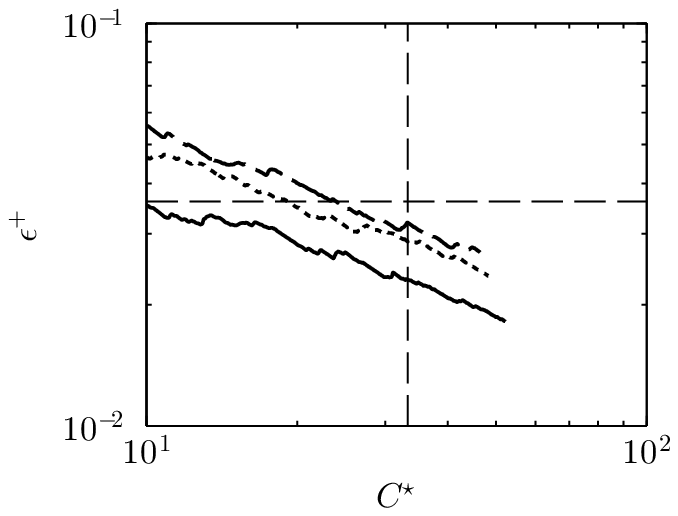}
		\label{fig:cdpcstar2}
	}
	\put(-13.5,4.7){(\emph{a})}
	\put(-6.75,4.7){(\emph{b})}
	\vspace{-2.3\baselineskip}
	\caption{
	Standard error of the velocity  as a function of the number of $z_c$-sized eddies eddies $C^\star$ (\ref{eqn:Cstar}) for
(\emph{a}) $z_c^+\approx 79$, and
(\emph{b}) $z_c^+\approx 158$.
Line styles:
\mbox{\protect\raisebox{0.8ex}{\linethickness{0.5mm}\line(1,0){0.5}}, smooth wall};
\mbox{\protect\raisebox{0.8ex}{\linethickness{0.5mm}\line(1,0){0.25}\hspace{0.15cm}\line(1,0){0.25}}, $k_t^+\approx40$ pyramid};
\mbox{\protect\raisebox{0.8ex}{\linethickness{0.5mm}\line(1,0){0.1}\hspace{0.15cm}\line(1,0){0.1}\hspace{0.15cm}\line(1,0){0.1}}, $k_t^+\approx80$ pyramid}.
Horizontal dashed line shows $\epsilon^+=0.1/2.77=0.036$, the desired tolerance level,
vertical dashed line shows  $C^\star=(91.4/(0.1z_c^+))^2$, the expected $C^\star$ value required to obtain this tolerance level (\ref{eqn:duerr}).
}
	\label{fig:cdpcstar}
\end{figure}

Figure \ref{fig:cdpcstar} shows the statistical uncertainty of $U^+(z=z_c)$ as a function of $C^\star$, the number of $z_c$-sized eddies. This is the same format as figure \ref{fig:convzc}(\emph{a}--\emph{c}). Importantly, the predicted number of $z_c$-sized eddies that need to be observed to obtain an error tolerance of $\Delta U^+\pm0.1$ using (\ref{eqn:duerr}) is indicated by the vertical dashed lines. The current data for both smooth and rough walls show reasonable agreement with this prediction, albeit the prediction tending to overestimate the necessary $C^\star$ value. While this means an overestimated number of CPU hours from the number given in \S\ref{sect:compsetup}, this is more desirable than underestimating the required number of CPU hours.
  It is worthwhile mentioning that  the current data is for a different $Re_\tau$ and roughness geometry then what the predicted $C^\star$ empirical relation was developed with, showing a certain robustness to the method.

The roughness function values of $8\lesssim\Delta U^+\lesssim11$ measured here place these pyramid cases in the fully rough regime, while the results and recommendations from the previous sections (\S\ref{sect:res}--\ref{sect:cost}) were conducted in the transitionally rough regime. Given that the pyramid cases were set up using these recommendations without any modifications
 and achieved reasonable agreement with the literature, this suggests that the results of this study can apply to both transitionally and fully rough flows.

%
%

\section{Conclusions}
A series of numerical experiments were conducted to investigate the fundamental dynamics of the minimal channel, with an emphasis on more efficiently simulating rough-wall flows in order to obtain the roughness function $\Delta U^+$. These experiments were performed using a finite difference code.

The streamwise length was investigated in \S\ref{sect:slength} for a minimal channel. 
It appears that the largest log-layer eddy scales as $L_x=3L_y$, where the spanwise width is obtained from the guidelines in \cite{Chung15}, $L_y\gtrsim\max(100\nu/U_\tau,k/0.4,\lambda_{r,y})$.
However, for very narrow channels this implies a very short streamwise domain length, which may not be able to sustain the turbulent motions. The small spatial domain also makes obtaining converged statistics a very time-consuming process. As such, minimal channels would benefit from a minimum streamwise domain length of around 1000 viscous units and so a guideline for setting $L_x$ is $L_x\gtrsim\max(3L_y,1000\nu/U_\tau,\lambda_{r,x})$, where $\lambda_{r,x}$ would be some characteristic streamwise length scale of the roughness.

Two alterations to the outer-layer flow were investigated to see whether they affected the healthy near-wall flow.
A half-height (open) channel was used in \S\ref{sect:half} which a slip-wall is positioned at the channel centre. This resulted in only a slight change to the mean velocity profile and turbulence intensities in both minimal and full-span channels, primarily affecting the wake region. 
The pressure fluctuations in the minimal channel were shown to be nearly an order of magnitude larger than in a full-span channel, however this was suggested to be as a result of the rapid pressure term which is related to the mean velocity gradient, $dU/dz$. When $dU/dz$ was instead artificially set to zero in the unphysical region above $z_c$, the pressure fluctuations from this altered velocity field were then in much better agreement with the full-span channel.
A benefit of using a half-height channel is that only half the number of gridpoints are required, at the cost of running the simulation for twice as long to reach the same level of statistical convergence. It is up to the user to balance this trade off between memory and time.

A forcing model was applied to the outer layer  in \S\ref{sect:outerForcing} to damp the fluctuations and reduce the centreline velocity. The near-wall flow was unaltered even though the Reynolds shear stress is zero above the location where the damping starts, $z_d$. This is similar to the filtering employed by \cite{Jimenez99}, although the present damping model forces $U(z>z_d)=U(z=z_d)$.
This allowed for an improvement in the computational time step of around 20--24\%, although this did not extend to higher friction Reynolds numbers in the present simulations due to the grid using a cosine mapping in the wall-normal direction.
The present data suggest that the location where the forcing starts should be $z_d\gtrsim \max(200\nu/U_\tau,2z_c)$. The first constraint is present for very narrow channels where $z_c^+$ is close to the wall, as the forcing will interfere with the near-wall flow otherwise.

A temporal sweep was conducted  in \S\ref{sect:sweepres} to see if the full roughness behaviour could be obtained by varying the roughness Reynolds number with time. A simple linear variation in the bulk velocity was attempted, with three different rates of change investigated. It was shown that the two slowest rates of change, with final pressure gradient parameter values of $\varDelta_{p,Re_\tau=180}=0.03$ and 0.07 could reasonably predict some of the $\Delta U^+$ vs $k_s^+$ curve. The largest value of  $\varDelta_p=0.15$ resulted in substantial flow changes compared to the steady flow and so are not feasible. The early work of \cite{Perry63} showed that  $\Delta U^+$ was independent of an applied pressure gradient, and the present work goes on to provide a bound on $\varDelta_p$ for when this holds.
 This also suggests that pressure gradients, normally applied by a spatial acceleration of the flow in experimental facilities, can equally be simulated by a temporal acceleration. This is similar to the approach of \cite{Kozul16} who represented a spatially developing boundary layer as temporally developing instead.

An eddy-counting argument was developed in \S\ref{sect:cost} to analyse the statistical uncertainties of numerical simulations in minimal channels. The total number of $z_c$-sized eddies, $C^\star$, are counted over the duration of the simulation, which involves estimating their characteristic length and time scales. 
For a desired level of uncertainty, $\zeta$, in the roughness function, $\Delta U^+\pm \zeta$, the expected number of $z_c$-sized eddies that need to exist over the course of the simulation was estimated as $C^\star=(91.4/(\zeta\cdot z_c^+))^{2}$ as in (\ref{eqn:duerr}).
This means that for a known $z_c^+$, the user can determine how many $z_c$-sized eddies need to be observed through this empirical relation. The simulation run time, and hence number of CPU hours required, can then be estimated \emph{a priori}. This would enable researchers to determine beforehand how best to allocate a limited number of CPU hours when studying rough-wall flows.

A case study of square-based pyramids (\S\ref{sect:pyramids}) was used to illustrate the minimal-span channel framework and the insights gained in this paper. The viscous dimensions of the pyramids were set to match those of \cite{Schultz09}. The roughness functions from these simulations were overestimated compared to those of \cite{Schultz09}, although underestimated compared to that of  \cite{DiCicca16} who had a similar roughness geometry. The predicted $C^\star$ value required to obtain a desired level of statistical uncertainty was shown to be in reasonable agreement, if not somewhat conservative, with the data from the pyramid simulations. As such, the estimated CPU hours required for the simulation can be predicted reasonably accurately before performing the simulation. For this pyramid roughness, the minimal-span channel used nearly 20 times less CPU hours than the estimated cost of a full-span channel, as well as using substantially fewer CPUs.

\section*{Acknowledgements}
This work was partly funded through the Multiflow program by the European Research Council.
Computational time was granted under the Victoria Life Sciences Computational Initiative, which is supported by the Victorian Government, Australia.

\bibliographystyle{jfm}
\bibliography{bibliography}

\end{document}